\documentclass[10pt]{ijnam}
\def\currentvolume{1}
\def\currentissue{1}
\def\currentyear{2004}
\def\month{March}
\pagespan{1}{18}
\copyrightinfo{2004}{} 

\usepackage{amssymb,amsmath}
\usepackage{graphicx}
\usepackage{subfigure}
\usepackage{float}

\begin{document}

\title[Approximate deconvolution large eddy simulations]
{A posteriori analysis of low-pass spatial filters for approximate deconvolution large eddy simulations of homogeneous incompressible flows}

\author[O. San]{Omer San}
\address{
  Interdisciplinary Center for Applied Mathematics,
  Virginia Tech,
  Blacksburg, VA 24061, USA
}
\email{omersan@vt.edu}
\urladdr{http://vt.academia.edu/OmerSan}

\author[A.~E. Staples]{Anne E. Staples}
\address{
  Department of Engineering Science and Mechanics,
  Virginia Tech,
  Blacksburg, VA 24061, USA
}
\email{aestaples@vt.edu}
\urladdr{http://www.esm.vt.edu/people/active/staplesa/staplesa-bio.html}

\author[T. Iliescu]{Traian Iliescu}
\address{
  Department of Mathematics,
  Virginia Tech,
  Blacksburg, VA 24061, USA
}
\email{iliescu@vt.edu}
\urladdr{http://www.math.vt.edu/people/iliescu/}




\date{June 1, 2013}



\abstract{
The goal of this paper is twofold: first, it investigates the effect of low-pass spatial filters for approximate deconvolution large eddy simulation (AD-LES) of turbulent incompressible flows. Second, it proposes the hyper-differential filter as a means of increasing the accuracy of the AD-LES model without increasing the computational cost. Box filters, Pad\'{e} filters, and differential filters with a wide range of parameters are studied in the AD-LES framework. The AD-LES model, in conjunction with these spatial filters, is tested in the numerical simulation of the three-dimensional Taylor-Green vortex problem. The numerical results are benchmarked against direct numerical simulation (DNS) data. An under-resolved numerical simulation is also used for comparison purposes. Four criteria are used to investigate the AD-LES model equipped with these spatial filters:
(i) the time series of the volume-averaged enstrophy;
(ii) the volume-averaged third-order structure function;
(iii) the $L^2$-norm of the velocity and vorticity errors; and
(iv) the volume-averaged velocity and vorticity correlation coefficients.
According to these criteria, the numerical results yield the following two conclusions: first, the AD-LES model equipped with any of these spatial filters yields accurate results at a fraction of the computational cost of DNS. Second, the most accurate results are obtained with the hyper-differential filter, followed by the differential filter. We demonstrate that the results highly depend on the selection of the filtering procedure. Although a careful parameter choice makes each class of filters used in this study competitive, it seems that filters whose transfer function resembles that of the Fourier cut-off filter (such as the hyper-differential filters) tend to perform best.
}


\maketitle

\section{Introduction}
\label{sec:intro}

Large eddy simulation (LES) is a widespread approach to performing accurate, relatively low cost computations of complex turbulent flows \cite{lesieur1996new,meneveau2000scale,bouffanais2010advances}. In this approach, a low pass filter is applied to the governing equations, separating the large, resolved scales from the unresolved, subfilter level scales. One of the main challenges in LES is the celebrated closure problem, which aims at modeling the subfilter-scale (SFS) stress tensor.
This tensor appears in LES as a result of the nonlinearity in the Navier-Stokes equations (NSE).
The SFS stress tensor depends on both the filtered and unfiltered flow variables.
Thus, to derive a practical LES model, one needs to address the closure problem, i.e., to approximate the SFS stress tensor in terms of the filtered flow variables only.

One of the recent closure modeling strategies used in LES is the {approximate deconvolution (AD)}.
The AD closure model was introduced in LES by Stolz and Adams  \cite{stolz1999approximate}.
The AD model uses repeated filtering on the available filtered flow variables to derive computable models for the SFS stress tensor.
The mathematical approach used in the derivation of the AD model distinguishes it from the standard closure modeling strategies employed in LES, such as eddy viscosity, which rely on phenomenological arguments.
The AD model has been used successfully in LES of three-dimensional turbulent engineering flows \cite{stolz2001approximate,stolz2001approximatec,stolz2004approximatei,schlatter2004transitional,domaradzki2002direct,habisreutinger2007coupled}, small scale geophysical flows, such as the atmospheric boundary layer \cite{chow2005explicit,chow2009evaluation,duan2010bridging,zhou2011large}, and large scale ocean circulation problems \cite{san2011approximate}.
The mathematical nature of the AD closure model has allowed the development of a strong mathematical support for the AD-LES model, including both well-posedness results and error analysis for the numerical discretization \cite{dunca2006stolz,layton2006residual,layton2007similarity,rebholz2007conservation,stanculescu2008existence,dunca2011existence}.
The applicability of AD method for computing the SFS stress tensor has been recently highlighted by Germano \cite{germano2009new} as well.

The need for low-pass filters arises in the numerical solution of partial differential equations. The ability to control the high-frequency content is an essential ingredient for many applications in computational fluid dynamics \cite{aldama1990filtering,raymond1994diffusion,mullen1999filtering}.
Low pass filters play a central role in LES, where they are used to define the large scales.
The most popular spatial filters used are the sharp cut-off, the Gaussian, and the box (or top-hat) filters.
The sharp cut-off filters operate in the spectral domain, whereas the box and the Gaussian filters are applied in the physical space.
The majority of LES calculations have employed sharp cut-off filters and Fourier spectral discretizations.
The remaining LES calculations have generally utilized box filters and finite difference discretizations.
A natural question is whether the particular spatial filter used has any effect on the LES results \cite{jordan1996large}.
A comparison of various discrete test filters within a finite difference discretization has been performed by Najjar and Tafti \cite{najjar1996study} using the dynamic subgrid-stress scale model.
Similar studies were performed using several other classes of filters for large eddy simulations of turbulent flows \cite{schumann1975subgrid,vasilyev1998general,sagaut1999discrete,brandt2006priori}.

In the AD-LES framework, the role of the spatial filter is central, since the SFS tensor is computed by the repeated use of the spatial filter.
The main goal of this study is to investigate the effect of the low-pass spatial filters in the AD-LES of turbulent incompressible flows.
Three classes of spatial filters are studied in conjunction with the AD-LES model.
The first class of filters consists of discrete representations \cite{schumann1975subgrid,jordan1996large,najjar1996study} based on the box filters with trapezoidal and Simpson's integration rules, and high-order constructions.
These filters are local and computationally efficient.
The second class of spatial filters that we examine comprises the Pad\'{e}-type low pass filters \cite{pruett2000priori,lele1992compact,visbal2002use} with various orders of accuracy.
The third class of spatial filter that we investigate consists of the Helmholtz-type elliptic differential filters \cite{germano1986differential,germano1986differential_2,mullen1999filtering}.
To our knowledge, this represents the first numerical investigation of the role of spatial filtering in AD-LES.
The second goal of this report is to propose the use of the hyper-differential filter as a means of increasing the physical accuracy of the AD-LES model without increasing its computational cost.

The AD-LES model, equipped with these spatial filters, is tested in the numerical simulation of the three-dimensional Taylor-Green vortex problem.
The Taylor-Green vortex problem is one of the simplest systems for the study of turbulent flows and has been extensively used to investigate the behavior of LES models for homogeneous flows \cite{berselli2005large,drikakis2007simulation,adams2007approximate}.
The numerical discretization employed throughout the paper is based on a vorticity-velocity formulation of the NSE \cite{fasel1976investigation,dennis1979finite,fasel1990numerical,quartapelle1993numerical,fasel2002numerical}.
This formulation is discretized in space using a finite difference discretization and in time using the total variation diminishing Runge-Kutta scheme.
The numerical results are benchmarked against a direct numerical simulation (DNS).
An under-resolved numerical simulation, denoted No-AD in the rest of the paper, is also used for comparison purposes.
The following four criteria are used to investigate the AD-LES model equipped with the spatial filters described above:
(i) the time series of the volume-averaged enstrophy;
(ii) the volume-averaged third-order structure function;
(iii) the $L^2$-norm of the velocity and vorticity errors; and
(iv) the volume-averaged velocity and vorticity correlation coefficients.

This paper is organized as follows. Section \ref{sec:ge} presents the governing equations for the incompressible flows. Section \ref{sec:ad} describes the AD methodology, and various low pass spatial filters and their transfer functions are given in Section \ref{sec:spatial_filters}. The temporal and spatial discretizations are briefly discussed in Section \ref{sec:numerics}. The Taylor-Green vortex problem, a benchmark test case for homogeneous isotropic flows, is introduced in Section \ref{sec:test_case}. The results of the AD-LES method are presented in Section \ref{sec:results}. Finally, the conclusions are summarized in Section \ref{sec:sum}.

\section{Governing Equations}
\label{sec:ge}
The dimensionless form of the NSE, which govern the incompressible viscous flows, is written as:
\begin{eqnarray}
\frac{\partial \boldsymbol u}{\partial t}+ \boldsymbol u \cdot \nabla \boldsymbol u &=&-\nabla \boldsymbol p + \frac{1}{Re}\nabla^2 \boldsymbol u
\label{eq:mom} \\
\nabla \cdot \boldsymbol u&=&0 ,
\label{eq:cont}
\end{eqnarray}
where $Re$ is the Reynolds number, $\boldsymbol u = (u, v, w)$ is the velocity vector, and $p$ is the pressure.
The NSE \eqref{eq:mom}-\eqref{eq:cont} have to be supplemented with appropriate boundary conditions and initial conditions.
In this study, we exclusively consider periodic boundary conditions.
This choice allows us to focus on the effect of the spatial filter on the SFS tensor, eliminating the potential complications introduced by the boundary conditions.
The initial conditions for the NSE \eqref{eq:mom}-\eqref{eq:cont} are specified in Section \ref{sec:test_case}.
In addition to the standard primitive variable formulation of the NSE  given in \eqref{eq:mom}-\eqref{eq:cont}, several alternative formulations are used in parctice.
For an overview of these alternative formulations, the reader is referred to the exquisite presentation given by Quartapelle \cite{quartapelle1993numerical}.
In this study, we employ the vorticity-velocity formulation of the NSE \cite{fasel1976investigation,dennis1979finite,fasel1990numerical,quartapelle1993numerical,fasel2002numerical}.
Next, we briefly describe this formulation; more details are given in Chapter 4 in Quartapelle \cite{quartapelle1993numerical}.

Taking the curl of \eqref{eq:mom}, one obtains the vorticity equation
\begin{equation}
\frac{\partial \boldsymbol \omega}{\partial t} + \boldsymbol u \cdot \nabla \boldsymbol \omega = \boldsymbol \omega \cdot \nabla \boldsymbol u + \frac{1}{Re} \nabla^2 \boldsymbol\omega ,
\label{eq:vv}
\end{equation}
where the vorticity field is defined as the curl of velocity field, $\boldsymbol \omega = \nabla \times \boldsymbol u$.
Next, taking the curl of the equation $\boldsymbol \omega = \nabla \times \boldsymbol u$ and using \eqref{eq:cont} yields the following Poisson equation for the velocity:
\begin{eqnarray}
- \nabla^2 \boldsymbol u = \nabla \times \boldsymbol \omega .
\label{eq:poisson_velocity}
\end{eqnarray}
Componentwise, the vector equation \eqref{eq:poisson_velocity} can also be written as \cite{fasel2002numerical}
\begin{eqnarray}
\frac{\partial^2 v}{\partial x^2} + \frac{\partial^2 v}{\partial y^2}  + \frac{\partial^2 v}{\partial z^2} &=&  \frac{\partial \omega_z}{\partial x} -\frac{\partial \omega_x}{\partial z} \label{eq:es1}\\
\frac{\partial^2 u}{\partial x^2} + \frac{\partial^2 u}{\partial z^2} &=&  \frac{\partial \omega_y}{\partial z} -\frac{\partial^2 v}{\partial x \partial y} \label{eq:es2}\\
\frac{\partial^2 w}{\partial x^2} + \frac{\partial^2 w}{\partial z^2} &=& -\frac{\partial \omega_y}{\partial x} -\frac{\partial^2 v}{\partial y \partial z} . \label{eq:es3}
\label{eq:psys}
\end{eqnarray}
Equations \eqref{eq:vv} and \eqref{eq:poisson_velocity} represent the vorticity-velocity formulation of the NSE.
In Theorem 4.2 in Quartapelle \cite{quartapelle1993numerical} it is shown that the primitive formulation of the NSE given in \eqref{eq:mom} and \eqref{eq:cont} and the vorticity-velocity formulation of the NSE given in \eqref{eq:vv} and \eqref{eq:poisson_velocity} are equivalent.
Both the advantages and disadvantages of the vorticity-velocity formulation over other formulations of the NSE are carefully discussed in Chapter 4 in Quartapelle \cite{quartapelle1993numerical}.
All the theoretical and computational developments in this report are presented for the vorticity-velocity formulation \eqref{eq:vv} and \eqref{eq:poisson_velocity}, which has been also used in LES for three-dimensional incompressible flows \cite{mansfield1998dynamic,tenaud2005large,cocle2009scale}.
We emphasize, however, that all these developments could equally well be presented for any other NSE formulation, such as the primitive variable one in \eqref{eq:mom} and \eqref{eq:cont}.

\section{Approximate Deconvolution Method}
\label{sec:ad}
To derive the equations for the filtered flow variables, (\ref{eq:vv}) and \eqref{eq:poisson_velocity} are first filtered with a low pass filter (to be specified later). Thus, using a bar to denote the filtered quantities, the filtered equations read:
\begin{eqnarray}
\frac{\partial \overline{\boldsymbol \omega}}{\partial t} + \overline{\boldsymbol u \cdot \nabla \boldsymbol \omega} = \overline{\boldsymbol \omega \cdot \nabla \boldsymbol u} + \frac{1}{Re} \nabla^2 \overline{\boldsymbol \omega}
\label{eq:fv} \\
- \Delta \overline{\boldsymbol u}
= \nabla \times \overline{\boldsymbol \omega} .
\label{eq:f_poisson_velocity}
\end{eqnarray}
The nonlinear equation \eqref{eq:fv} can also be written as
\begin{equation}
\frac{\partial \overline{\boldsymbol \omega}}{\partial t} + \overline{\boldsymbol u} \cdot \nabla \overline{\boldsymbol \omega} = \overline{\boldsymbol \omega} \cdot \nabla \overline{\boldsymbol u} + \frac{1}{Re} \nabla^2 \overline{\boldsymbol \omega} + \textbf{S} ,
\label{eq:fvv}
\end{equation}
where $\textbf{S}$ is the subfilter-scale term, given by
\begin{equation}
\textbf{S} =  \overline{\boldsymbol u} \cdot \nabla \overline{\boldsymbol \omega} - \overline{\boldsymbol u \cdot \nabla \boldsymbol \omega} - \overline{\boldsymbol \omega} \cdot \nabla \overline{\boldsymbol u} - \overline{\boldsymbol \omega \cdot \nabla \boldsymbol u} .
\label{eq:sfs}
\end{equation}
It is precisely at this point in the LES model derivation that the celebrated closure problem must be addressed. In order to close the filtered \eqref{eq:fvv}, the subfilter-scale term $\textbf{S}$ in \eqref{eq:sfs} needs to be modeled in terms of the filtered flow variables, $\overline{\boldsymbol \omega}$ and $\overline{\boldsymbol u}$.

The goal in AD is to use repeated filtering in order to obtain approximations of the unfiltered unresolved flow variables when approximations of the filtered resolved flow variables are available.
These approximations of the unfiltered flow variables are then used in the SFS tensor to close the LES system.
To derive the new AD model, we start by denoting by $G$ the spatial filtering operator: $Gf=\bar{f}$, $G\bar{f}=\bar{\bar{f}}$ and so on, where $f$ represents any flow variable (i.e., vorticity or velocity components in this study) and a bar denotes the application of one filtering operation.
Since $G=I-(I-G)$, an inverse to $G$ can be written formally as the non-convergent Neumann series:
\begin{equation}
G^{-1} \sim \sum_{i=0}^{\infty}(I-G)^i .
\label{eq:2}
\end{equation}
Truncating the series gives the Van Cittert approximate deconvolution operator, $Q_N$ \cite{bertero1998introduction,layton2012approximate}.
We truncate the series at $N$ and obtain $Q_N$ as an approximation of $G^{-1}$:
\begin{equation}
Q_N = \sum_{i=1}^{N}(I-G)^{i-1} ,
\label{eq:3}
\end{equation}
where $I$ is the identity operator.
The approximations $Q_N$ are not convergent as $N$ goes to infinity, but rather are asymptotic as the filter radius, $\Delta$, approaches zero \cite{berselli2006mathematics}.
An approximate deconvolution of any variable $f$ can now be obtained as follows:
\begin{equation}
f^*= Q_N f ,
\label{eq:4}
\end{equation}
where an asterisk represents the approximated value for the unfiltered (unresolved) quantities.
For higher values of $N$, we get increasingly more accurate approximations of $f$:
\begin{eqnarray}
Q_1 &=& I \\
Q_2 &=& 2I -G \\
Q_3 &=& 3I-3G + G^2 \\
Q_4 &=& 4I-6G + 4G^2 -G^3 \\
Q_5 &=& 5I-10G + 10G^2 - 5G^3 + G^4 \\
\vdots \nonumber
\label{eq:q5}
\end{eqnarray}
Following the same approach as that used by Dunca and Epshteyn \cite{dunca2006stolz}, one can prove that these
models are highly accurate and stable. Error estimates and convergence studies of approximate deconvolution approach have been recently investigated in \cite{dunca2012error,berselli2012convergence}.
For example, if we choose $N=5$, we can find an AD approximation of the resolved variable $f$ as
\begin{equation}
f \approx f^*=5 \bar{f} - 10 \bar{\bar{f}} + 10 \bar{\bar{\bar{f}}} -5 \bar{\bar{\bar{\bar{f}}}} + \bar{\bar{\bar{\bar{\bar{f}}}}} \ .
\label{eq:6}
\end{equation}
Using \eqref{eq:6}, we can now approximate the SFS tensor \eqref{eq:sfs} by applying a filter to each flow variable:
\begin{equation}
\textbf{S} =  \overline{\boldsymbol u} \cdot \nabla \overline{\boldsymbol \omega} - \overline{\boldsymbol \omega} \cdot \nabla \overline{\boldsymbol u} - \overline{\boldsymbol u^{*} \cdot \nabla \boldsymbol \omega^{*}}  - \overline{\boldsymbol \omega^{*} \cdot \nabla \boldsymbol u^{*}}
\label{eq:ss}
\end{equation}
The {\it AD-LES model} that we investigate in this study consists of \eqref{eq:fv}, \eqref{eq:f_poisson_velocity}, and \eqref{eq:ss}.
To completely specify the AD-LES model, we need to choose a computationally efficient filtering operator.

\section{Spatial Filters}
\label{sec:spatial_filters}
In this section, we describe the three classes of spatial filters that we use in our LES investigation: box filters (Section \ref{sec:dbf}), Pad\'{e}-type filters (Section \ref{sec:pf}), and differential filters (Section \ref{sec:diff}).
For each type of filter, we study its transfer function in the wavenumber space and compare it to the transfer function of the Fourier cut-off filter.
A particular emphasis is placed on how much each filter attenuates the high and low wavenumber components of the function being filtered.
For each class of filters, a wide range of parameters is considered.

\subsection{Box filters}
\label{sec:dbf}
The box filter, which is also known as the top-hat filter, is commonly used in finite difference discretizations of LES models (see, e.g., Balaras et al. \cite{balaras1995finite}, Jordan and Ragab \cite{jordan1996large}, and Najjar and Tafti \cite{najjar1996study}).

Formally, any filter operation in the three-dimensional physical space is defined by the convolution integral:
\begin{equation}
\bar{f}(\textbf{x}) = \int f(\acute{\textbf{x}})G(\textbf{x}, \acute{\textbf{x}})d\acute{\textbf{x}} ,
\label{eq:cint}
\end{equation}
where $G(\textbf{x}, \acute{\textbf{x}})$ is the filter kernel.
For the box filter, the filter kernel is given by the following formula:
\begin{equation}
G(\textbf{x}, \acute{\textbf{x}}) =
\bigg\{ \begin{array}{l} 1/\Delta \\0 \end{array} \begin{array}{l} \quad  \mbox{if} \quad |\textbf{x}_i - \acute{\textbf{x}}|< \Delta /2  \\ \quad \mbox{otherwise} . \end{array}
\label{eq:ffi}
\end{equation}

Next, we briefly sketch the derivation of the discrete form of the box filter (for details, see Sagaut \cite{sagaut2006large} and Garnier et al. \cite{garnier2009large}).
We start with the one-dimensional case.
Using grid point averaging in \eqref{eq:cint}, we get
\begin{equation}
\bar{f}(x) = \frac{1}{2\Delta}\int^{\Delta}_{-\Delta} f(\acute{x})d\acute{x} .
\label{eq:box_filter_1}
\end{equation}
Using numerical integration to approximate the integral in \eqref{eq:box_filter_1} yields the following discrete box filters:
The trapezoidal rule yields the {\it trapezoidal filter (TF)}
\begin{equation}
\bar{f}_j = \frac{1}{4}(f_{j+1} + 2f_{j} + f_{j-1})
\label{eq:trap}
\end{equation}
and using Simpson's rule yields the {\it Simpson's filter (SF)}
\begin{equation}
\bar{f}_j = \frac{1}{6}(f_{j+1} + 4f_{j} + f_{j-1}) ,
\label{eq:simp}
\end{equation}
where $\bar{f}_j$ is the filtered quantity at discrete point $j$.
Extending this procedure to a three-dimensional grid is straightforward.
For example, in the three-dimensional physical space, the TF \eqref{eq:trap} and the SF \eqref{eq:simp}  result in a 27-point operator
\begin{eqnarray}
\bar{f}_{i,j,k} &=& f_{i,j,k} + c_1(f_{i\pm1,j,k} + f_{i,j\pm1,k} + f_{i,j,k\pm1}) \nonumber \\
&+& c_2(f_{i\pm1,j\pm1,k} + f_{i\pm1,j,k\pm1} + f_{i,j\pm1,k\pm1})  + c_3(f_{i\pm1,j\pm1,k\pm1}) ,
\label{eq:trap3D}
\end{eqnarray}
where $c_1=\frac{1}{16}$, $c_2=\frac{1}{32}$, and $c_3=\frac{1}{64}$ for the TF, and $c_1=\frac{2}{27}$, $c_2=\frac{1}{54}$, and $c_3=\frac{1}{216}$ for the SF.
The box filters can also be constructed by using high-order numerical integration schemes that include more neighboring points.
For example, in the one-dimensional physical space, a {\it seven-point filter (7PF)} is given by \cite{najjar1996study}:
\begin{equation}
\bar{f}_j = \frac{1}{256}(f_{j+3} -18f_{j+2} + 63f_{j+1} + 164f_{j} + 63f_{j-1} -18f_{j-2} + f_{j-3}) .
\label{eq:spoint}
\end{equation}
The 7PF \eqref{eq:spoint} can easily be extended to the three-dimensional grid \cite{najjar1996study}.

\begin{figure}
\centering
\includegraphics[width=0.75\textwidth]{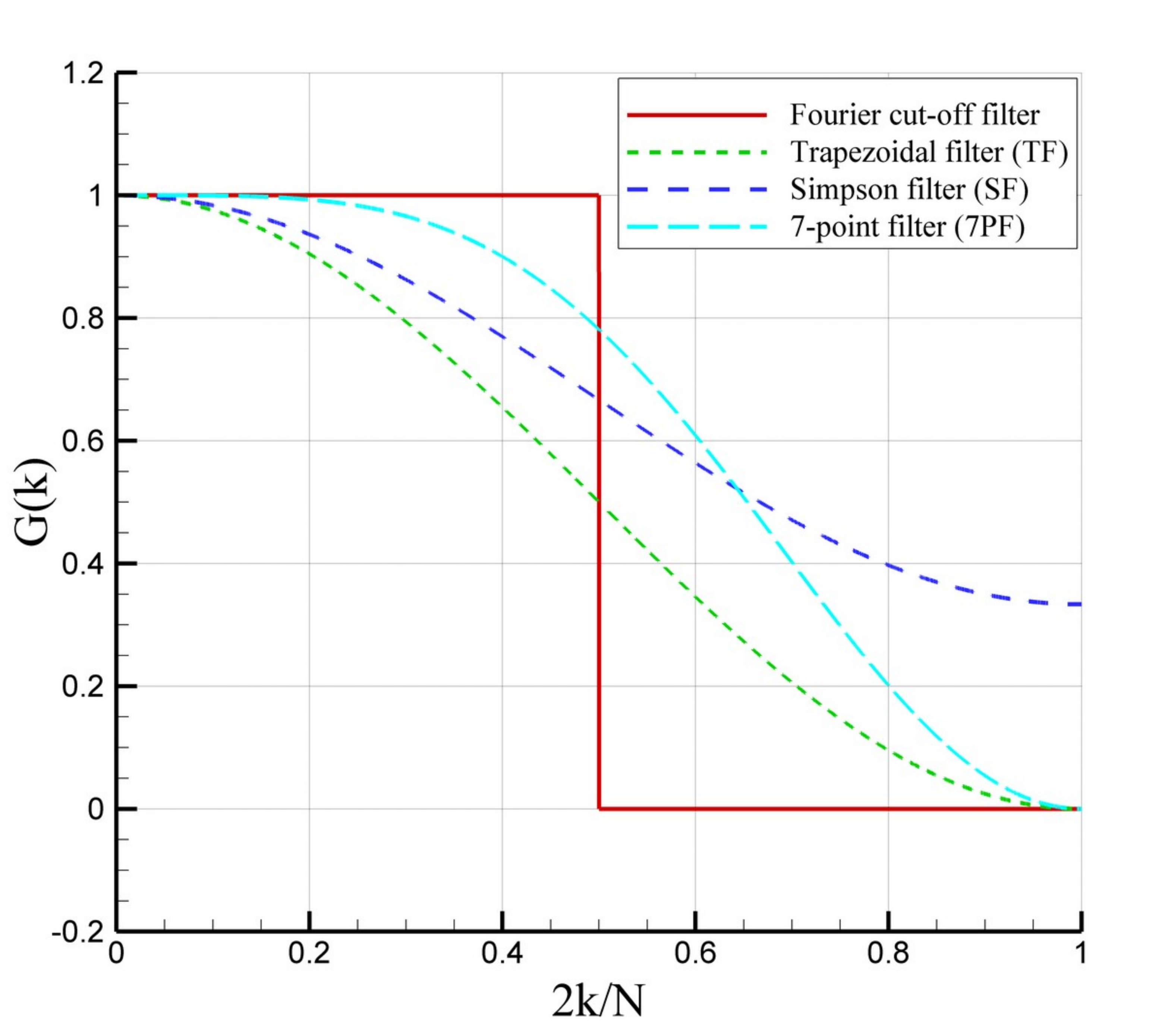}
\caption{
Transfer functions for the TF \eqref{eq:trap}, the SF \eqref{eq:simp}, and the 7PF \eqref{eq:spoint}.
The transfer function of the Fourier cut-off filter is also included for comparison purposes.
}
\label{fig:discf}
\end{figure}

Since the box filters have been constructed in the physical space, a Fourier analysis is applied to study their characteristics in the wavenumber space.
This analysis leads to the transfer function, $G(k)$, that correlates the Fourier coefficients of the filtered variable to those of the unfiltered variable as follows:
\begin{equation}
\hat{\bar{f}} = G(k) \hat{f} ,
\label{eq:traf}
\end{equation}
where $\hat{\bar{f}}$ and $\hat{f}$ are the corresponding Fourier coefficients of the filtered and unfiltered variables, respectively.
The transfer function of the TF is
\begin{equation}
G^{(TF)}(k)=\frac{1}{2}\big(1+\cos(k\Delta)\big) ,
\label{eq:traptf}
\end{equation}
where $\Delta = 2 \pi /N$ and $N$ is the number of grid points in the corresponding direction.
Similarly, the transfer function of the SF is
\begin{equation}
G^{(SF)}(k)=\frac{1}{3}\big(2+\cos(k\Delta)\big) ,
\label{eq:simptf}
\end{equation}
and the transfer function of the 7PF is
\begin{equation}
G^{(7PF)}(k)=\frac{1}{128}\big(82 + 63\cos(k\Delta) - 18\cos(2k\Delta)+\cos(3k\Delta) \big) .
\label{eq:simptf}
\end{equation}
Fig.~\ref{fig:discf} illustrates the transfer functions for the three box filters that we consider: the TF, SF, and 7PF.
The transfer functions of the Fourier cut-off filter is also shown for comparison purposes.
It is known that the Fourier cut-off filter removes the small scales with wavenumbers $2k/N > 1/2$, while retaining the larger scales with wavenumbers $2k/N < 1/2$.
The box filters, however, attenuate the wavenumber components differently, as shown in Fig.~\ref{fig:discf}.
Ranking the three box filters in the decreasing  order of wavenumber attenuation, the TF is consistently the first.
For low wavenumbers ($2k/N < 1/2$), the SF is the second and the 7PF is the third.
For high wavenumbers ($2k/N > 1/2$), the ranking starts to change: the 7PF increasingly attenuates more than the SF.

\subsection{Pad\'{e}-type filters}
\label{sec:pf}

The implicit Pad\'{e}-type discrete filters, which have been introduced by Lele \cite{lele1992compact}, have been successfully used in LES (see, e.g., Stolz and Adams \cite{stolz1999approximate}, Stolz \emph{et al.} \cite{stolz2001approximate}, Pruett and Adams \cite{pruett2000priori} and San \emph{et al.} \cite{san2011approximate}).
In this study, we consider the following one-parameter second-order {\it Pad\'{e}-type filter (PF)}, proposed by Stolz and Adams \cite{stolz1999approximate}:
\begin{equation}
\alpha \bar{f}_{j-1}
+ \bar{f}_{j}
+ \alpha \bar{f}_{j+1}
= \left(\frac{1}{2} + \alpha\right)\left(f_{j}+\frac{ f_{j-1} + f_{j+1}}{2}\right) ,
\label{eq:119}
\end{equation}
where $\bar{f}_{j}$ represents the filtered value of a discrete quantity $f_j$.
This results in a tridiagonal system of equations, which can be solved efficiently by using, e.g., the well-known Thomas-algorithm.
The transfer function of the PF given in \eqref{eq:119} can be written as
\begin{equation}
G^{(PF)}(k) = \left(\frac{1}{2}+\alpha\right) \, \frac{1+\cos(k\Delta)}{1+2\alpha\cos(k\Delta)}.
\label{eq:tfun-1}
\end{equation}

\begin{figure}
\centering
\includegraphics[width=0.75\textwidth]{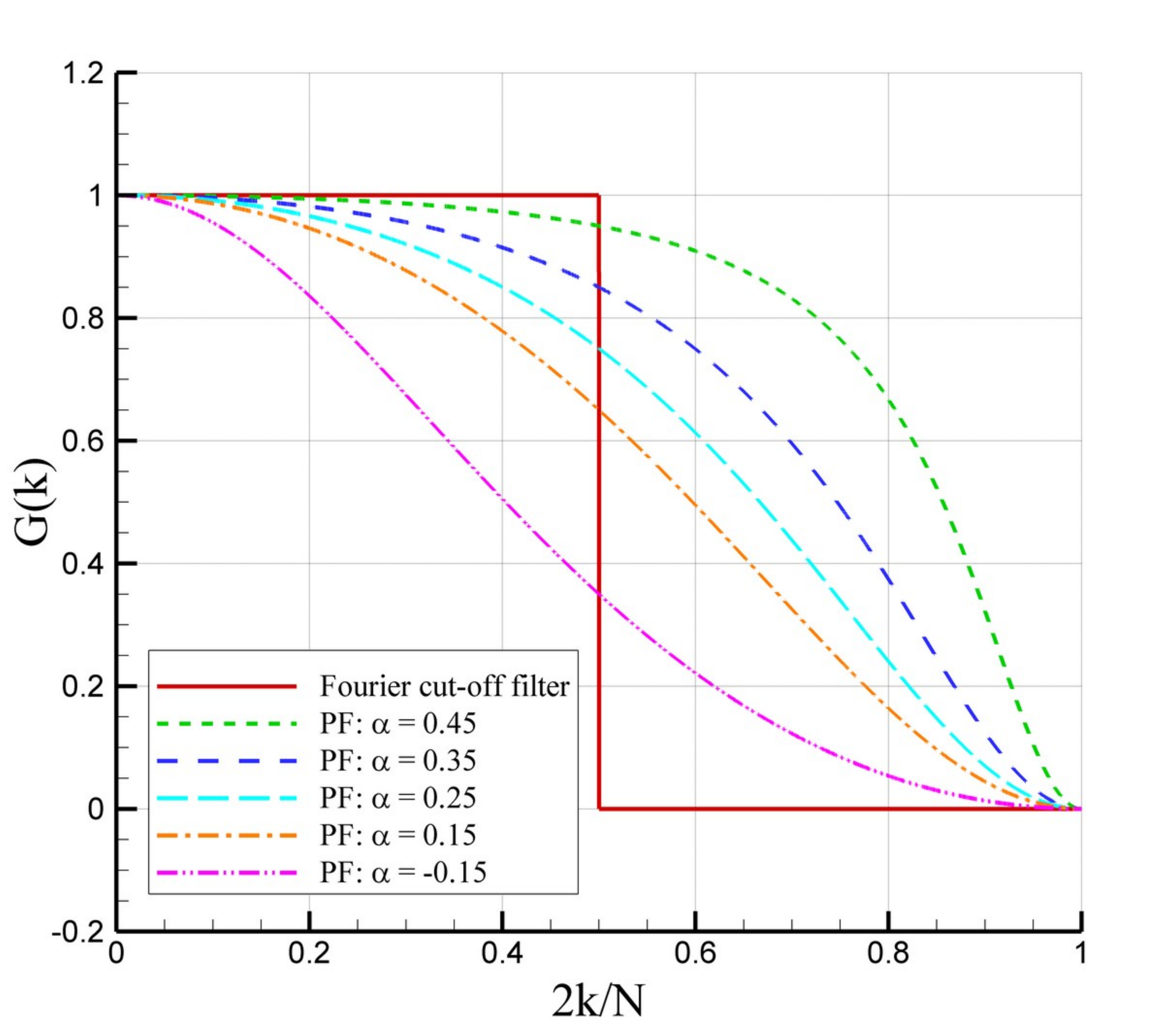}
\caption{
Transfer functions for the PF \eqref{eq:119} for different values of the parameter $\alpha$.
The transfer function of the Fourier cut-off filter is also included for comparison purposes.
}
\label{fig:padef}
\end{figure}
The free parameter, $\alpha$, determines the filtering properties of the PF, with high values of $\alpha$ yielding less dissipative results.
The transfer function $G^{(PF)}$ is positive when the parameter lies in the interval of $0 \leq |\alpha| \leq 0.5$.
This, in turn, ensures the well-posedness of the AD-LES model, as shown in literature \cite{stanculescu2008existence,layton2012approximate}.
Thus, in this report, we follow Lele \cite{lele1992compact} and Pruett and Adams \cite{pruett2000priori}, and use parameter values in the interval $0 \leq |\alpha| \leq 0.5$. More details on the PF can be found in Pruett and Adams \cite{pruett2000priori}.

To study the characteristics of the PF \eqref{eq:119}, we plot in Fig.~\ref{fig:padef} its transfer function $G^{(PF)}$ (which is given by \eqref{eq:tfun-1}) for different values of the parameter $\alpha$.
As pointed out in Pruett and Adams \cite{pruett2000priori} (see also Lele \cite{lele1992compact}), rewriting the filter parameter as $\alpha = -\cos(\beta_c)/2$ allows an easier interpretation of the transfer function $G^{(PF)}$ in terms of the new cut-off parameter $\beta_c$, which lies in the range $0\leq\beta_c\leq\pi$.
It is clear from Fig.~\ref{fig:padef} that $\beta_c$ plays the role of a cut-off number for the PF: $\beta_c = \pi$ turns off the filter, whereas low $\beta_c$ values result in extreme dissipation (i.e., high attenuation of all the wavenumber components).
Increasing the value of the cut-off parameter results in a corresponding increase in the dissipation levels over the entire range of wavenumbers.
Similar conclusions were drawn by Pruett and Adams \cite{pruett2000priori} (see Fig. 2 in Pruett and Adams \cite{pruett2000priori}).

\subsection{Elliptic differential filters}
\label{sec:diff}
The concept of differential filters was introduced in LES by Germano \cite{germano1986differential}.
Since then, it has been successfully used in LES of both engineering and geophysical flows
\cite{iliescu2003large,ozgokmen2009large}.
Solid mathematical foundations were also developed \cite{dunca2006stolz,layton2006residual,layton2007similarity,rebholz2007conservation,stanculescu2008existence,layton2012approximate}.

\begin{figure}
\centering
\includegraphics[width=0.75\textwidth]{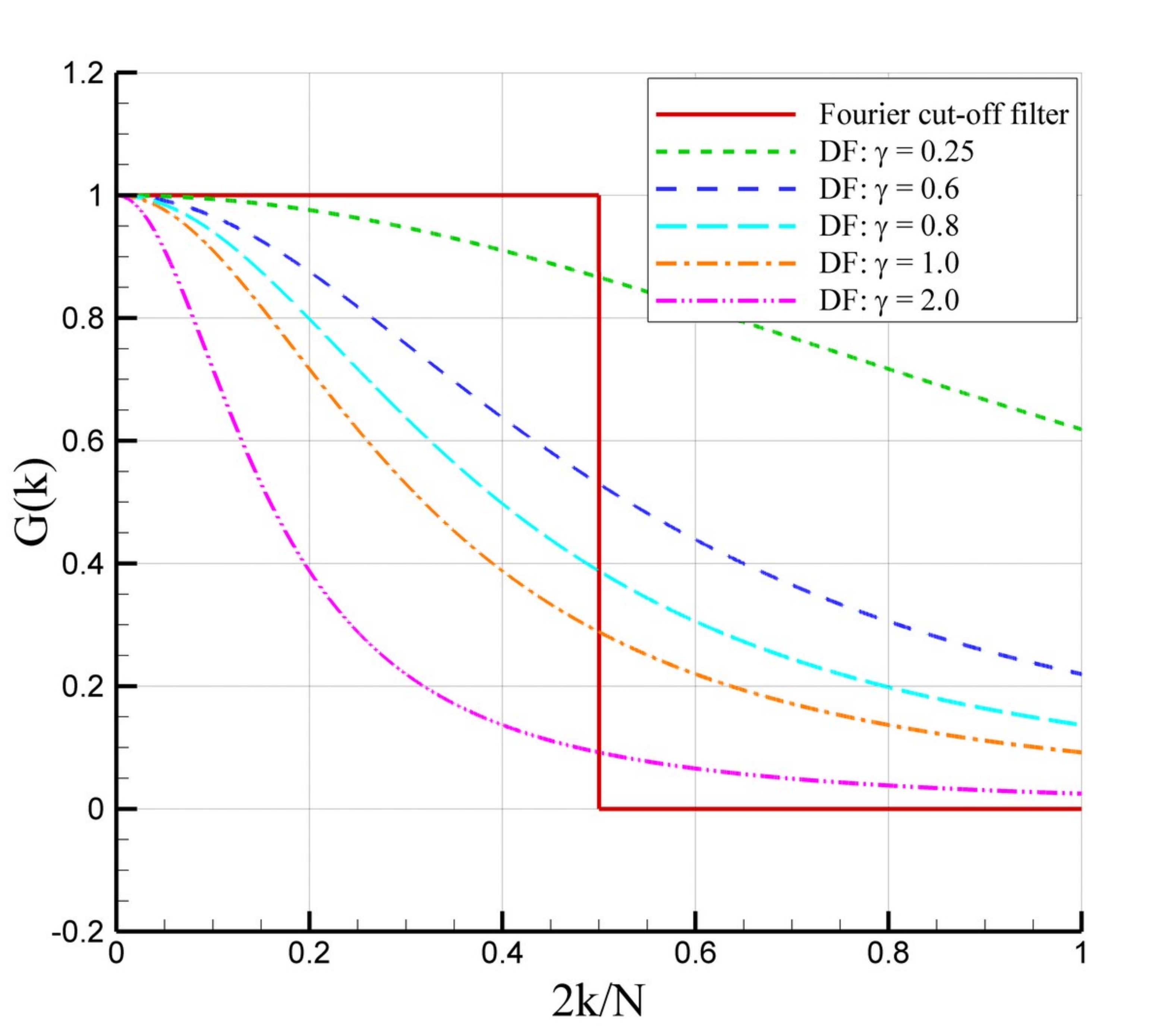}
\caption{
Transfer functions for the DF \eqref{eq:10a} for different values of the parameter $\gamma$.
The transfer function of the Fourier cut-off filter is also included for comparison purposes.
}
\label{fig:ediff}
\end{figure}

The elliptic {\it differential filter (DF)}, also called Helmholtz filter, can be written as:
\begin{equation}
\bar{f} - \lambda^2 \, \left(\frac{\partial^2 \bar{f}}{\partial x^2} + \frac{\partial^2 \bar{f}}{\partial y^2} + \frac{\partial^2 \bar{f}}{\partial z^2} \right) = f ,
\label{eq:10a}
\end{equation}
where $\lambda$ determines the effective width of the filter.
The filtered value $\bar{f}$ is obtained by applying the inverse Helmholtz operator to the unfiltered flow variable $f$.
This inversion is done efficiently by using the {\it fast Fourier transform (FFT)} techniques \cite{press1992numerical}.
The transfer function of the DF is:
\begin{equation}
G^{(DF)}(k)=\frac{1}{1+\lambda^2 k^2} .
\label{eq:tfunh}
\end{equation}
It is obvious that the transfer function $G^{(DF)}$ in \eqref{eq:tfunh} is positive, which ensures the well-posedness of the AD-LES model \cite{stanculescu2008existence,layton2012approximate}.
To study the characteristics of the DF, we plot in Fig.~\ref{fig:ediff} its transfer function, $G^{(DF)}$, for different values of the parameter $\gamma$.
This parameter is defined as $\gamma = \lambda / \Delta$, where $\Delta$ is the grid spacing.
That is, the filter parameter $\gamma$ represents the ratio of the filter width $\lambda$ to the grid spacing $\Delta$.
Thus, increasing the value of $\gamma$ in Fig.~\ref{fig:ediff} amounts to increasing the filter width while keeping the grid spacing fixed.
Fig.~\ref{fig:ediff} clearly shows that increasing $\gamma$ (i.e., increasing the filter radius) results in a significant increase of the dissipation of the DF (i.e., the attenuation of the wavenumber components of the filtered variable).

\begin{figure}
\centering
\includegraphics[width=0.75\textwidth]{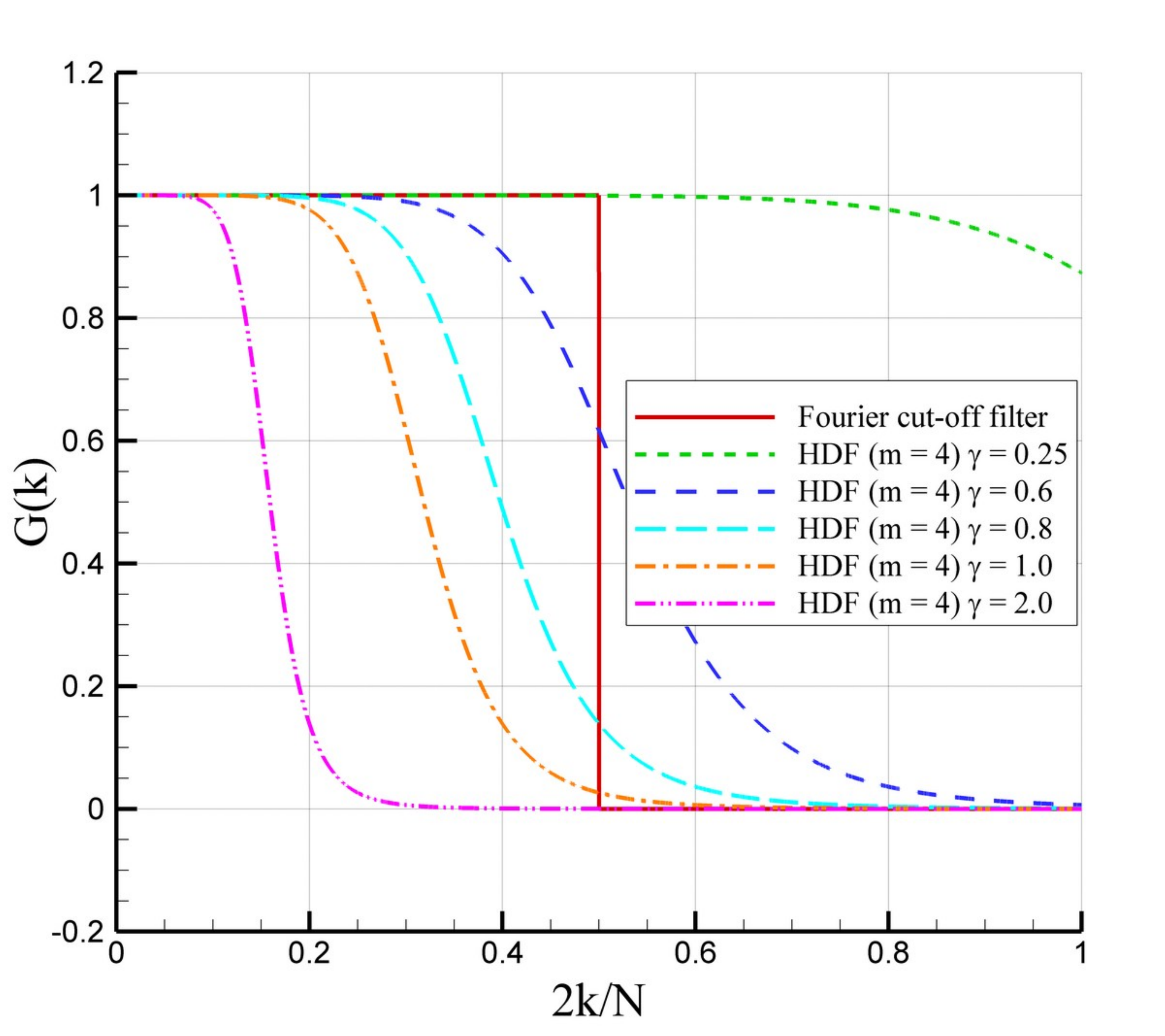}
\caption{
Transfer functions for the HDF \eqref{eq:10b} for $m=4$ and different values of the parameter $\gamma$.
The transfer function of the Fourier cut-off filter is also included for comparison purposes.
}
\label{fig:hdiff4}
\end{figure}

\begin{figure}
\centering
\includegraphics[width=0.75\textwidth]{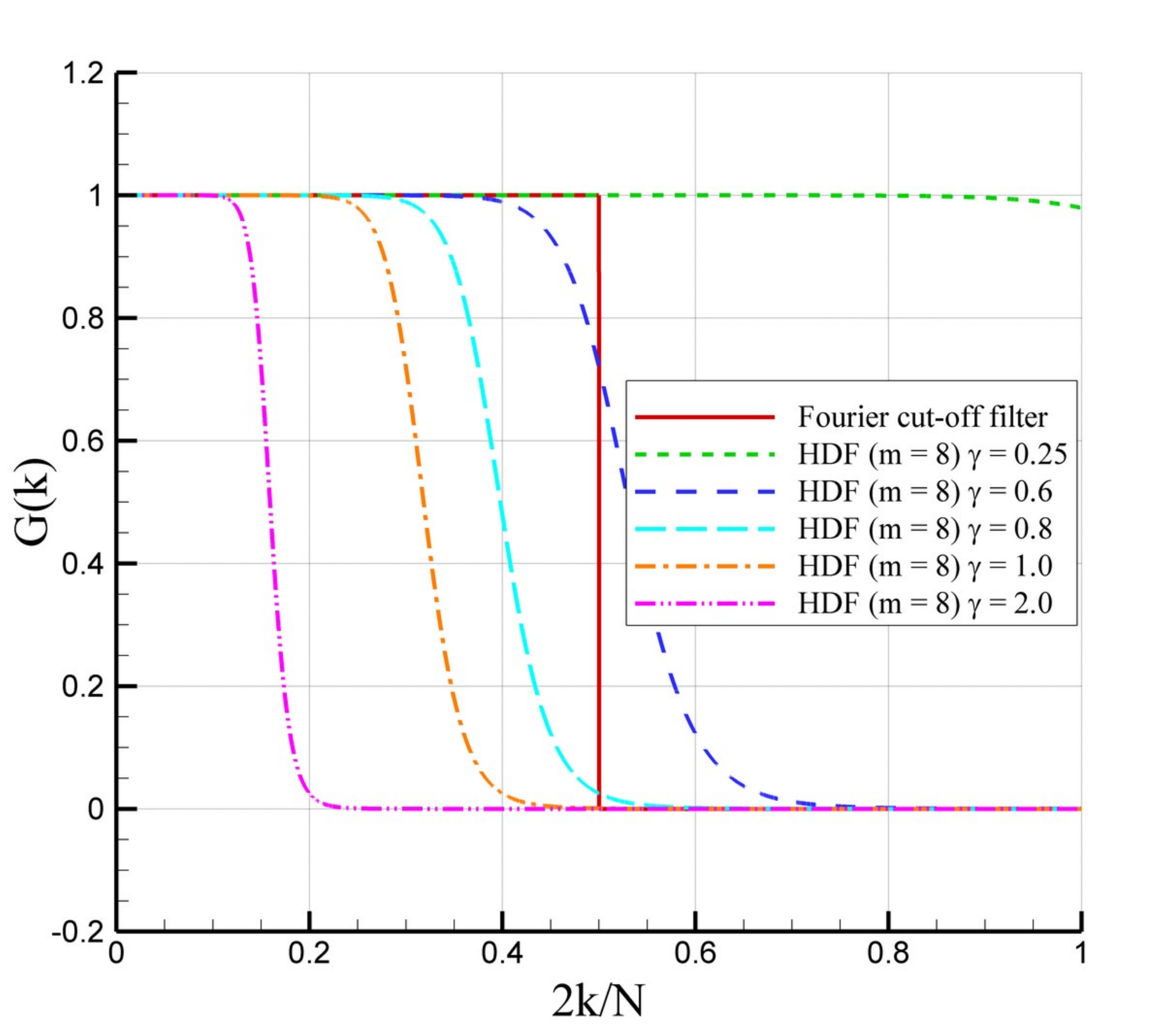}
\caption{
Transfer functions for the HDF \eqref{eq:10b} for $m=8$ and different values of the parameter $\gamma$.
The transfer function of the Fourier cut-off filter is also included for comparison purposes.
}
\label{fig:hdiff8}
\end{figure}

\begin{figure}
\centering
\includegraphics[width=0.75\textwidth]{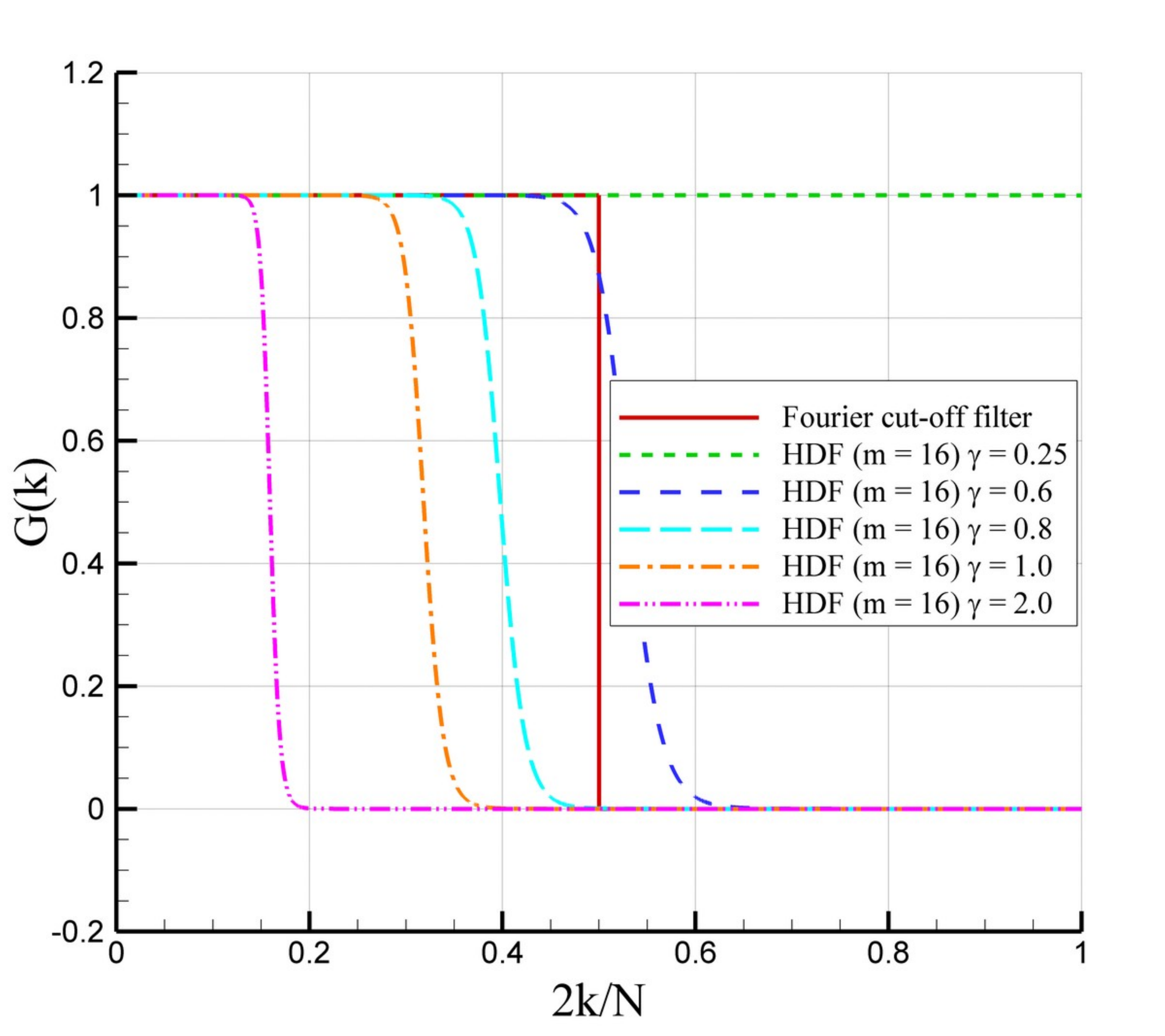}
\caption{
Transfer functions for the HDF \eqref{eq:10b} for $m=16$ and different values of the parameter $\gamma$.
The transfer function of the Fourier cut-off filter is also included for comparison purposes.
}
\label{fig:hdiff16}
\end{figure}

In order to provide a more rapid decay of the high wavenumber components of the filtered variables, a generalized form of the Helmholtz filter has been suggested by Mullen and Fischer \cite{mullen1999filtering}.
This filter, which we call in this report the {\it hyper-differential elliptic filter (HDF)}, is defined as follows:
\begin{equation}
\bar{f} - \lambda^{2m} \, \left(\frac{\partial^{2m} \bar{f}}{\partial x^{2m}} + \frac{\partial^{2m} \bar{f}}{\partial y^{2m}}  + \frac{\partial^{2m} \bar{f}}{\partial z^{2m}} \right) = f ,
\label{eq:10b}
\end{equation}
where $m$ is any positive integer.
Note that we recover the DF when we let $m=1$ in the HDF \eqref{eq:10b}.
We emphasize that, to the best of our knowledge,this study represents the first use of the HDF in an AD-LES framework.
The transfer function of the HDF is
\begin{equation}
G^{(HDF)}(k)=\frac{1}{1+\lambda^{2m} k^{2m}} .
\label{eq:tfunhd}
\end{equation}
The transfer functions of the HDF for $m=4$, $m=8$, $m=16$, and for different values of $\gamma = \lambda / \Delta$ are plotted in Figs.~\ref{fig:hdiff4}, \ref{fig:hdiff8}, and \ref{fig:hdiff16}, respectively.
These figures yield the following general conclusions.
First, increasing the parameter $\gamma$ results in a significant increase of the dissipation of the HDF, just like it did for the DF in Fig.~\ref{fig:ediff}.
Compared with the transfer function of the DF, however, the transfer function of the HDF has a much sharper transition between small and large wavenumbers.
Indeed, for all values of $m$, the HDF with $0.5 \leq \gamma \leq 1.0$ satisfactorily captures the contribution of the scales with wavenumbers $2 K / N < 1 / 2$ and attenuates the contribution of the scales with wavenumbers $2 K / N > 1 / 2$, just like the Fourier cut-off filter.
Figs. \ref{fig:hdiff4}-\ref{fig:hdiff16} also show that increasing the parameter $m$ in the HDF makes the transition region between small and large wavenumbers if the corresponding transfer functions even sharper.
We also note that, since we utilize FFT techniques to invert the elliptic subproblems, the computational overhead of the HDF is negligible.

\subsection{Summary}
\label{sec:filter_summary}

In this section, we presented the following classes of spatial filters: the box filters (Section \ref{sec:dbf}), the Pad\'{e}-type filters (Section \ref{sec:pf}), and the differential filters (Section \ref{sec:diff}).
For each class of filters, we considered several model parameters.
The characteristics of these filters were illustrated by plotting their corresponding transfer functions.
The effect of the parameters on the transfer functions was discussed for each class of filters.
Comparing the three classes of spatial filters among them, it seems that no general conclusion that is valid for all parameters can be drawn.
The one exception, however, is the HDF \eqref{eq:10b}, whose transfer function resembles that of a Fourier cut-off for all parameter values.

Since the main goal of this study is to investigate the effect of the spatial filters on the AD-LES model, the following natural question arises: ``Which spatial filter is the most appropriate for AD-LES?"
Based on the discussion in this section, the answer to this question is not clear.
Indeed, {\it a priori} one cannot decide whether the spatial filter used in the AD-LES model should resemble the Fourier cut-off filter (like the HDF), or, e.g., be more like the Pad\'{e}-type filters.
Thus, to answer the question above, in Section \ref{sec:results} we carry out an {\it a posteriori} testing of the AD-LES model \eqref{eq:fvv}-\eqref{eq:sfs} equipped with each type of spatial filter that we considered in this section.

\section{Numerical Methods}
\label{sec:numerics}
%

In this section, we briefly describe the numerical discretization used for all the tests in Section \ref{sec:results}.
To describe the time discretization, we rewrite the AD-LES model \eqref{eq:fvv}-\eqref{eq:sfs} as follows:
\begin{equation}
\frac{d \overline{\boldsymbol \omega}}{d t} = \pounds (\overline{\boldsymbol \omega}, \overline{\boldsymbol u}) ,
\end{equation}
where $\pounds (\overline{\boldsymbol \omega}, \overline{\boldsymbol u})$ is the discrete operator of spatial derivatives for the nonlinear convective terms, linear diffusive terms, and SFS term:
\begin{equation}
\pounds (\overline{\boldsymbol \omega},\overline{\boldsymbol u})=- \overline{\boldsymbol u} \cdot \nabla \overline{\boldsymbol \omega} + \overline{\boldsymbol \omega} \cdot \nabla \overline{\boldsymbol u} + \frac{1}{Re} \nabla^2 \overline{\boldsymbol \omega} + \textbf{S} .
\label{eq:numerics_1}
\end{equation}
All the spatial derivatives in \eqref{eq:numerics_1} are computed by using standard second-order finite difference approximations except some validation cases which will be explained further in the following section. The details of the numerical discretization can also be found in \cite{san2012high}.
To discretize the resulting system of ordinary differential equations, we assume that the numerical approximation for time level $n$ is known, and we seek the numerical approximation for time level $n+1$, after the time step $\Delta t$.
To this end, we utilize the third-order accurate total variation diminishing Runge-Kutta scheme \cite{gottlieb1998total}:
\begin{eqnarray}
\overline{\boldsymbol \omega}^{(1)} &=& \overline{\boldsymbol \omega}^{n} + \Delta t \pounds(\overline{\boldsymbol \omega}^{n},\overline{\boldsymbol u}^{n}) \nonumber \\
\overline{\boldsymbol \omega}^{(2)} &=& \frac{3}{4}  \overline{\boldsymbol \omega}^{n} + \frac{1}{4} \overline{\boldsymbol \omega}^{(1)} + \frac{1}{4}\Delta t \pounds (\overline{\boldsymbol \omega}^{(1)},\overline{\boldsymbol u}^{(1)}) \nonumber \\
\overline{\boldsymbol \omega}^{n+1} &=& \frac{1}{3} \overline{\boldsymbol \omega}^{n} + \frac{2}{3} \overline{\boldsymbol \omega}^{(2)} + \frac{2}{3}\Delta t \pounds (\overline{\boldsymbol \omega}^{(2)},\overline{\boldsymbol u}^{(2)}).
\label{eq:TVDRK}
\end{eqnarray}
Finally, to recover the velocity field $\overline{\boldsymbol u}^{n+1}$ from the vorticity field $\overline{\boldsymbol \omega}^{n+1}$ at time level $n+1$, a direct solver based on the FFT is used to solve the elliptic subproblems given in (\ref{eq:es1})-(\ref{eq:es3}).
The FFT based direct elliptic solvers are also utilized for computing the filtered variables used in the definition of the SFS tensor $\textbf{S}$ in \eqref{eq:numerics_1}.
More details on the FFT based direct elliptic solvers can be found in Press \emph{et al.} \cite{press1992numerical} and Moin \cite{moin2001fundamentals}.

\section{Test Case}
\label{sec:test_case}
The fundamental mechanism involved in isotropic, homogeneous turbulent flows is the enhancement of vorticity by vortex stretching and the consequent production of small eddies.
Energy is transferred forward in spectral space, from low wavenumbers (large scales) to high wavenumbers (smaller scales).
This process controls the turbulent energy dynamics and hence the global structure of the evolution of the turbulent flow.
A prototype of this process is given by the generalized Taylor-Green vortex problem \cite{taylor1937mechanism,brachet1983small,brachet1991direct,brachet1992numerical, shu2005numerical,adams2007approximate}, which models the decay of isotropic, homogeneous, turbulent incompressible flow that develops from the single mode initial condition:
\begin{eqnarray}
u(x,y,z,t=0)&=&\frac{2}{\sqrt{3}} \mbox{sin}\left(\theta+\frac{2 \pi}{3}\right) \mbox{sin}(x)\mbox{cos}(y)\mbox{cos}(z) \label{eq:tgv3a}\\
v(x,y,z,t=0)&=&\frac{2}{\sqrt{3}} \mbox{sin}\left(\theta-\frac{2 \pi}{3}\right) \mbox{cos}(x)\mbox{sin}(y)\mbox{cos}(z)\\
w(x,y,z,t=0)&=&\frac{2}{\sqrt{3}} \mbox{sin}\left(\theta \right) \mbox{cos}(x)\mbox{cos}(y)\mbox{sin}(z) .
\label{eq:tgv3}
\end{eqnarray}
All the numerical tests conducted in this study are for the Taylor-Green vortex flow problem.
The computational domain used in all the numerical tests is a cubic box whose edge has a length of $2\pi$.
Periodic boundary conditions are used in all directions.
We set $\theta=0$ in \eqref{eq:tgv3a}-\eqref{eq:tgv3}.
In this case, the initial flow has two-dimensional streamlines, but the flow is three-dimensional for all $t>0$.

\begin{figure}
\centering
\includegraphics[width=1.0\textwidth]{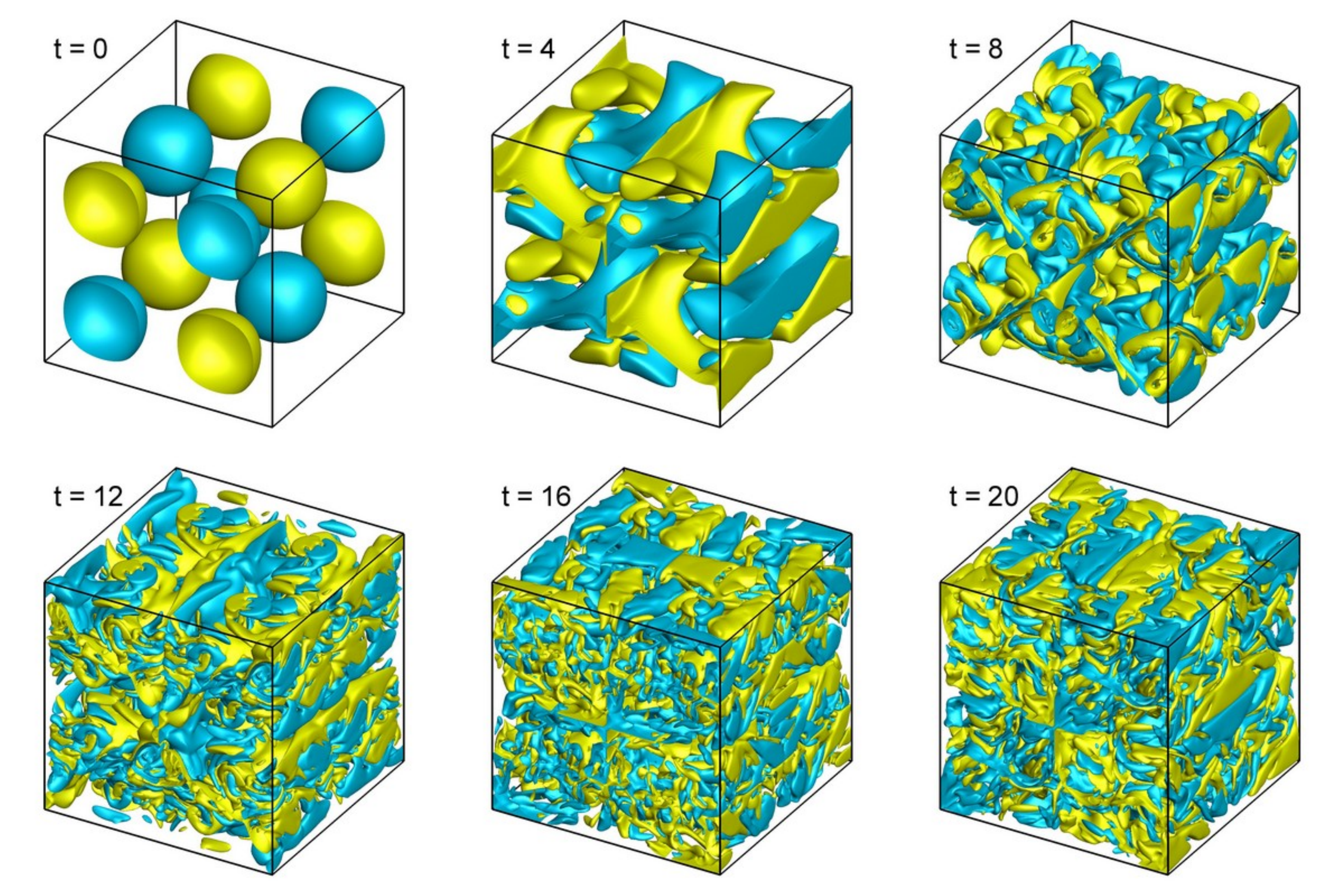}
\caption{
Evolution of the $x$-component of the vorticity on a $256^3$ resolution grid for $Re=1000$. Iso-surfaces of $\omega_x = \pm \ 0.5$ are shown.
}
\label{fig:wx-1000}
\end{figure}

\begin{figure}
\centering
\includegraphics[width=1.0\textwidth]{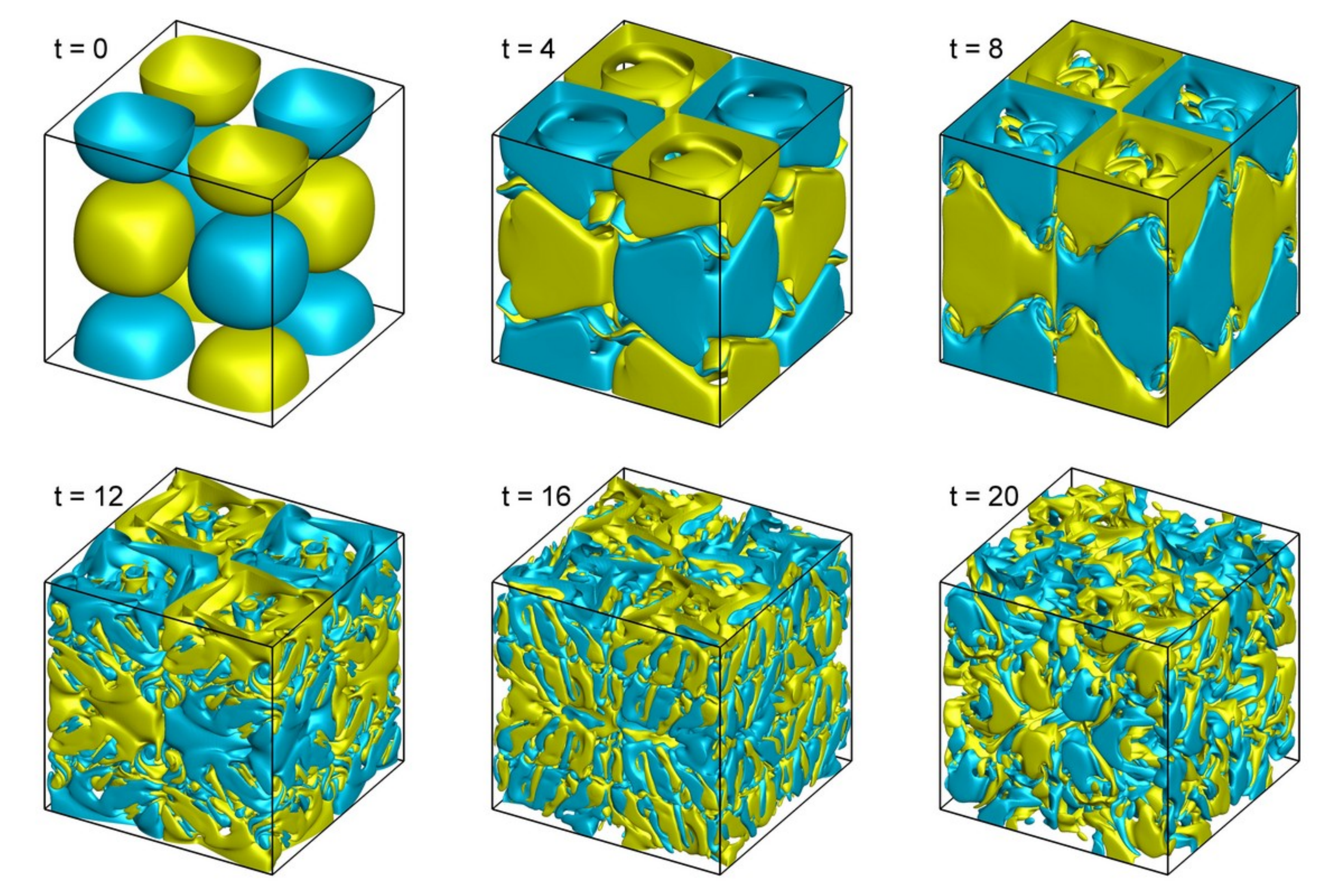}
\caption{
Evolution of the $z$-component of the vorticity on a $256^3$ resolution grid for $Re=1000$. Iso-surfaces of $\omega_x = \pm \ 0.5$ are shown.
}
\label{fig:wz-1000}
\end{figure}

\begin{figure}
\centering
\includegraphics[width=0.75\textwidth]{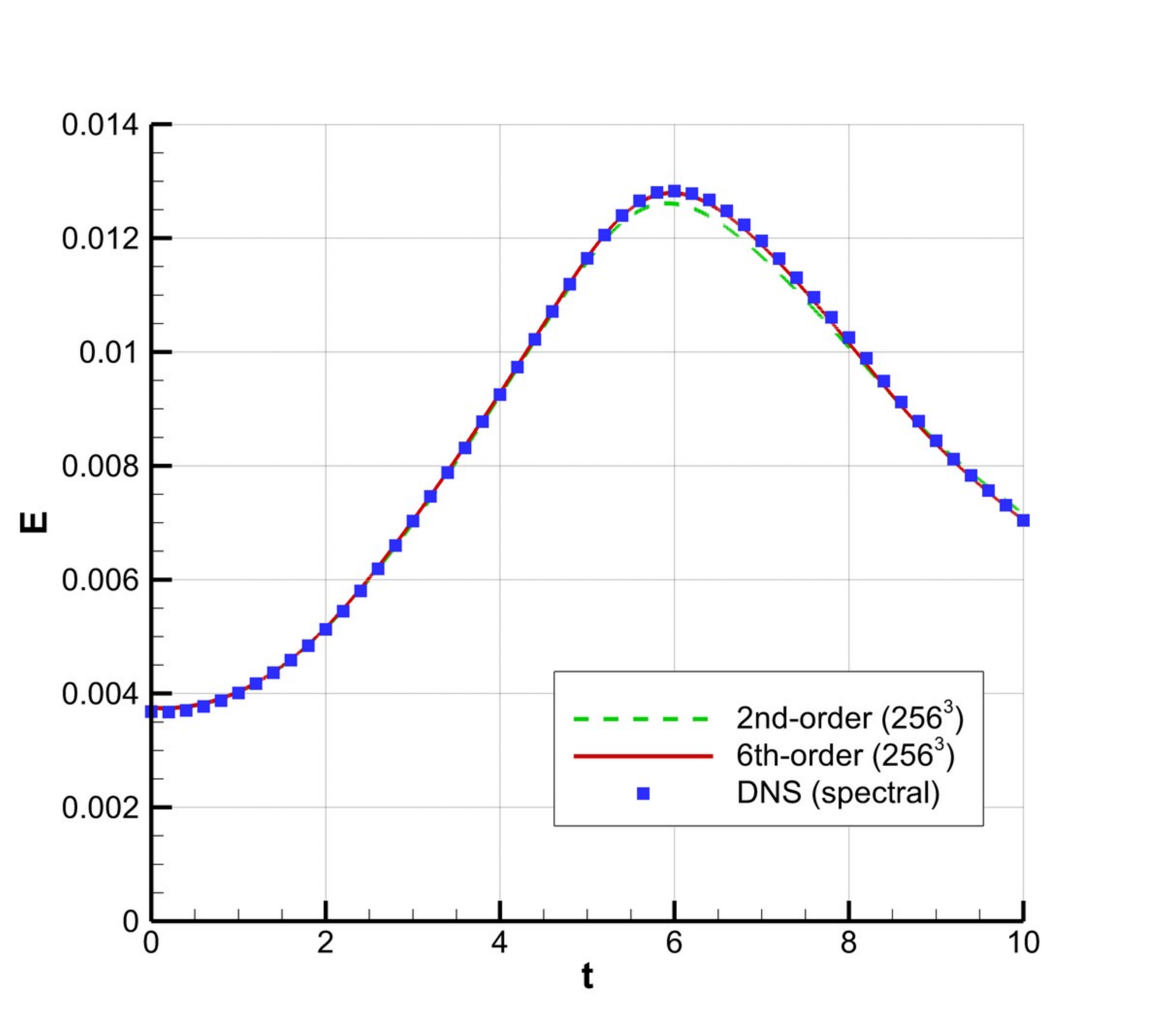}
\caption{
Time series of volume averaged total kinetic energy for the Taylor-Green vortex problem at $Re=200$. DNS data computed by a pseudo-spectral method is available in Brachet \emph{et al}. \cite{brachet1983small}, Hickel \emph{et al}. \cite{hickel2006adaptive}, and Adams \emph{et al}. \cite{adams2007approximate}.
}
\label{fig:comp-dns}
\end{figure}

\begin{figure}
\centering
\mbox{
\subfigure{\includegraphics[width=0.33\textwidth]{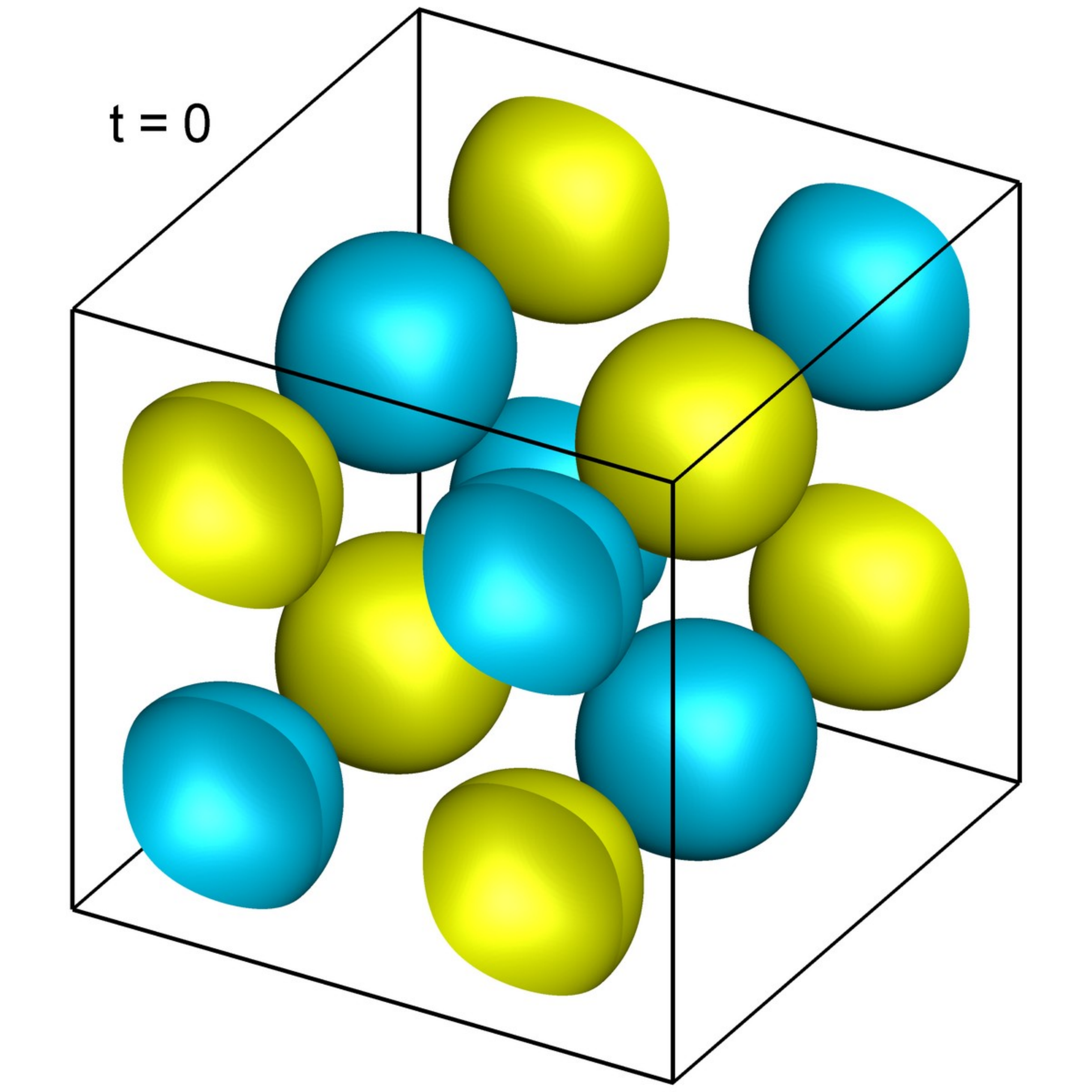}}
\subfigure{\includegraphics[width=0.33\textwidth]{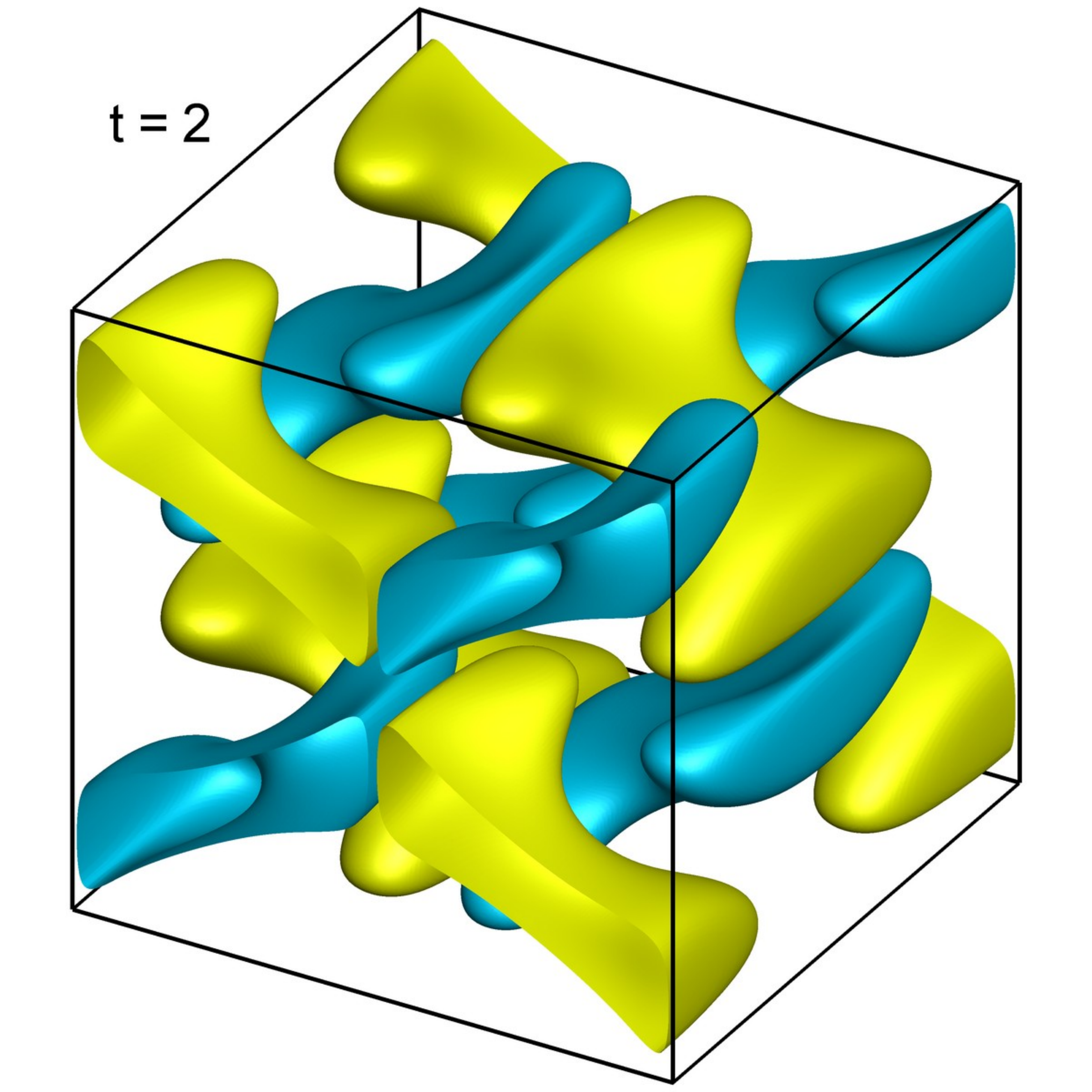}}
\subfigure{\includegraphics[width=0.33\textwidth]{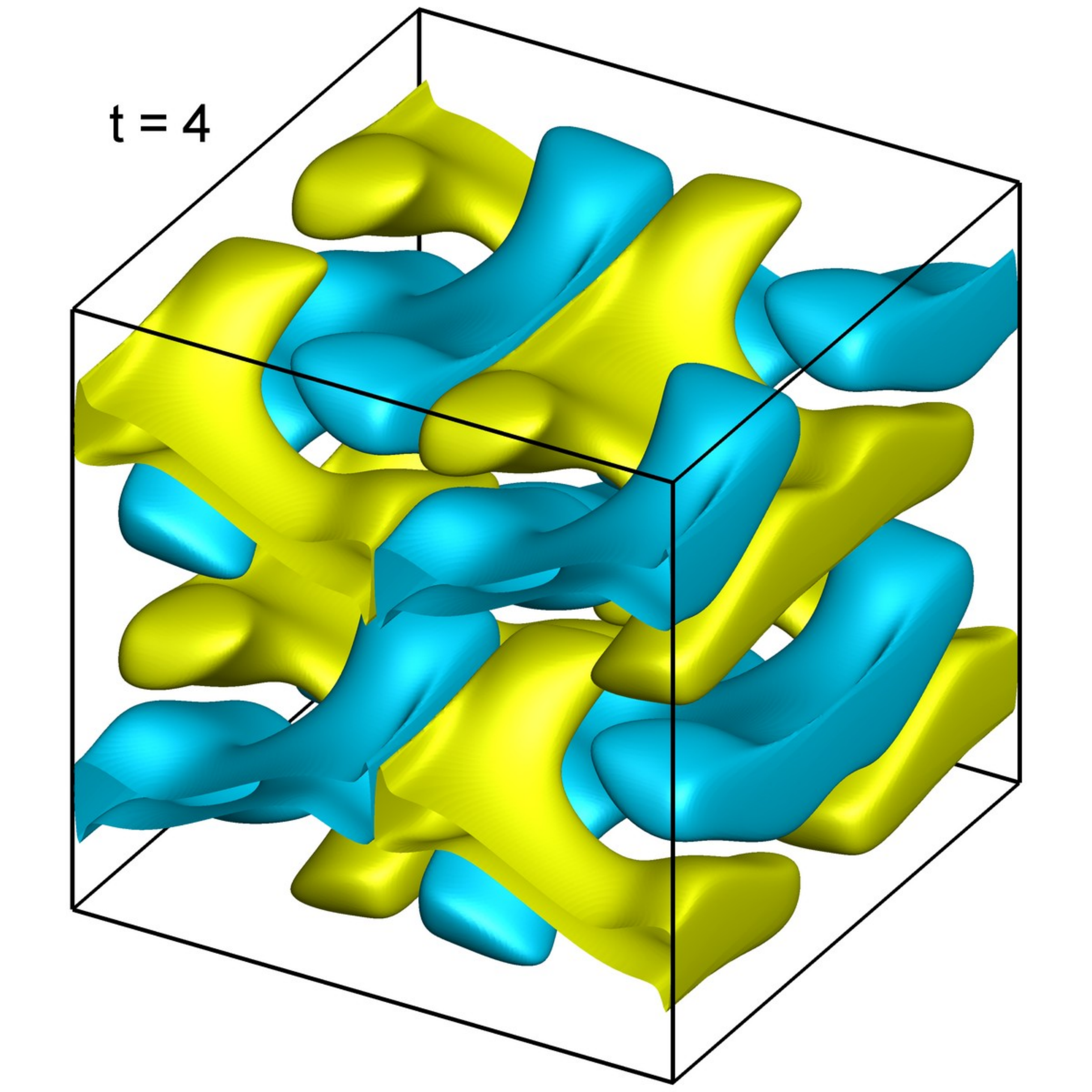}}
}\\
\mbox{
\subfigure{\includegraphics[width=0.33\textwidth]{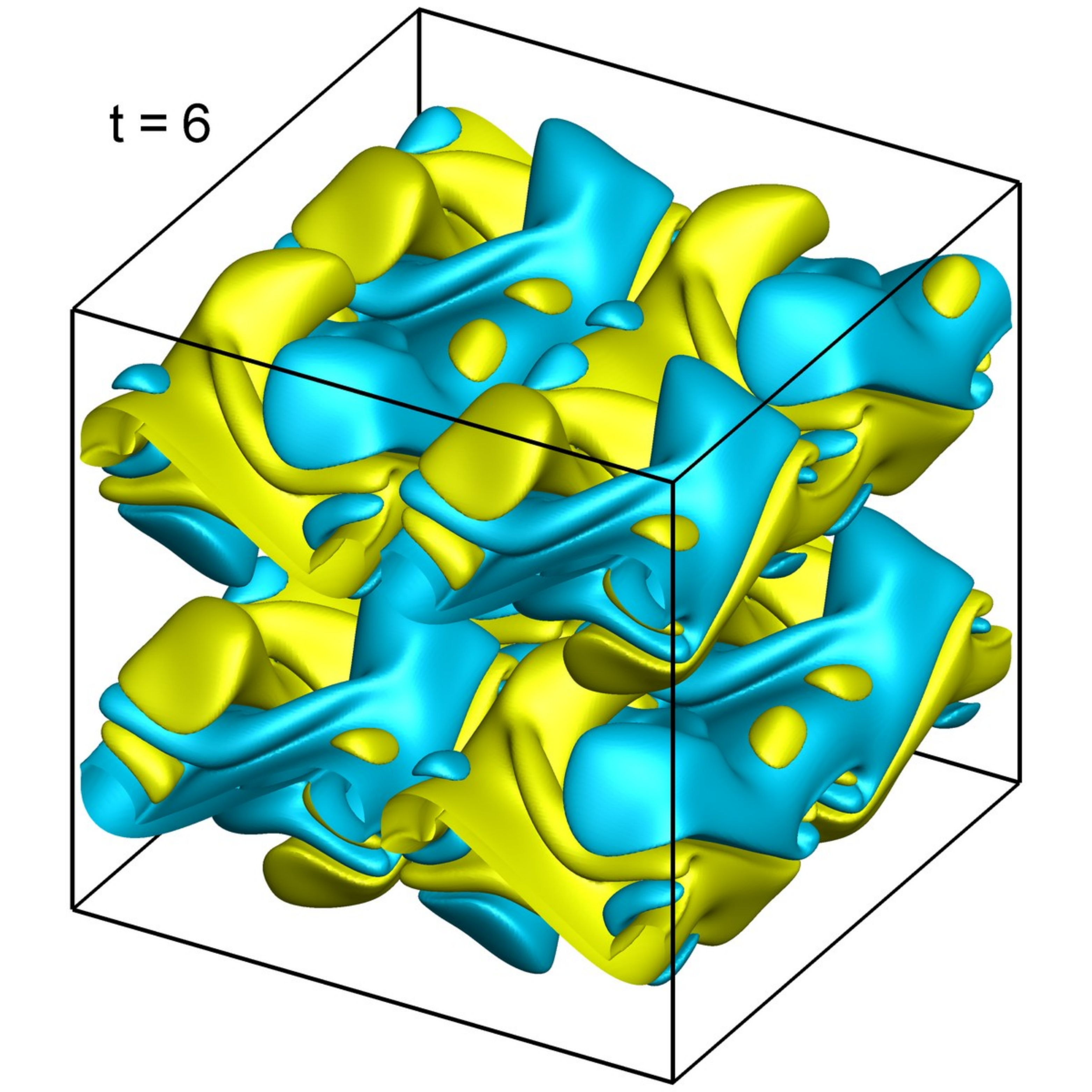}}
\subfigure{\includegraphics[width=0.33\textwidth]{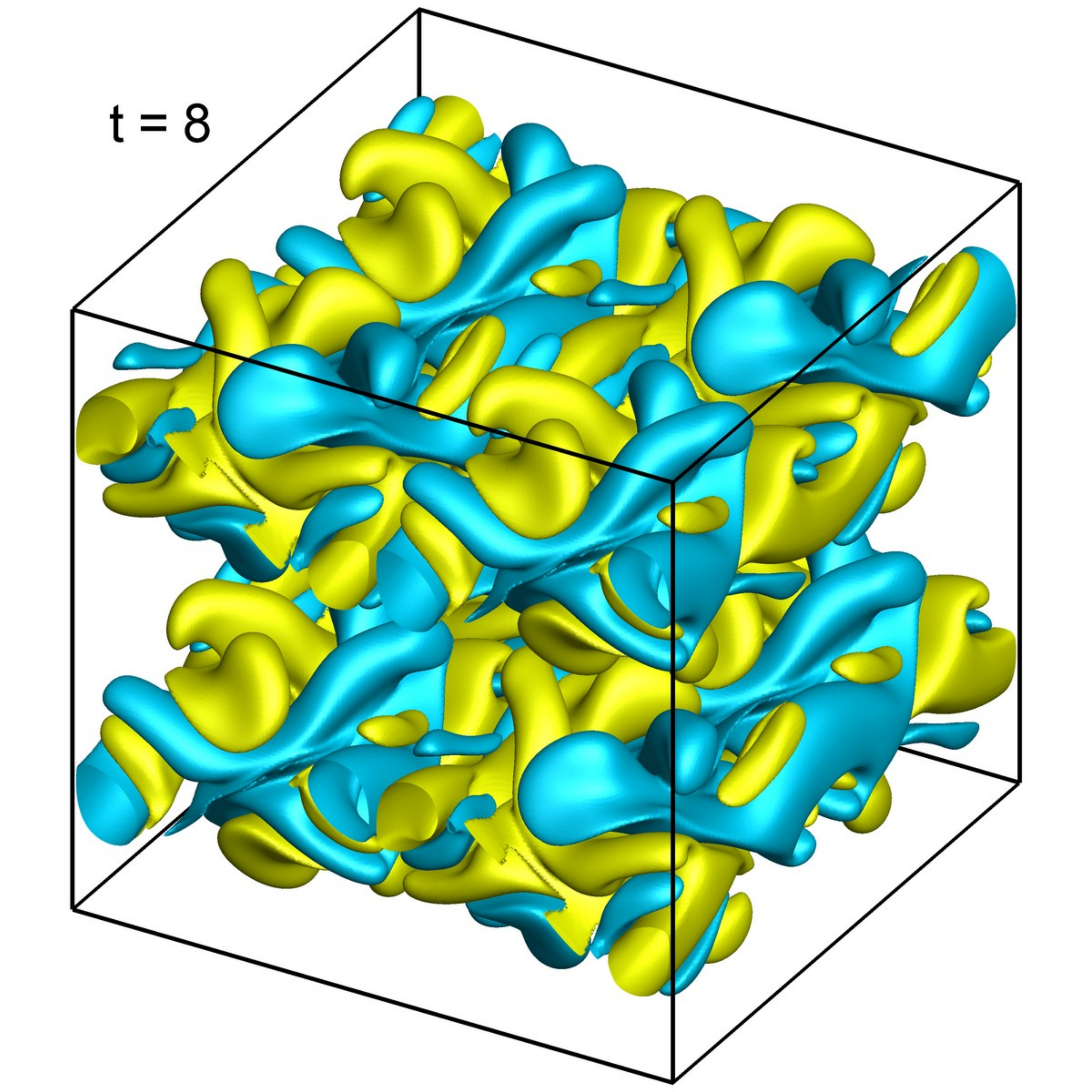}}
\subfigure{\includegraphics[width=0.33\textwidth]{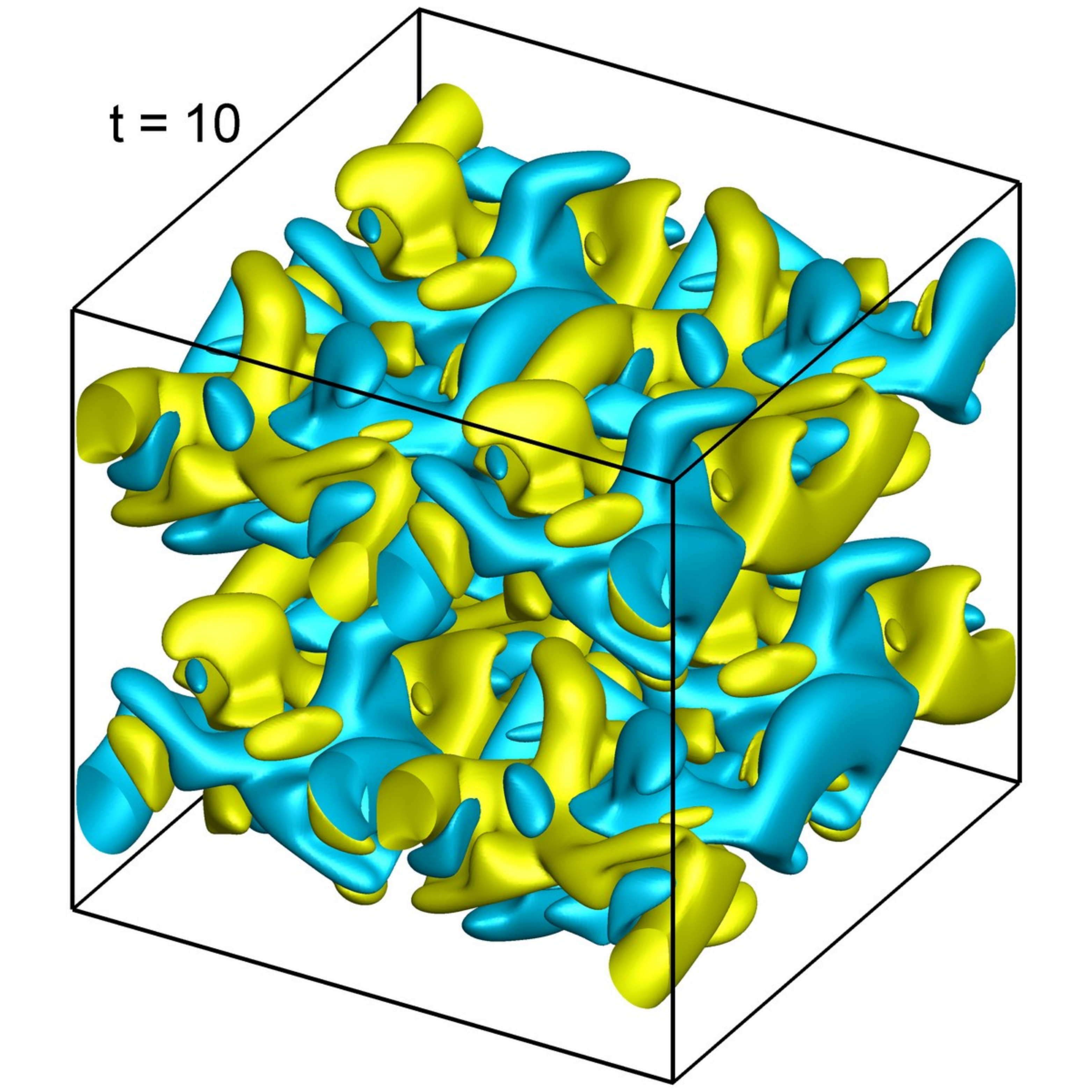}}
}
\caption{Evolution of the $x$-component of the vorticity on a $256^3$ resolution grid for $Re=200$. Iso-surfaces of $\omega_x = \pm \ 0.5$ are shown.}
\label{fig:time-tgv3}
\end{figure}

This flow configuration is perhaps the simplest system in which to study the generation of smaller scale motions and the resulting turbulence. For example, the time evolution of the $x$ and $z$ components of vorticity field for $Re=1000$ are shown in Fig.~\ref{fig:wx-1000} and Fig.~\ref{fig:wz-1000}, respectively using a sixth-order compact difference scheme at a resolution of $256^3$. This is the maximum level of resolution in the present study because of the available computational resources.
Although we were able to carry our a direct numerical simulation for this value of the Reynolds number (i.e., $Re = 1000$), this required the use of a sixth-order compact finite difference scheme.  Since one of our objectives in the present study is to analyze the AD-LES methodology for the standard second-order finite difference scheme, we will use a Reynolds number Re = 200 in all the numerical tests.
The same Reynolds number was used by Brachet \emph{et al}. \cite{brachet1983small}, Hickel \emph{et al}. \cite{hickel2006adaptive}, and Adams \emph{et al}. \cite{adams2007approximate}.
At a resolution of $256^3$, in Fig.~\ref{fig:comp-dns} we compare our results obtained with the standard second-order and compact sixth-order finite difference schemes with the DNS data obtained with a pseudo-spectral simulation \cite{brachet1983small,hickel2006adaptive,adams2007approximate} showing the evolution of the volume averaged total kinetic energy.
Thus, we conclude that the underlying second-order finite difference scheme yields an appropriate reference solution that can be used as a benchmark in our numerical tests.
The effects of low-pass filters on the high-order discretization schemes will be investigated in future studies for higher Reynolds numbers.
For $Re = 200$, the time evolution of the $x$-component of vorticity field, $\omega_x$, is shown in Fig.~\ref{fig:time-tgv3}.
The instantaneous vorticity iso-surfaces for $\omega_x= \pm 0.5$ at six time instances are plotted in Fig.~\ref{fig:time-tgv3}.
These iso-surfaces clearly demonstrate that small scale structures are generated at this Reynolds number as well.

\section{Numerical Results}
\label{sec:results}

The goal of this section is twofold:
First, we numerically investigate the effect of the spatial filters described in Section \ref{sec:spatial_filters} on the AD-LES model \eqref{eq:fvv}-\eqref{eq:sfs}.
As mentioned in Section \ref{sec:spatial_filters}, a simple visual inspection of the corresponding  transfer functions would not allow us to decide which spatial filter is the most appropriate for the AD-LES model.
The second goal of this section is to test the AD-LES model \eqref{eq:fvv}-\eqref{eq:sfs} with the HDF \eqref{eq:10b}.

There are several model parameters that need to be monitored in this numerical investigation.
First, the order of the AD filter, $N$, should be considered.
In this study, we use two values: $N=2$ and $N=5$.
Furthermore, for each fixed $N$, each filter introduces new parameter choices, e.g., $\alpha$ for the PF and $\gamma$ for the DF and the HDF.

For clarity, we use the following approach in presenting the numerical results.
In this section, we investigate the sensitivity of the AD-LES model with respect to parameters for each class of filters separately.
In Section \ref{sec:sum}, we compare the results for all the spatial filters and draw some general conclusions.

All the numerical tests are carried out on the Taylor-Green vortex decay problem described in Section \ref{sec:test_case} and employ the numerical discretization outlined in Section \ref{sec:numerics} with a time-step $\Delta t=5\times10^{-3}$.
In our numerical investigation, we employ the standard LES methodology:
We first run a DNS computation on a fine mesh with a resolution of $256^3$.
A mesh refinement study clearly shows that the DNS resolution has been achieved.
We note that the same DNS spatial resolution was used by Brachet \emph{et al}. \cite{brachet1983small} and Adams \emph{et al}. \cite{adams2007approximate}.
Furthermore, our DNS results are qualitatively similar to the DNS results of Brachet \emph{et al}. \cite{brachet1983small} and Adams \emph{et al}. \cite{adams2007approximate}: All three numerical datasets display a fairly consistent dissipation peak at $t\approx6$.
Next, we run an under-resolved numerical simulation on a much coarser mesh with a resolution of $64^3$ (denoted in what follows as No-AD), which does not employ any SFS model.
Finally, we employ the AD-LES model on the same coarse mesh utilized in No-AD (i.e., with a resolution of $64^3$).
We expect that the AD-LES model should yield results that are significantly better than those obtained with No-AD and are close to the DNS results, at a fraction of the computational cost.

Four criteria are used in evaluating the numerical results.
The first criterion is the time series of integrated enstrophy, which is defined as follows:
\begin{equation}
Q(t)= \frac{1}{2} \iint \left( \omega_{x}^{2} + \omega_{y}^{2} + \omega_{z}^{2} \right) dx \, dy \, dz .
\label{eq:41}
\end{equation}
The second criterion is the correlation coefficient between the DNS and the LES data.
For any two fields $f$ and $g$, which in our case will be velocity or vorticity components, the standard correlation coefficient \cite{pruett2000priori} is given by the following formula:
\begin{equation}
C(f,g) = \frac{\langle fg \rangle -\langle f \rangle \langle g \rangle }{[(\langle f^2 \rangle -\langle f \rangle^2)(\langle g^2 \rangle -\langle g \rangle^2)]^{1/2}}
\label{eq:corre}
\end{equation}
where the angle brackets denote the volume averages over the entire domain (e.g., $\langle f \rangle = \int f dV/V $, where $dV=dx dy dz$).
The third criterion is the $L^2$-norms of the flow variables. The reference solution for computing the $L^2$ norms is the DNS, which is obtained at a resolution of $256^3$, while all other LES computations are performed on a coarser resolution of $64^3$.
The same coarse resolution results without the AD procedure are also included for comparison purposes.
The fourth criterion is the third-order structure function, which is defined as
\begin{equation}
\langle \delta \boldsymbol u (r)^{3}\rangle = \langle | \boldsymbol u(\textbf{x} + \textbf{r}) - \boldsymbol u(\textbf{x})|^{3} \rangle
\label{eq:str}
\end{equation}
with $r=|\textbf{r}|$ being the spatial separation.
Note that, since the turbulence is assumed isotropic, the velocity increment depends only on the modulus of the vector $\textbf{r}$.
According to the Kolmogorov theory of turbulence \cite{frisch1995turbulence,davidson2004turbulence}, the nth-order structure function scales as $\langle \delta \boldsymbol u (r)^{n}\rangle \sim r^{n/3}$ in the inertial range ($\eta \ll r \ll L$).

\begin{figure}
\centering
\mbox{
\subfigure[~$N=2$]{\includegraphics[width=0.5\textwidth]{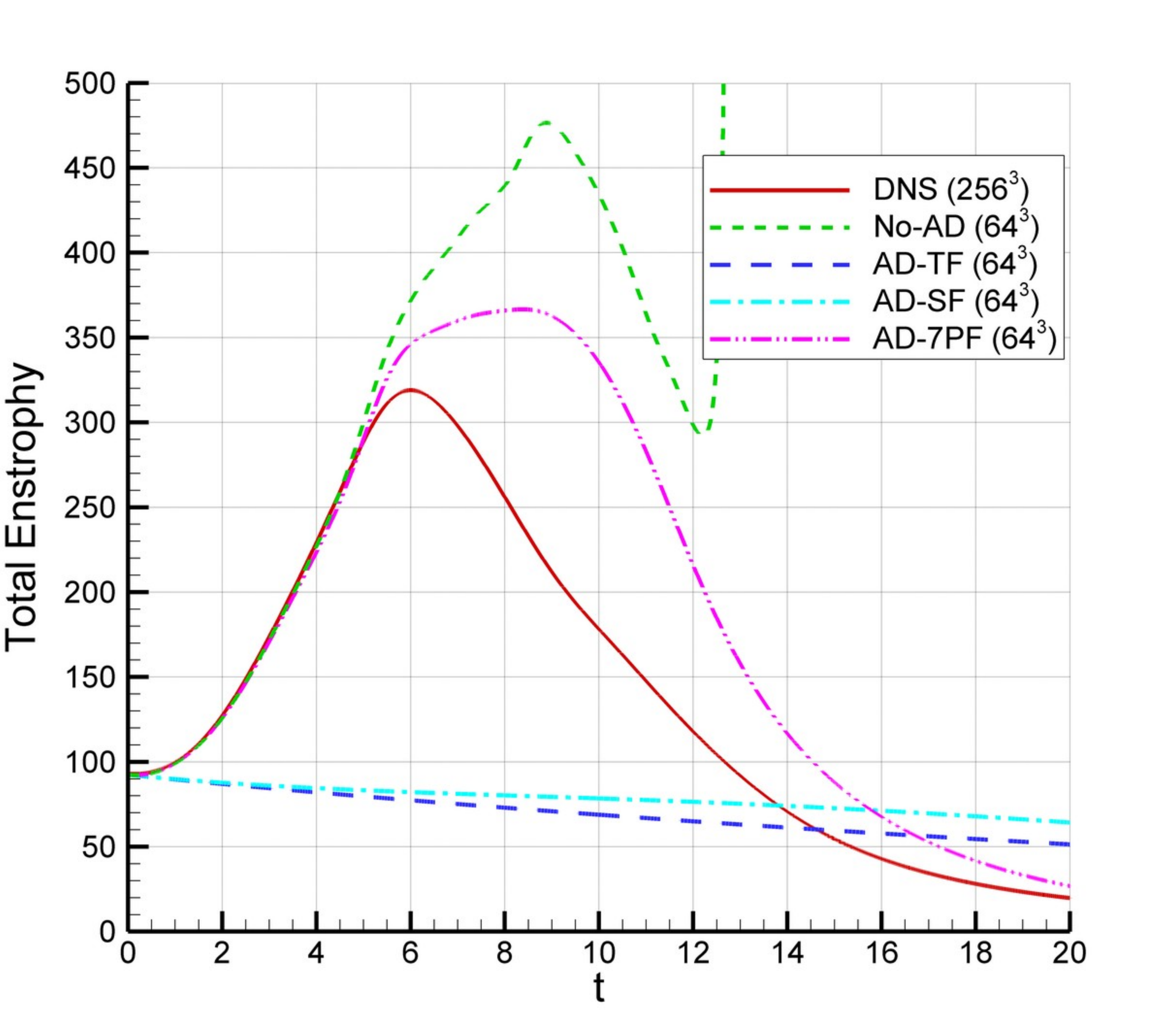}}
\subfigure[~$N=5$]{\includegraphics[width=0.5\textwidth]{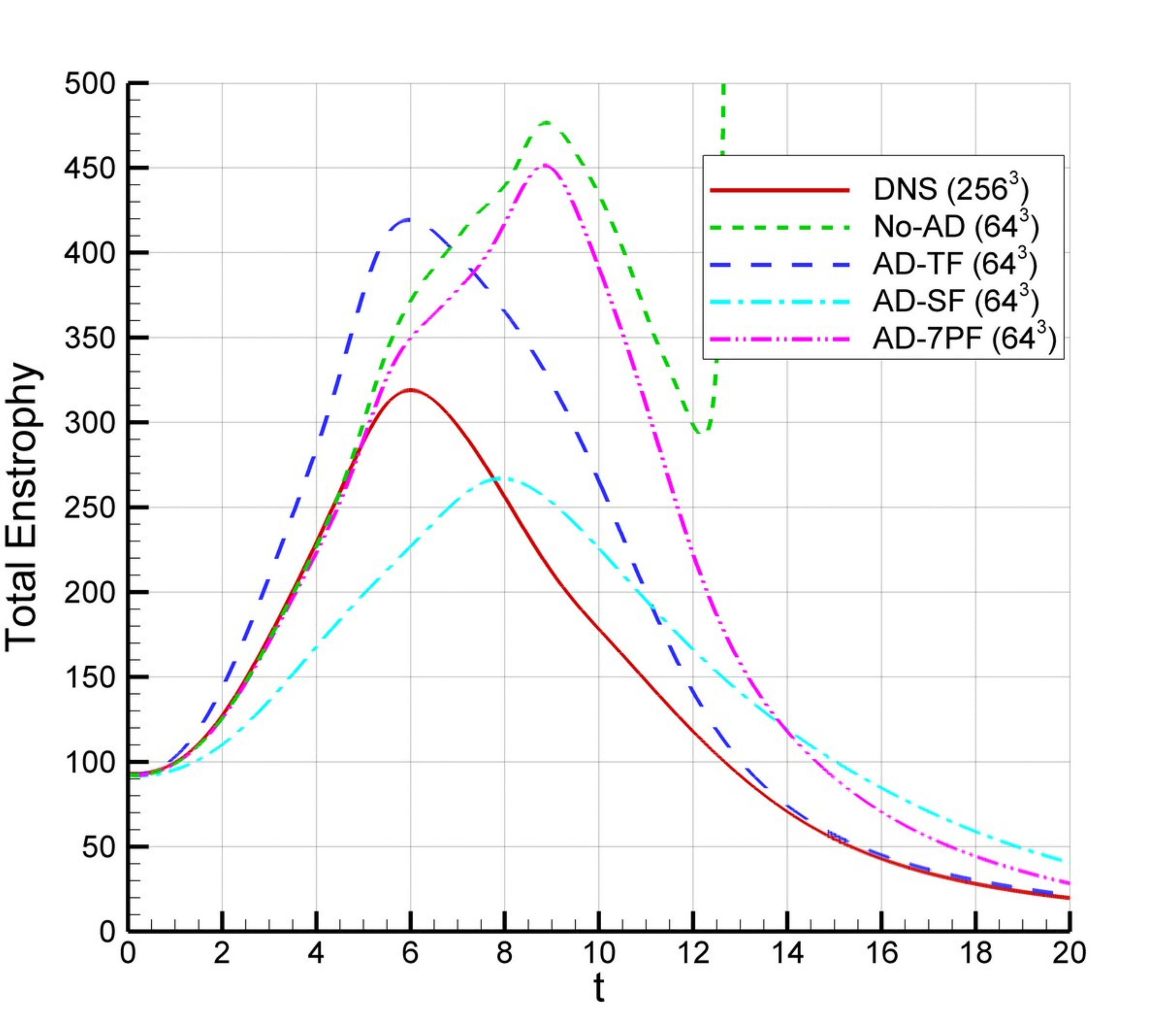}}
}
\caption{Time series of total enstrophy for the box filters.}
\label{fig:q-box}
\end{figure}

\begin{figure}
\centering
\mbox{
\subfigure[~$N=2$]{\includegraphics[width=0.5\textwidth]{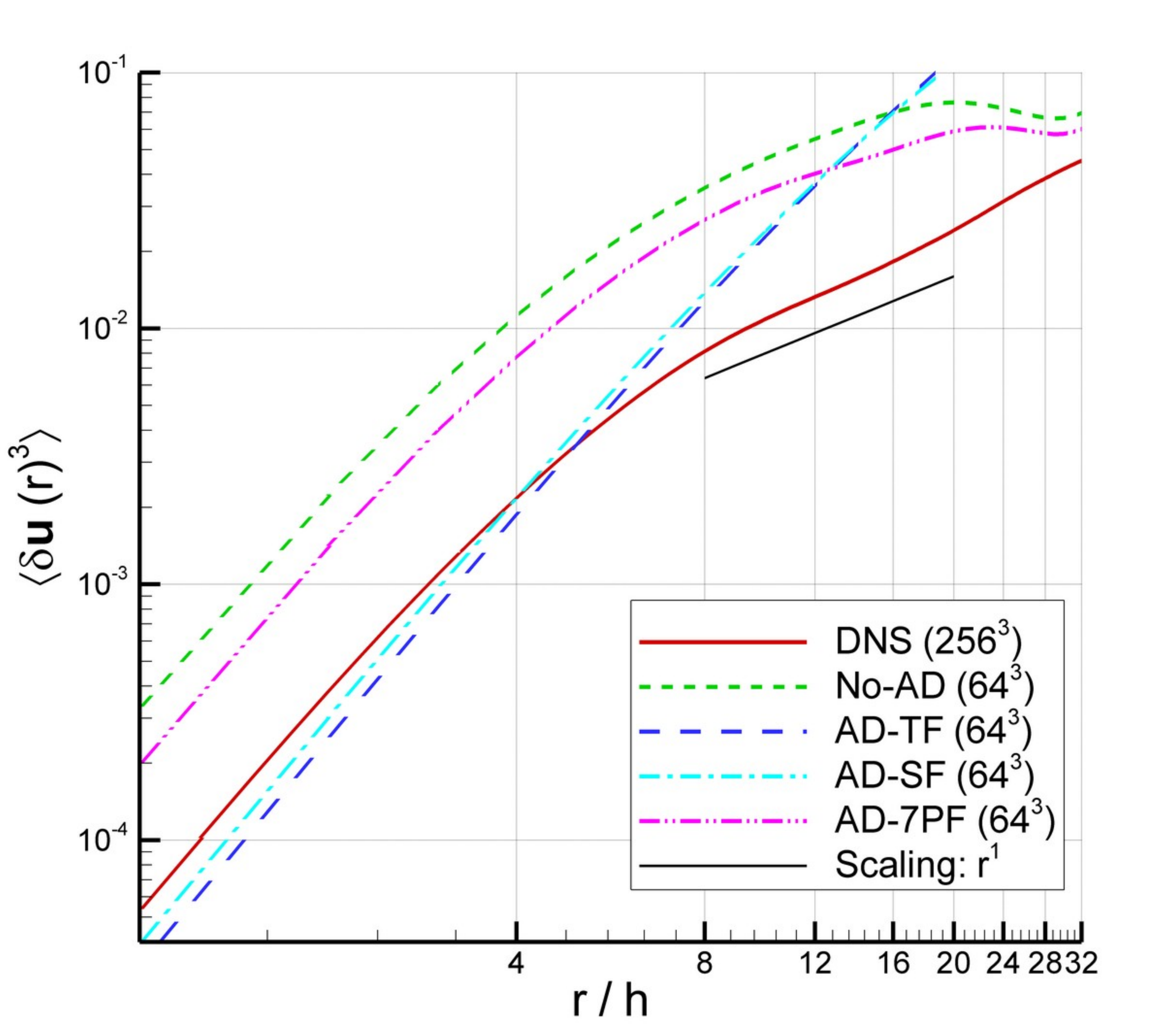}}
\subfigure[~$N=5$]{\includegraphics[width=0.5\textwidth]{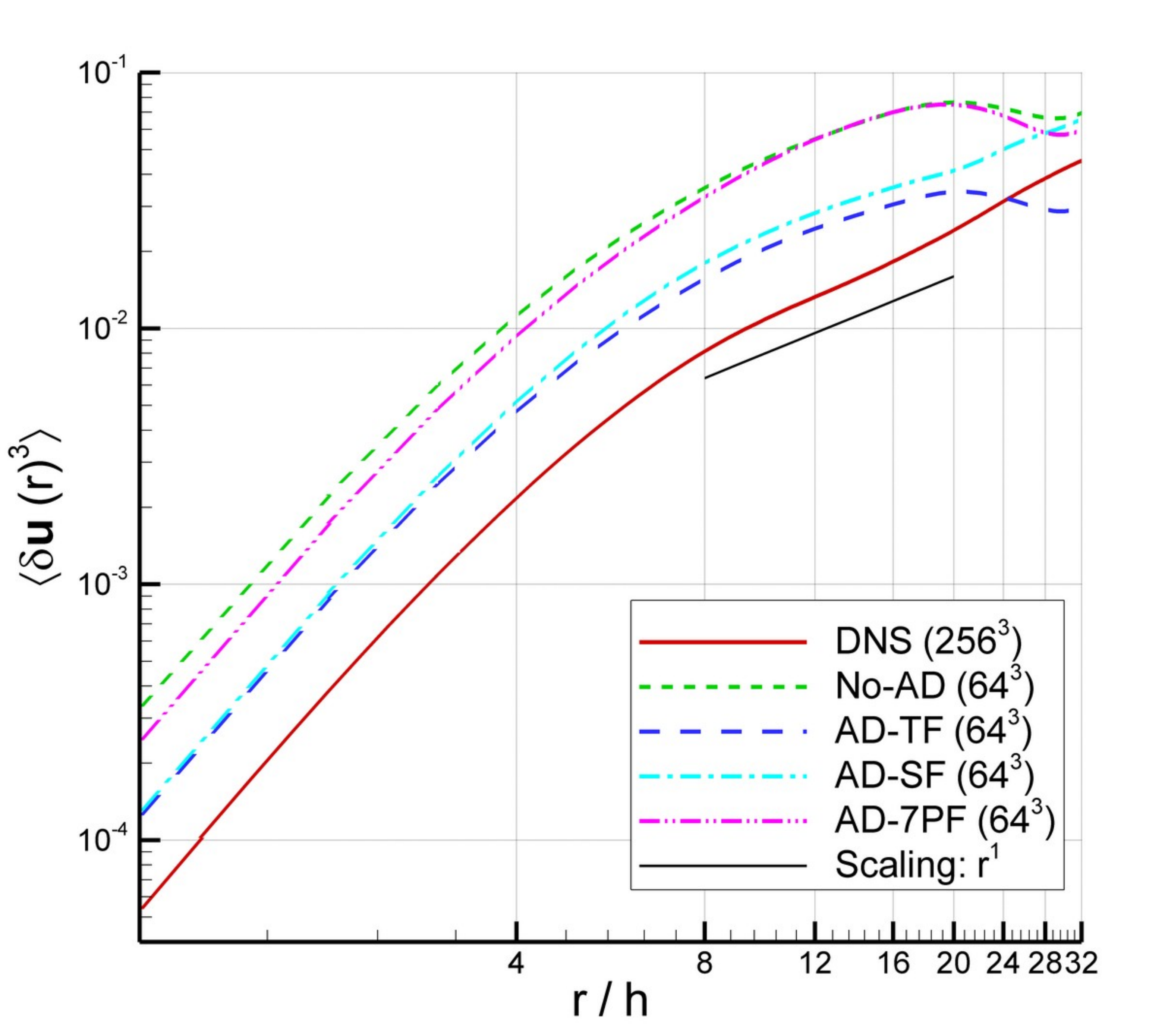}}
}
\caption{The third-order structure functions at $t=10$ for the box filters.}
\label{fig:s-box}
\end{figure}


\begin{table}
\centering
\caption{Discrete $L^2$-norms using the explicit box filters (with resolutions of $64^3$) for ensemble averaging the data on a time interval between $t=8$ and $t=12$. The reference solution for computing the $L^2$-norm is the DNS data obtained with a resolution of $256^3$.}
\begin{tabular}{llllllll}
\hline
Field & No-AD & \multicolumn{2}{l}{\underline{Trapezoidal filter}} & \multicolumn{2}{l}{\underline{Simpson filter}} & \multicolumn{2}{l}{\underline{7-point filter}} \\
 &  & $N=2$ & $N=5$ & $N=2$ & $N=5$ & $N=2$ & $N=5$ \\
\hline
$u$         & 0.151 & 0.214 & 0.132 & 0.199 & 0.083 & 0.131 & 0.152 \\
$v$         & 0.169 & 0.216 & 0.144 & 0.200 & 0.083 & 0.139 & 0.169 \\
$w$         & 0.166 & 0.124 & 0.134 & 0.125 & 0.094 & 0.146 & 0.170 \\
$\omega_x$  & 1.233 & 0.739 & 0.844 & 0.727 & 0.586 & 0.875 & 0.995 \\
$\omega_y$  & 1.001 & 0.730 & 0.840 & 0.721 & 0.623 & 0.810 & 0.898 \\
$\omega_z$  & 1.307 & 0.639 & 0.801 & 0.611 & 0.468 & 0.806 & 0.937 \\
\hline
\end{tabular}
\label{tab:L2-box-a}
\end{table}


\begin{table}
\centering
\caption{Correlation coefficient between the DNS data (with a resolution of $256^3$) and AD results with explicit box filters (with resolutions of $64^3$) for ensemble averaging on a time interval between $t=8$ and $t=12$. Correlation coefficients between the DNS and No-AD model (with a resolution of $64^3$) are also listed for comparison purposes.}
\begin{tabular}{llllllll}
\hline
Field & $C(\mbox{DNS, No-AD})$ & \multicolumn{2}{l}{\underline{$C(\mbox{DNS, TF} )$}} & \multicolumn{2}{l}{$\underline{C(\mbox{DNS, SF} )}$} & \multicolumn{2}{l}{$\underline{C(\mbox{DNS, 7PF} )}$} \\
 &                  &  $N=2$ & $N=5$ & $N=2$ & $N=5$ & $N=2$ & $N=5$ \\
\hline
$u$         & 0.692 & 0.732 & 0.700 & 0.760 & 0.922 & 0.751 & 0.664 \\
$v$         & 0.528 & 0.730 & 0.570 & 0.761 & 0.899 & 0.615 & 0.500 \\
$w$         & 0.552 & 0.305 & 0.577 & 0.392 & 0.825 & 0.661 & 0.565 \\
$\omega_x$  & 0.342 & 0.190 & 0.440 & 0.280 & 0.751 & 0.493 & 0.378 \\
$\omega_y$  & 0.410 & 0.206 & 0.405 & 0.287 & 0.696 & 0.503 & 0.440 \\
$\omega_z$  & 0.277 & 0.433 & 0.378 & 0.461 & 0.703 & 0.442 & 0.350 \\
\hline
\end{tabular}
\label{tab:corr-box-a}
\end{table}

\subsection{The AD-LES model with box filters}
\label{sec:numerics_box}

In this section, we numerically investigate the AD-LES model in conjunction with the three box filters discussed in Section \ref{sec:dbf}: the TP \eqref{eq:trap}, the SF \eqref{eq:simp}, and the 7PF \eqref{eq:spoint}.
The resulting LES models are denoted as AD-TF, AD-SF, and AD-7PF, respectively.

Fig.~\ref{fig:q-box} presents the time series of the integrated enstrophy $Q(t)$ defined in \eqref{eq:41} for the AD-TF, AD-SF and AD-7PF with $N=2$ and $N=5$.
Results for the DNS and No-AD are also included for comparison purposes.
For $N=2$, the AD-7PF performs the best, and AD-TF and AD-SF perform badly.
For $N=5$, the AD-SF performs the best.
Comparing the $N=2$ plot with the $N=5$ plot, the AD-SF with $N=5$ performs the best.
As expected, the No-AD performs the worst for both $N=2$ and $N=5$.
In fact, the numerical simulation with the No-AD blows up around $t=12$.

Fig.~\ref{fig:s-box} presents the third-order structure function defined in \eqref{eq:str} for the AD-TF, AD-SF and AD-7PF with $N=2$ and $N=5$ at $t=10$.
Results for the DNS and No-AD are also included for comparison purposes.
For $N=2$, the AD-TF and AD-SF perform the best.
For $N=5$, the AD-TF and AD-SF again perform the best, with a plus for the former.
Comparing the $N=2$ plot with the $N=5$ plot, the AD-TF and AD-SF with $N=2$ perform the best for the small values of the ratio $r / h$, and the AD-TF with $N=5$ performs the best for the large values of the ratio $r / h$.
As expected, the No-AD performs the worst for both $N=2$ and $N=5$.

Table \ref{tab:L2-box-a} presents the $L^2$-norm of the error of the AD-TF, AD-SF, and AD-7PF for $N=2$ and $N=5$.
Results for the No-AD are also included for comparison purposes.
The errors are averaged over the time interval $8 \leq t \leq 12$.
For $N=2$, the AD-SF performs the best.
For $N=5$, the AD-SF again performs the best.
Comparing the $N=2$ results with the $N=5$ results, the AD-SF with $N=5$ consistently performs the best.
As expected, the No-AD performs the worst.
Table \ref{tab:L2-box-a} also shows that there is a significant sensitivity of the numerical results with respect to $N$.
For the AD-TF, increasing $N$ does not yield a consistent qualitative change - for some flow variables the error increases, for others it decreases.
For the AD-SF, increasing $N$ yields a consistent dramatic decrease in the error.
Finally, for the AD-7PF, increasing $N$ results in a consistent significant increase in the error.
We also note that in general the velocity components have lower errors than the vorticity components.
This is true for the No-AD run, for the AD-LES models, and for both $N=2$ and $N=5$.
We attribute this behavior to the fact that the vorticity requires first order derivatives of the velocity components and, thus, is less accurately approximated than the velocity.

Table \ref{tab:corr-box-a} presents the correlation coefficients for the AD-TF, AD-SF, and AD-7PF for $N=2$ and $N=5$.
The correlation coefficients are averaged over the time interval $8 \leq t \leq 12$.
Results for the No-AD are also included for comparison purposes.
For $N=2$, the AD-SF and the AD-7PF perform the best.
For $N=5$, the AD-SF performs the best.
Comparing the $N=2$ results with the $N=5$ results, the AD-SF with $N=5$ consistently performs the best.
As expected, the No-AD performs the worst.
Increasing $N$ has the following effect on the correlation coefficients: no consistent trend for the AD-TF, a consistent dramatic increase for the AD-SF, and a consistent significant decrease for the AD-7PF.
In general, the correlation coefficients of the velocity components are higher than those of the vorticity components.

Based on the results in Figs.~\ref{fig:q-box} and \ref{fig:s-box} and Tables~\ref{tab:L2-box-a} and \ref{tab:corr-box-a}, we conclude that the AD-SF with $N=5$ yields the best overall results.

\begin{figure}
\centering
\mbox{
\subfigure[~$N=2$]{\includegraphics[width=0.5\textwidth]{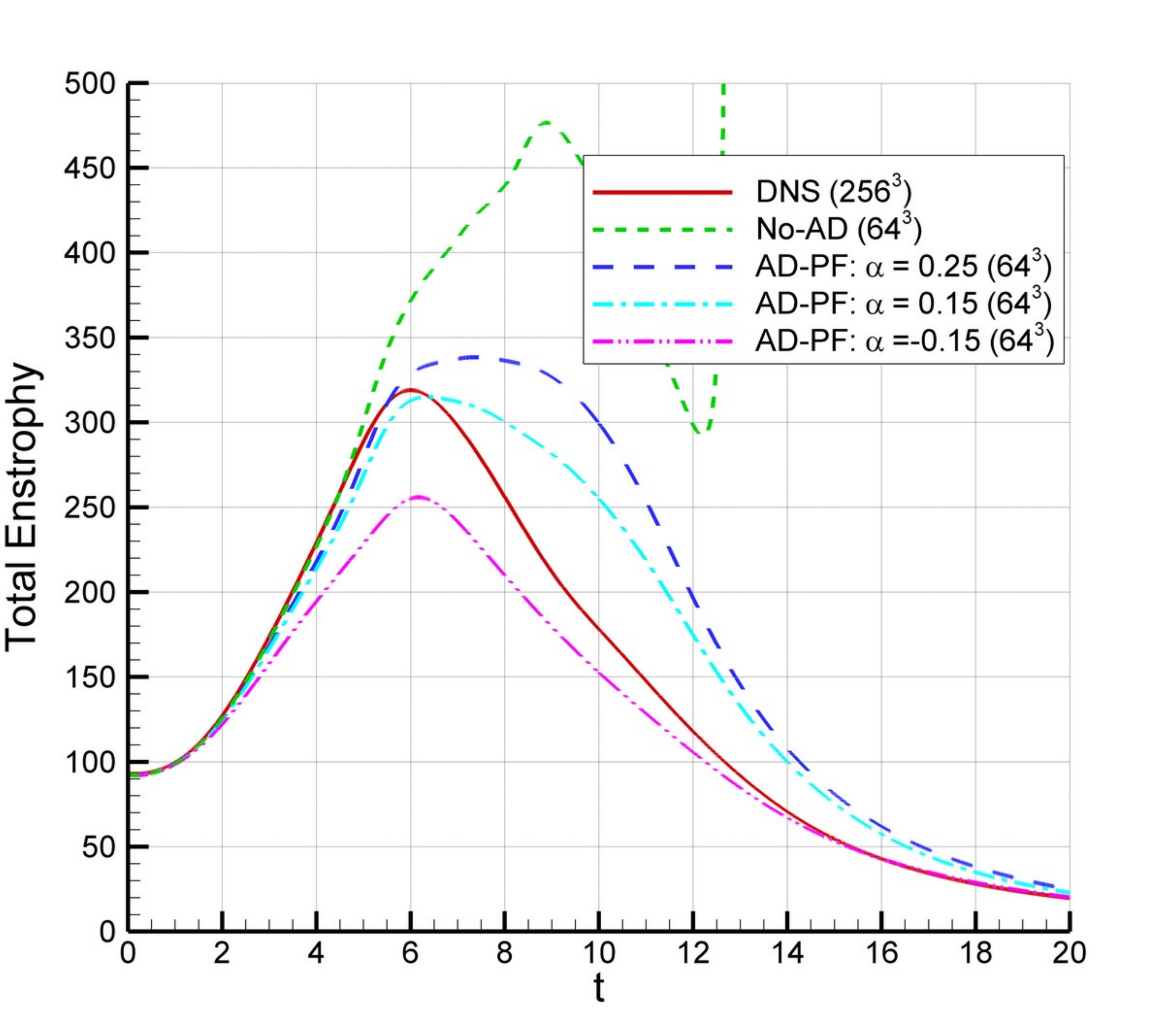}}
\subfigure[~$N=5$]{\includegraphics[width=0.5\textwidth]{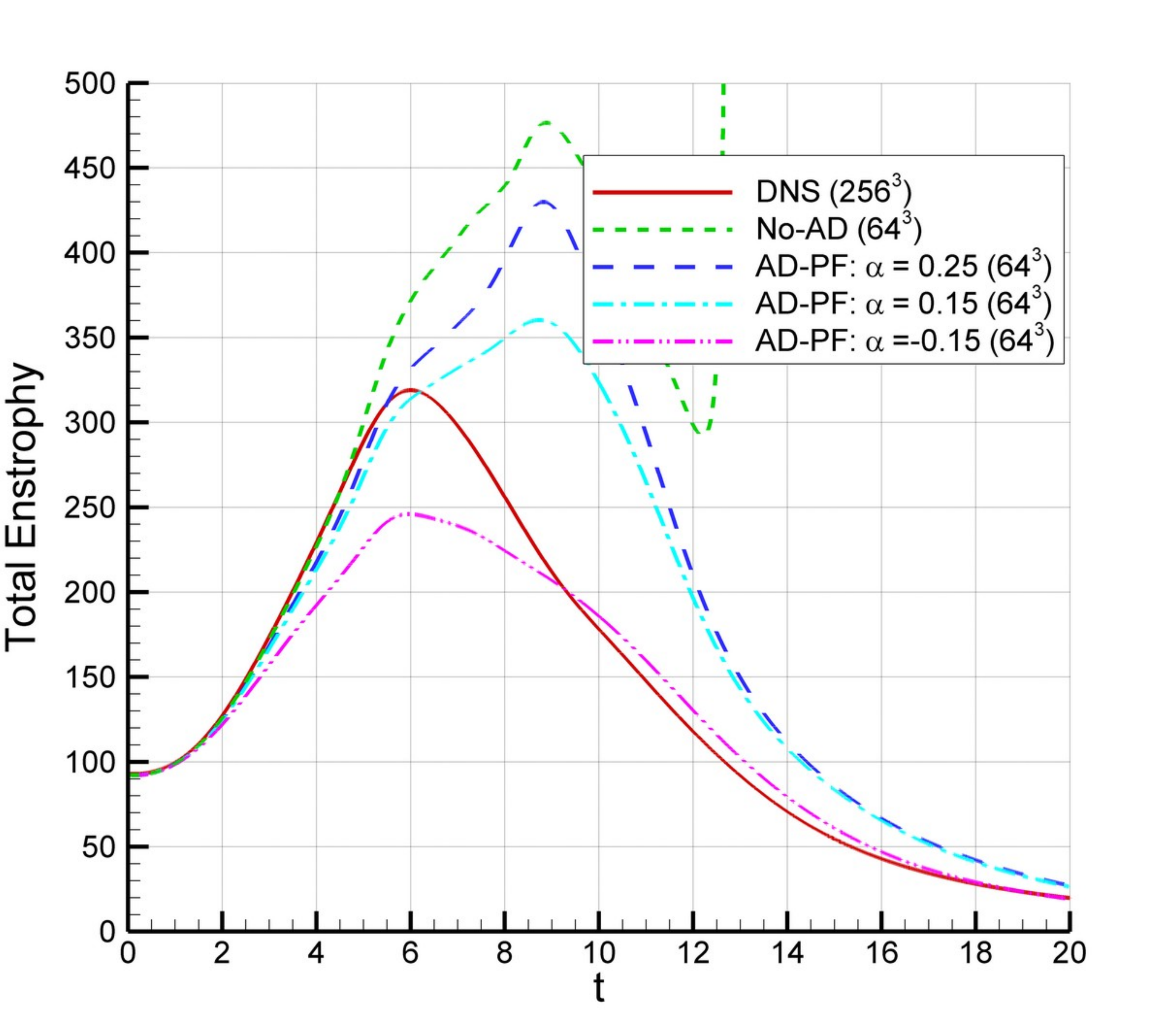}}
}
\caption{Time series of total enstrophy for the second-order Pad\'{e}-type filters.}
\label{fig:q-pade}
\end{figure}

\begin{figure}
\centering
\mbox{
\subfigure[~$N=2$]{\includegraphics[width=0.5\textwidth]{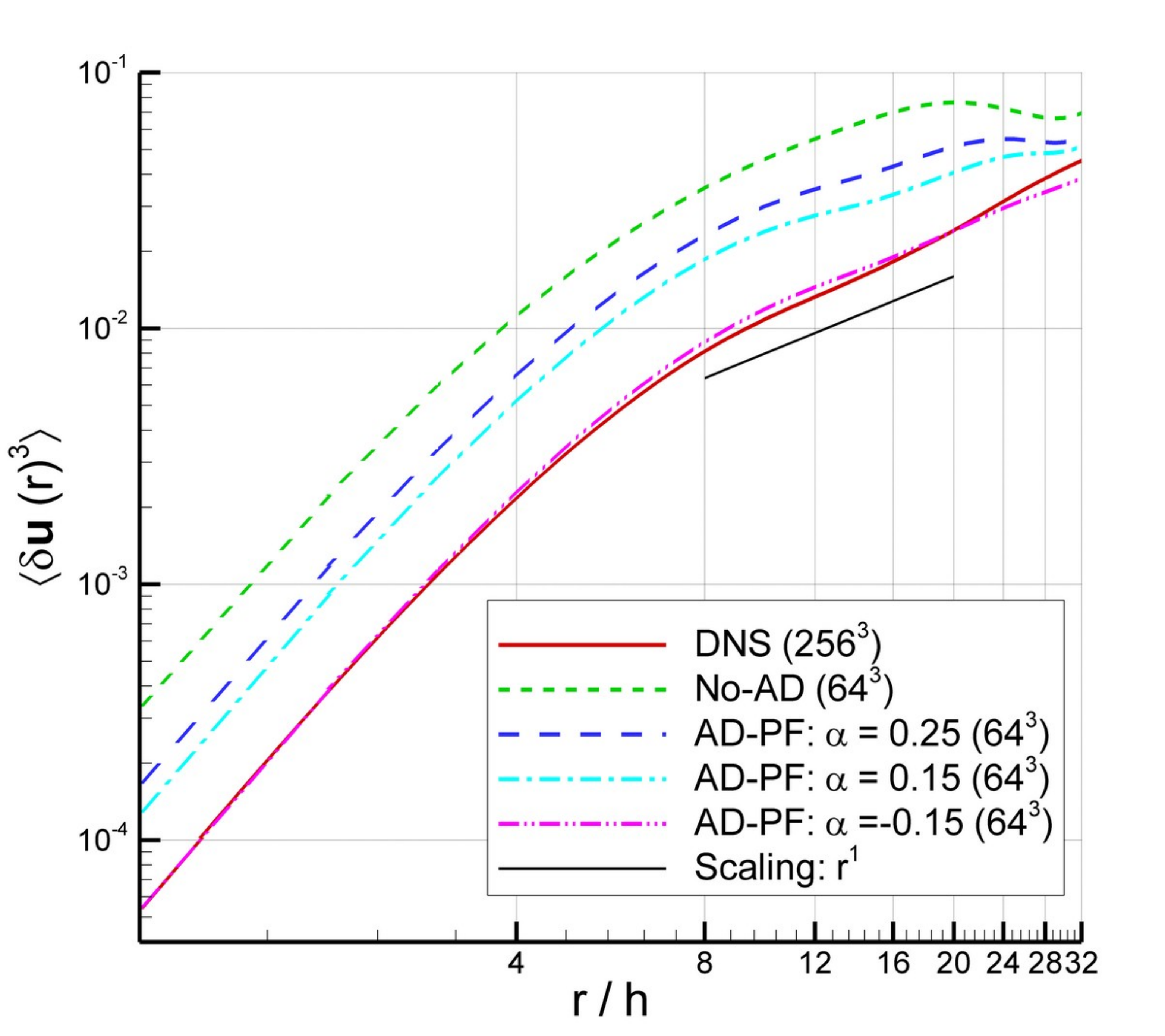}}
\subfigure[~$N=5$]{\includegraphics[width=0.5\textwidth]{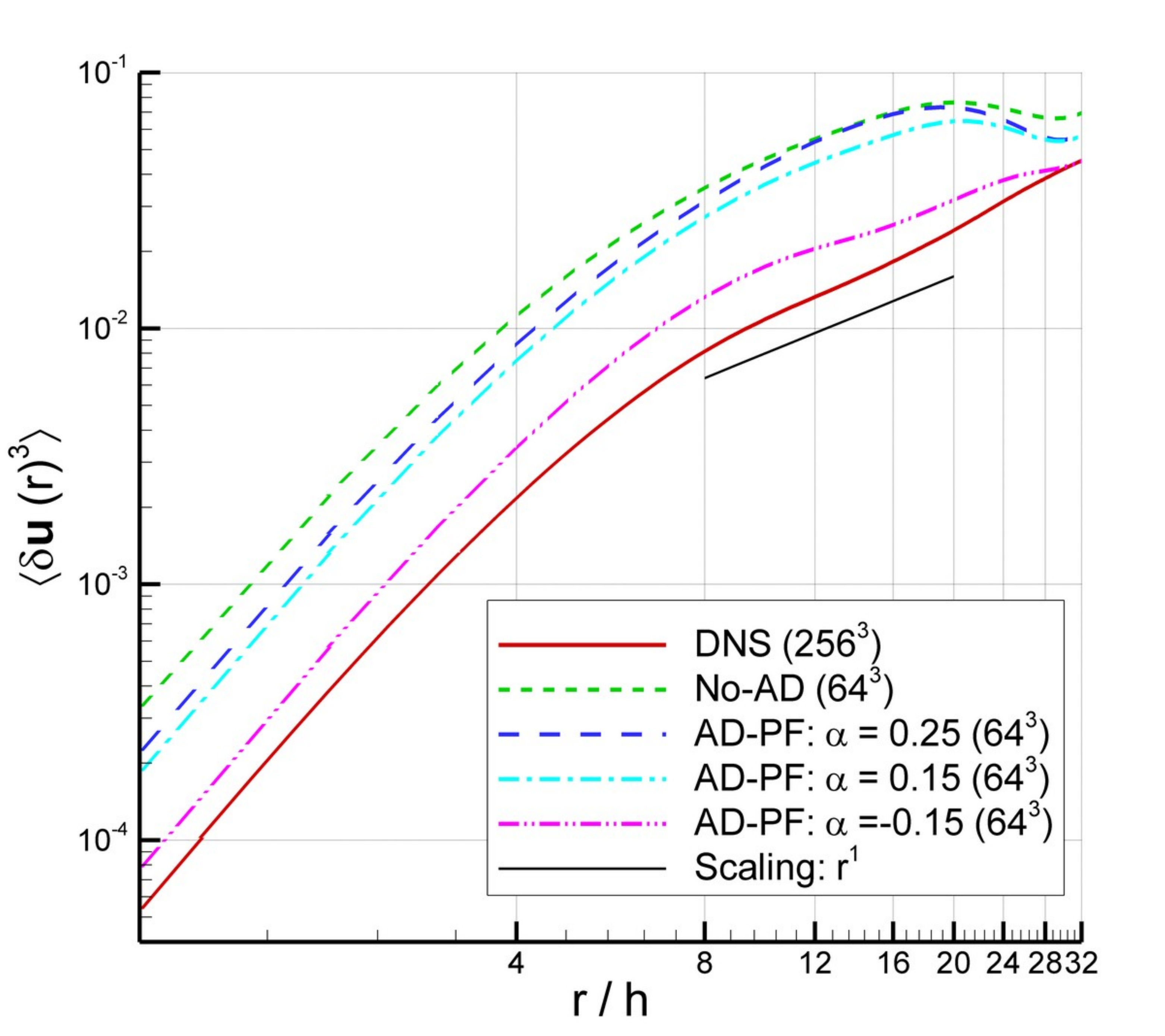}}
}
\caption{The third-order structure functions at $t=10$ for the second-order Pad\'{e}-type filters.}
\label{fig:s-pade}
\end{figure}


\begin{table}
\centering
\caption{Discrete $L^2$-norms using the second-order Pad\'{e} filters (with resolutions of $64^3$) for ensemble averaging the data on a time interval between $t=8$ and $t=12$. The reference solution for computing the $L^2$-norm is the DNS data obtained with a resolution of $256^3$.}
\begin{tabular}{llllllll}
\hline
Field & No-AD & \multicolumn{2}{l}{\underline{PF($\alpha=0.25$)}} & \multicolumn{2}{l}{\underline{PF($\alpha=0.15$)}} & \multicolumn{2}{l}{\underline{PF($\alpha=-0.15$)}} \\
 &  & $N=2$ & $N=5$ & $N=2$ & $N=5$ & $N=2$ & $N=5$ \\
\hline
$u$         & 0.151 & 0.124 & 0.152 & 0.107 & 0.141 & 0.065 & 0.096 \\
$v$         & 0.169 & 0.130 & 0.170 & 0.113 & 0.153 & 0.073 & 0.101 \\
$w$         & 0.166 & 0.137 & 0.170 & 0.118 & 0.157 & 0.075 & 0.106 \\
$\omega_x$  & 1.233 & 0.820 & 0.980 & 0.722 & 0.895 & 0.478 & 0.615 \\
$\omega_y$  & 1.001 & 0.762 & 0.873 & 0.682 & 0.796 & 0.493 & 0.581 \\
$\omega_z$  & 1.307 & 0.747 & 0.915 & 0.639 & 0.810 & 0.398 & 0.518 \\
\hline
\end{tabular}
\label{tab:L2-pade-a}
\end{table}


\begin{table}
\centering
\caption{Correlation coefficient between the DNS data (with a resolution of $256^3$) and AD results with the second-order Pad\'{e} filters (with resolutions of $64^3$) for ensemble averaging on a time interval between $t=8$ and $t=12$. Correlation coefficients between the DNS and No-AD model (with a resolution of $64^3$) are also listed for comparison purposes.}
\begin{tabular}{llllllll}
\hline
Field & $C(\mbox{DNS, No-AD})$ & \multicolumn{2}{l}{\underline{$C(\mbox{DNS}, \alpha=0.25)$}} & \multicolumn{2}{l}{$\underline{C(\mbox{DNS}, \alpha=0.15)}$} & \multicolumn{2}{l}{$\underline{C(\mbox{DNS}, \alpha=-0.15)}$} \\
 &                  & $N=2$ & $N=5$ & $N=2$ & $N=5$ & $N=2$ & $N=5$ \\
\hline
$u$         & 0.692 & 0.771 & 0.657 & 0.821 & 0.705 & 0.924 & 0.848 \\
$v$         & 0.528 & 0.648 & 0.492 & 0.721 & 0.556 & 0.887 & 0.768 \\
$w$         & 0.552 & 0.678 & 0.554 & 0.723 & 0.606 & 0.816 & 0.725 \\
$\omega_x$  & 0.342 & 0.521 & 0.368 & 0.590 & 0.434 & 0.756 & 0.627 \\
$\omega_y$  & 0.410 & 0.521 & 0.439 & 0.571 & 0.480 & 0.717 & 0.608 \\
$\omega_z$  & 0.277 & 0.465 & 0.348 & 0.535 & 0.404 & 0.729 & 0.589 \\
\hline
\end{tabular}
\label{tab:corr-pade-a}
\end{table}

\subsection{The AD-LES model with Pad\'{e}-type filters}
\label{sec:numerics_pade}

In this section, we numerically investigate the AD-LES model in conjunction with the Pad\'{e}-type filter given in \eqref{eq:119} and discussed in Section \ref{sec:pf}.
The resulting LES model is denoted as AD-PF.
The following values for the parameter $\alpha$ are considered: $\alpha = 0.25$, $\alpha = 0.15$, and $\alpha = -0.15$.

Fig.~\ref{fig:q-pade} presents the time series of the integrated enstrophy $Q(t)$ defined in \eqref{eq:41} for the AD-PF with $N=2$ and $N=5$.
Results for the DNS and No-AD are also included for comparison purposes.
For $N=2$, $\alpha = -0.15$ yields the best results.
For $N=5$, $\alpha = -0.15$ yields again the best results.
Comparing the $N=2$ plot with the $N=5$ plot, $\alpha = -0.15$ yields similar (good) results for both cases.
As expected, the No-AD performs the worst for both $N=2$ and $N=5$.

Fig.~\ref{fig:s-pade} presents the third-order structure function defined in \eqref{eq:str} for the AD-PF with $N=2$ and $N=5$ at $t=10$.
Results for the DNS and No-AD are also included for comparison purposes.
For $N=2$, $\alpha = -0.15$ yields the best results, just as it did in Fig.~\ref{fig:q-pade}.
For $N=5$, $\alpha = -0.15$ yields again the best results..
Comparing the $N=2$ plot with the $N=5$ plot, $\alpha = -0.15$ with $N=2$ consistently performs the best.
As expected, the No-AD performs the worst for both $N=2$ and $N=5$.

Table \ref{tab:L2-pade-a} presents the $L^2$-norm of the error of the AD-PF for $N=2$ and $N=5$.
Results for the No-AD are also included for comparison purposes.
The errors are averaged over the time interval $8 \leq t \leq 12$.
For $N=2$, $\alpha = -0.15$ yields the best results.
For $N=5$, $\alpha = -0.15$ yields again the best results.
Comparing the $N=2$ results with the $N=5$ results, $\alpha = -0.15$ with $N=2$ consistently performs the best.
As expected, the No-AD performs the worst.
Increasing $N$ results in a consistent significant increase for all values of the parameter $\alpha$.
In general, the errors of the velocity components are lower than those of the vorticity components.

Table \ref{tab:corr-pade-a} presents the correlation coefficients for the AD-PF for $N=2$ and $N=5$.
Results for the No-AD are also included for comparison purposes.
The correlation coefficients are averaged over the time interval $8 \leq t \leq 12$.
For $N=2$, $\alpha = -0.15$ yields the best results.
For $N=5$, $\alpha = -0.15$ yields again the best results.
Comparing the $N=2$ results with the $N=5$ results, $\alpha = -0.15$ with $N=2$ consistently performs the best.
As expected, the No-AD performs the worst for both $N=2$ and $N=5$.
Increasing $N$ results in a consistent significant decrease for all values of the parameter $\alpha$.
In general, the correlation coefficients of the velocity components are higher than those of the vorticity components.

Based on the results in Figs.~\ref{fig:q-pade} and \ref{fig:s-pade} and Tables~\ref{tab:L2-pade-a} and \ref{tab:corr-pade-a}, we conclude that the AD-PF with $\alpha = -0.15$ and $N=2$ yields the best overall results.

\begin{figure}
\centering
\mbox{
\subfigure[~$N=2$]{\includegraphics[width=0.5\textwidth]{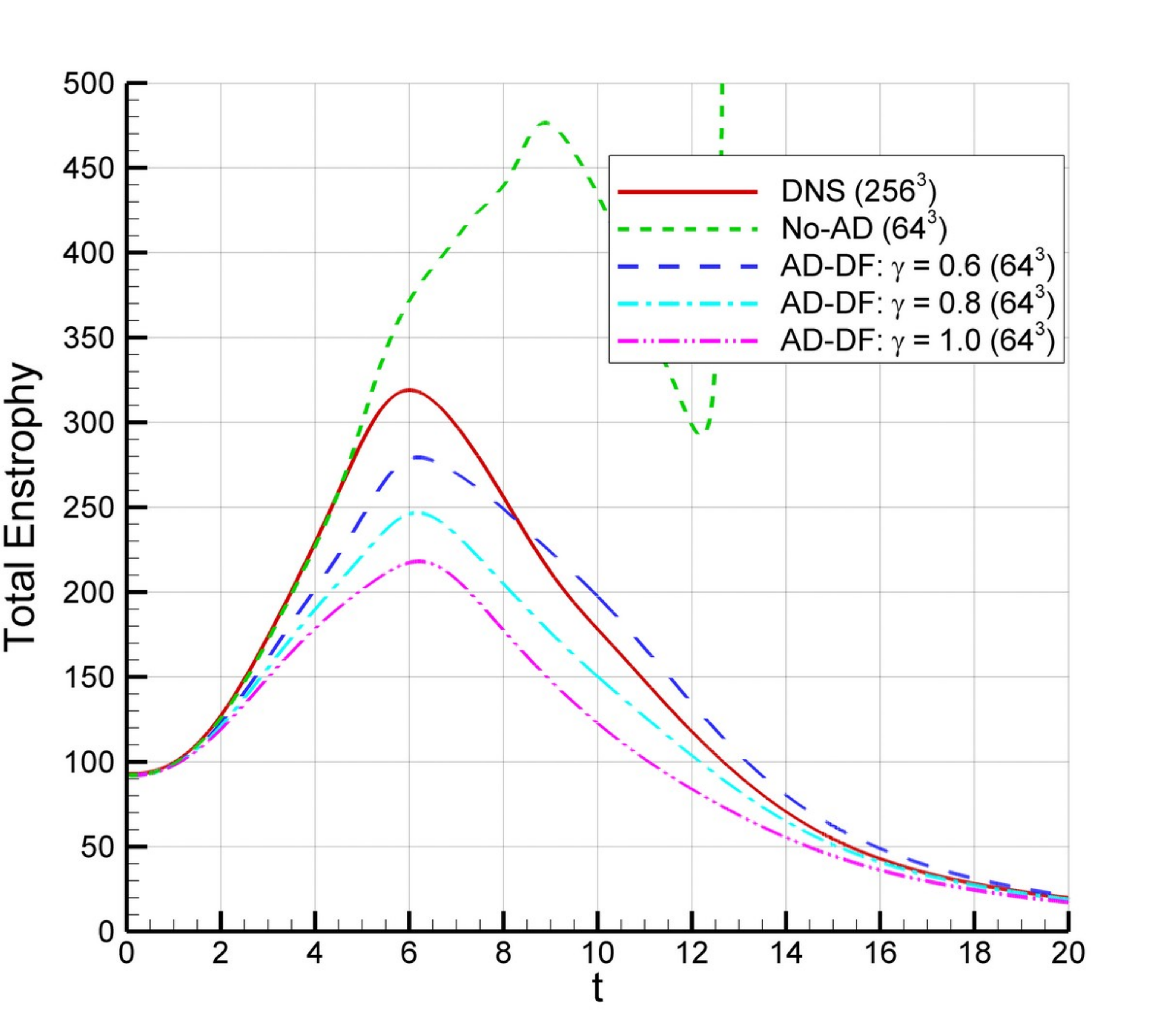}}
\subfigure[~$N=5$]{\includegraphics[width=0.5\textwidth]{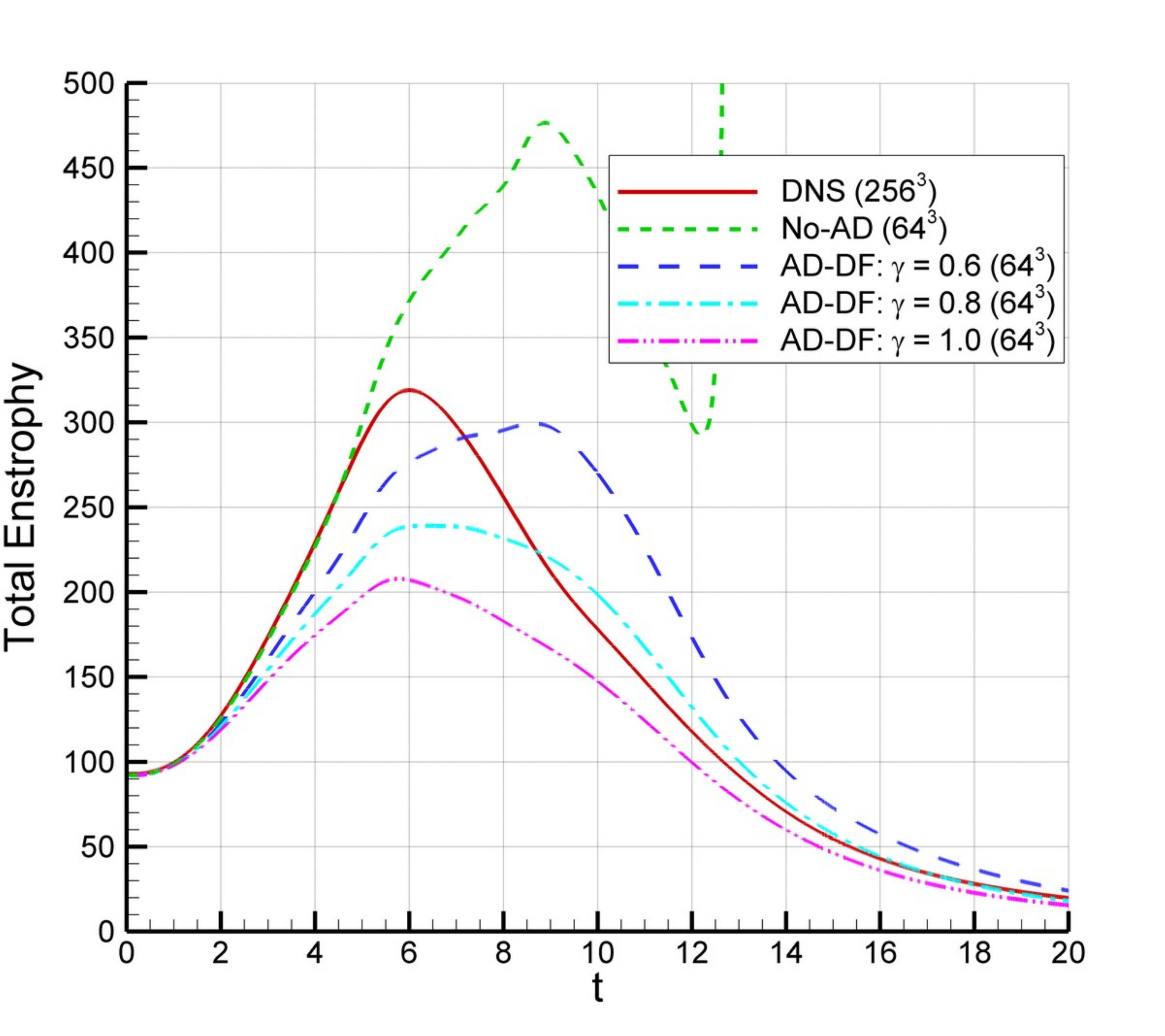}}
}
\caption{Time series of total enstrophy for the differential filters.}
\label{fig:q-helm}
\end{figure}

\begin{figure}
\centering
\mbox{
\subfigure[~$N=2$]{\includegraphics[width=0.5\textwidth]{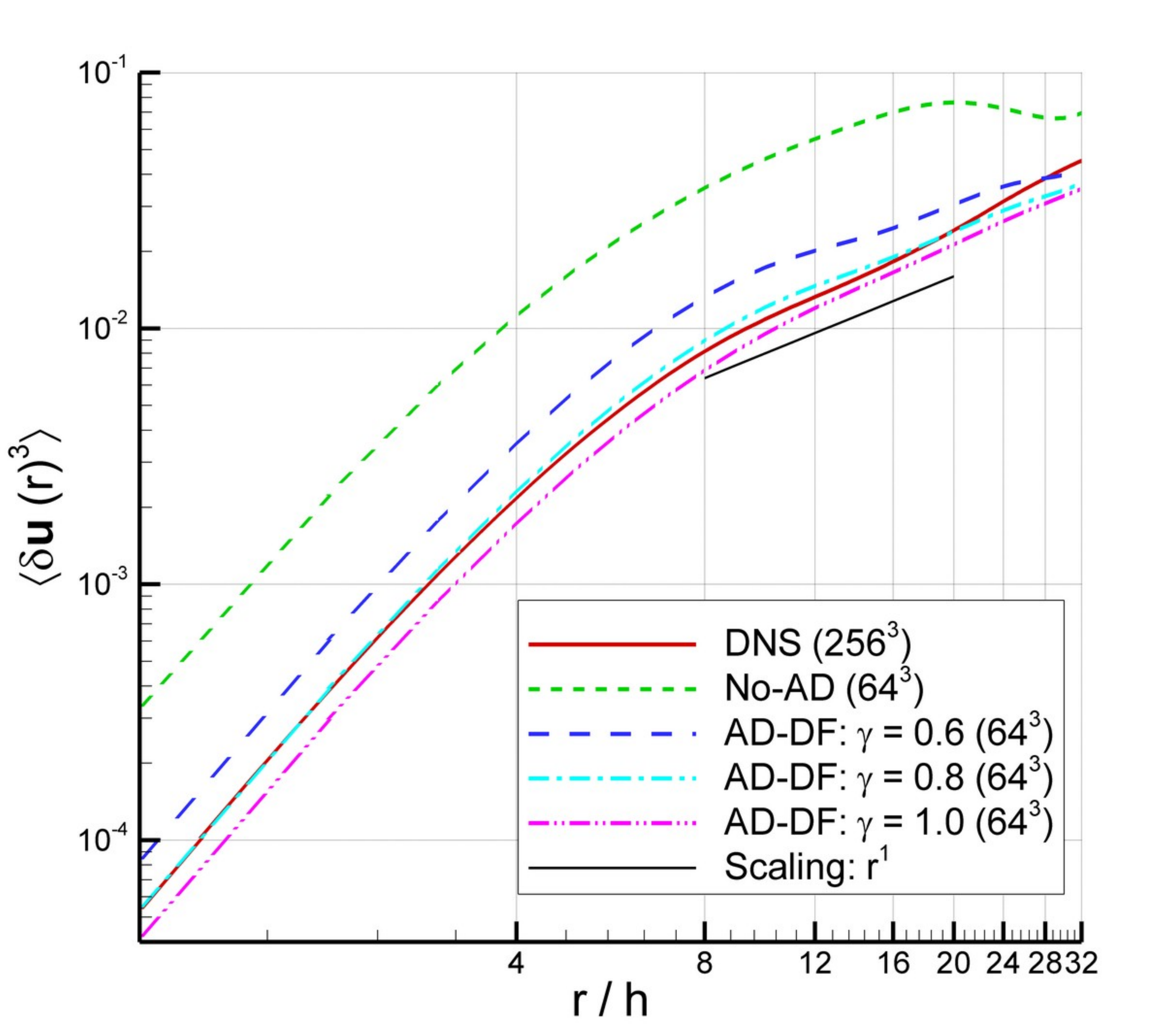}}
\subfigure[~$N=5$]{\includegraphics[width=0.5\textwidth]{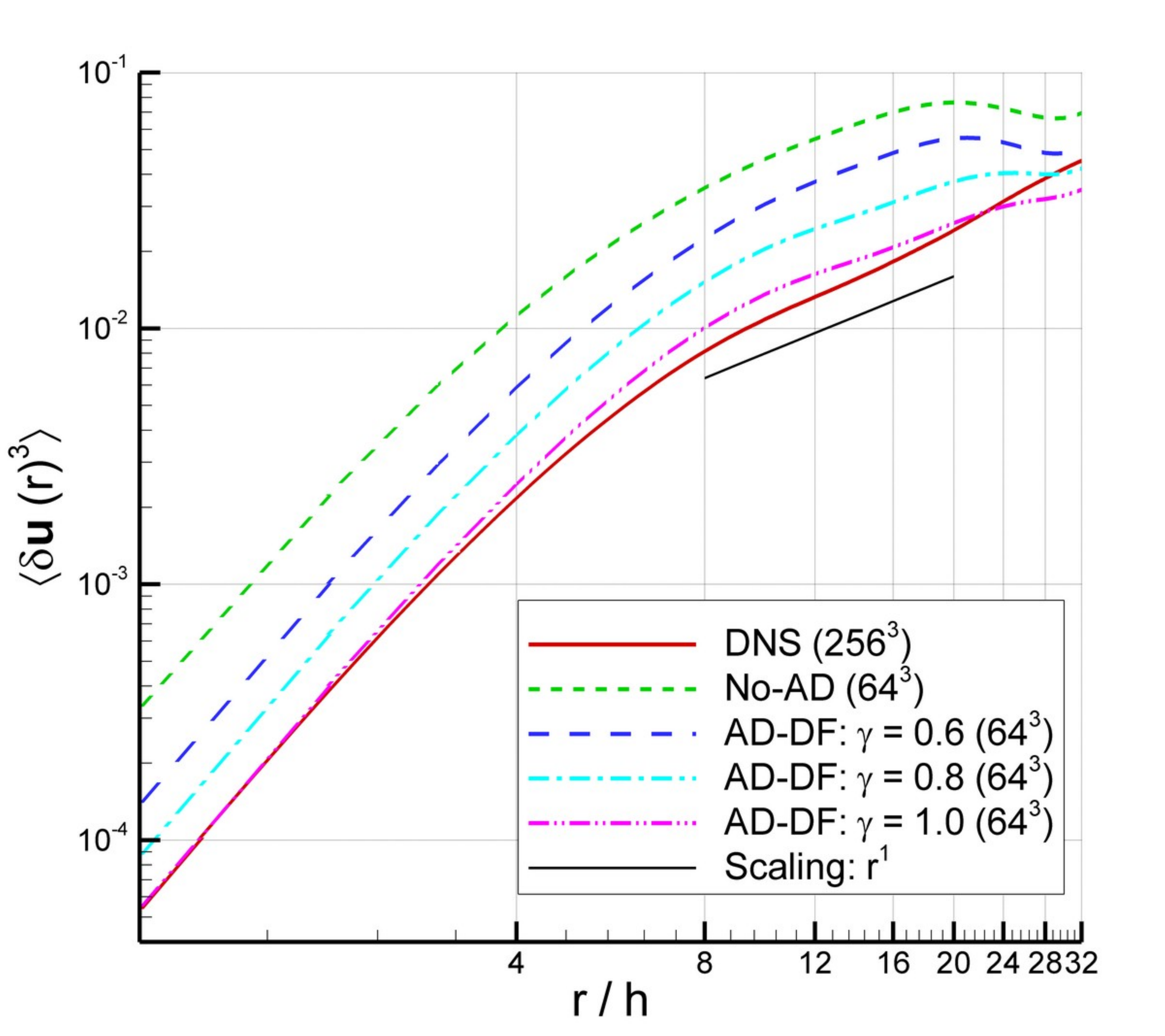}}
}
\caption{The third-order structure functions at $t=10$ for the differential filters.}
\label{fig:s-helm}
\end{figure}


\begin{table}
\centering
\caption{Discrete $L^2$-norms using the differential filters (with resolutions of $64^3$) for ensemble averaging the data on a time interval between $t=8$ and $t=12$. The reference solution for computing the $L^2$-norm is the DNS data obtained with a resolution of $256^3$.}
\begin{tabular}{llllllll}
\hline
Field & No-AD & \multicolumn{2}{l}{\underline{HF($\gamma=0.6$)}} & \multicolumn{2}{l}{\underline{HF($\gamma=0.8$)}} & \multicolumn{2}{l}{\underline{HF($\gamma=1.0$)}} \\
 &  & $N=2$ & $N=5$ & $N=2$ & $N=5$ & $N=2$ & $N=5$ \\
\hline
$u$         & 0.151 & 0.089 & 0.135 & 0.070 & 0.115 & 0.060 & 0.098 \\
$v$         & 0.169 & 0.097 & 0.147 & 0.078 & 0.123 & 0.067 & 0.106 \\
$w$         & 0.166 & 0.098 & 0.149 & 0.079 & 0.125 & 0.070 & 0.104 \\
$\omega_x$  & 1.233 & 0.608 & 0.843 & 0.490 & 0.705 & 0.426 & 0.606 \\
$\omega_y$  & 1.001 & 0.591 & 0.731 & 0.508 & 0.637 & 0.464 & 0.573 \\
$\omega_z$  & 1.307 & 0.532 & 0.752 & 0.421 & 0.607 & 0.367 & 0.516 \\
\hline
\end{tabular}
\label{tab:L2-helm-a}
\end{table}


\begin{table}
\centering
\caption{Correlation coefficient between the DNS data (with a resolution of $256^3$) and AD results with the differential filters (with resolutions of $64^3$) for ensemble averaging on a time interval between $t=8$ and $t=12$. Correlation coefficients between the DNS and No-AD model (with a resolution of $64^3$) are also listed for comparison purposes.}
\begin{tabular}{llllllll}
\hline
Field & $C(\mbox{DNS, No-AD})$ & \multicolumn{2}{l}{\underline{$C(\mbox{DNS}, \gamma=0.6)$}} & \multicolumn{2}{l}{$\underline{C(\mbox{DNS}, \gamma=0.8)}$} & \multicolumn{2}{l}{$\underline{C(\mbox{DNS}, \gamma=1.0)}$} \\
 &                  & $N=2$ & $N=5$ & $N=2$ & $N=5$ & $N=2$ & $N=5$ \\
\hline
$u$         & 0.692 & 0.862 & 0.715 & 0.911 & 0.779 & 0.936 & 0.832 \\
$v$         & 0.528 & 0.786 & 0.572 & 0.867 & 0.670 & 0.913 & 0.747 \\
$w$         & 0.552 & 0.757 & 0.607 & 0.798 & 0.665 & 0.817 & 0.700 \\
$\omega_x$  & 0.342 & 0.650 & 0.440 & 0.737 & 0.530 & 0.794 & 0.595 \\
$\omega_y$  & 0.410 & 0.625 & 0.491 & 0.694 & 0.541 & 0.743 & 0.585 \\
$\omega_z$  & 0.277 & 0.598 & 0.414 & 0.701 & 0.493 & 0.758 & 0.558 \\
\hline
\end{tabular}
\label{tab:corr-helm-a}
\end{table}

\subsection{The AD-LES model with differential filters}
\label{sec:numerics_differential}

In this section, we numerically investigate the AD-LES model in conjunction with the differential filter given in \eqref{eq:10a} and discussed in Section \ref{sec:diff}.
The resulting LES model is denoted as AD-DF.
The following values for the parameter $\alpha$ are considered: $\gamma = 0.6$, $\gamma = 0.8$, and $\gamma = 1.0$.

Fig.~\ref{fig:q-helm} presents the time series of the integrated enstrophy $Q(t)$ defined in \eqref{eq:41} for the AD-DF with $N=2$ and $N=5$.
Results for the DNS and No-AD are also included for comparison purposes.
For $N=2$, $\gamma = 0.6$ yields the best results.
For $N=5$, $\gamma = 0.8$ yields the best results.
Comparing the $N=2$ plot with the $N=5$ plot, the combination $\gamma = 0.6$ and $N=2$ yields the best results.
As expected, the No-AD performs the worst for both $N=2$ and $N=5$.

Fig.~\ref{fig:s-helm} presents the third-order structure function defined in \eqref{eq:str} for the AD-DF with $N=2$ and $N=5$ at $t=10$.
Results for the DNS and No-AD are also included for comparison purposes.
For $N=2$, $\gamma = 0.8$ yields the best results.
For $N=5$, $\gamma = 1.0$ yields the best results..
Comparing the $N=2$ plot with the $N=5$ plot, the combination $\gamma = 0.8$ and $N=2$ yields the best results.
As expected, the No-AD performs the worst for both $N=2$ and $N=5$.

Table \ref{tab:L2-helm-a} presents the $L^2$-norm of the error of the AD-DF for $N=2$ and $N=5$.
Results for the No-AD are also included for comparison purposes.
The errors are averaged over the time interval $8 \leq t \leq 12$.
For $N=2$, $\gamma = 1.0$ yields the best results.
For $N=5$, $\gamma = 1.0$ yields again the best results.
Comparing the $N=2$ results with the $N=5$ results, the combination $\gamma = 1.0$ and $N=2$ yields the best results.
As expected, the No-AD performs the worst.
Increasing $N$ results in a consistent significant increase of the error for all values of the parameter $\gamma$.
In general, the errors of the velocity components are lower than those of the vorticity components.

Table \ref{tab:corr-helm-a} presents the correlation coefficients for the AD-DF for $N=2$ and $N=5$.
Results for the No-AD are also included for comparison purposes.
The correlation coefficients are averaged over the time interval $8 \leq t \leq 12$.
For $N=2$, $\gamma = 1.0$ yields the best results.
For $N=5$, $\gamma = 1.0$ yields again the best results.
Comparing the $N=2$ results with the $N=5$ results, the combination $\gamma = 1.0$ and $N=2$ yields the best results.
As expected, the No-AD performs the worst for both $N=2$ and $N=5$.
Increasing $N$ results in a consistent significant decrease of the correlation coefficients for all values of the parameter $\gamma$.
In general, the correlation coefficients of the velocity components are higher than those of the vorticity components.

Based on the results in Figs.~\ref{fig:q-helm} and \ref{fig:s-helm} and Tables~\ref{tab:L2-helm-a} and \ref{tab:corr-helm-a}, we conclude that the AD-DF with $\gamma = 1.0$ and $N=2$ yields the best overall results.

\begin{figure}
\centering
\mbox{
\subfigure[~$N=2$]{\includegraphics[width=0.5\textwidth]{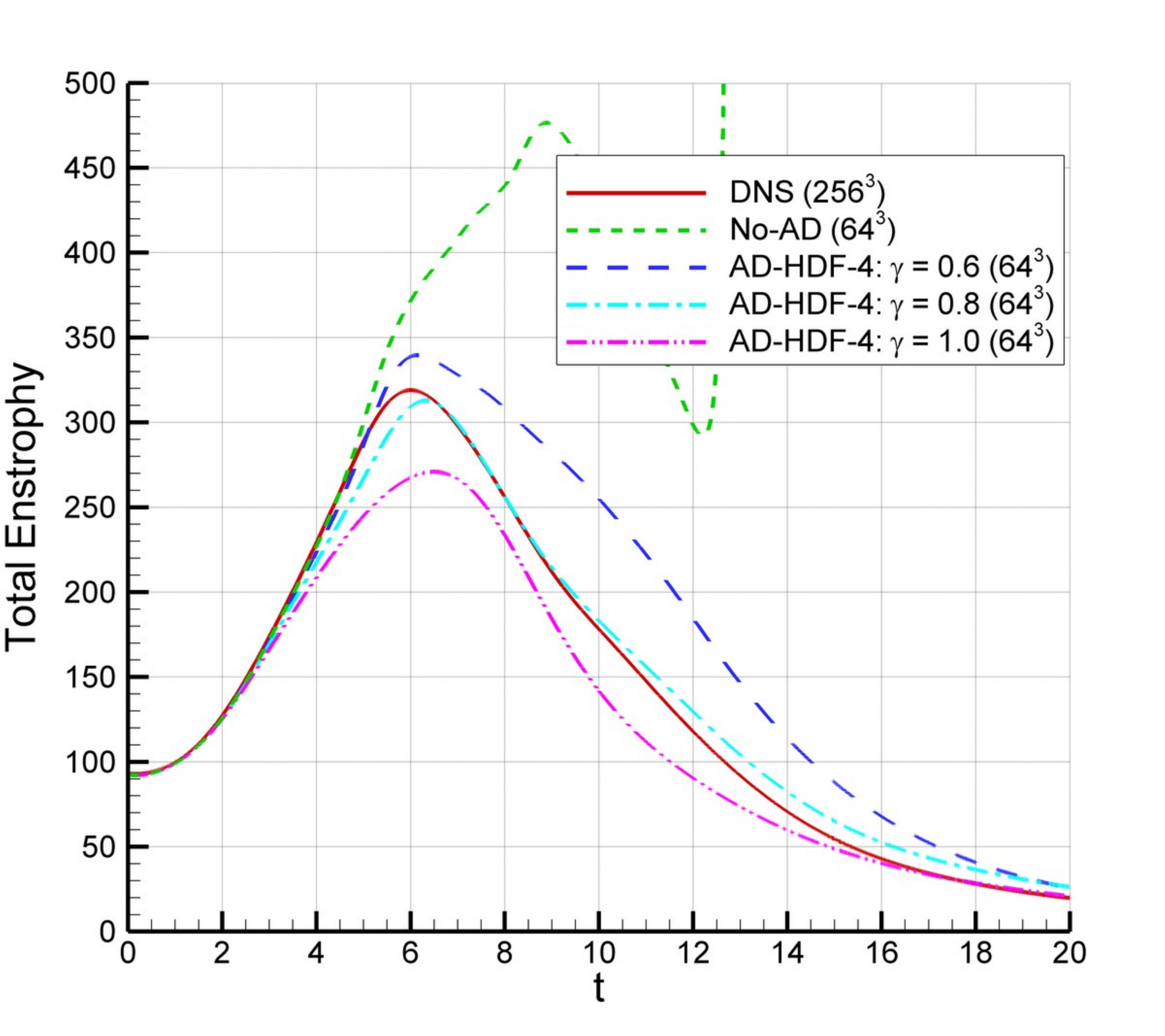}}
\subfigure[~$N=5$]{\includegraphics[width=0.5\textwidth]{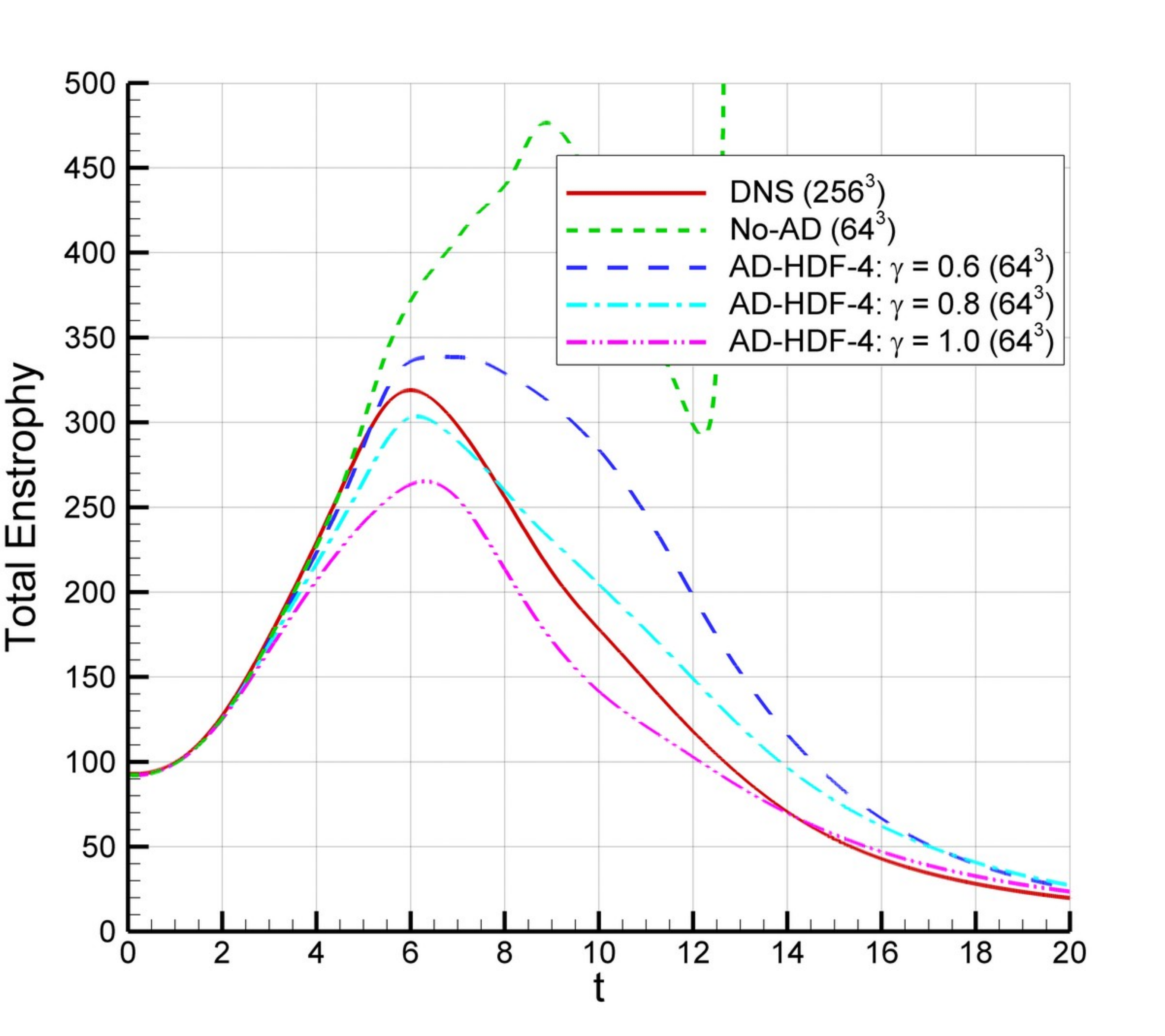}}
}
\caption{Time series of total enstrophy for the hyper-differential filters ($m=4$).}
\label{fig:q-hyper4}
\end{figure}

\begin{figure}
\centering
\mbox{
\subfigure[~$N=2$]{\includegraphics[width=0.5\textwidth]{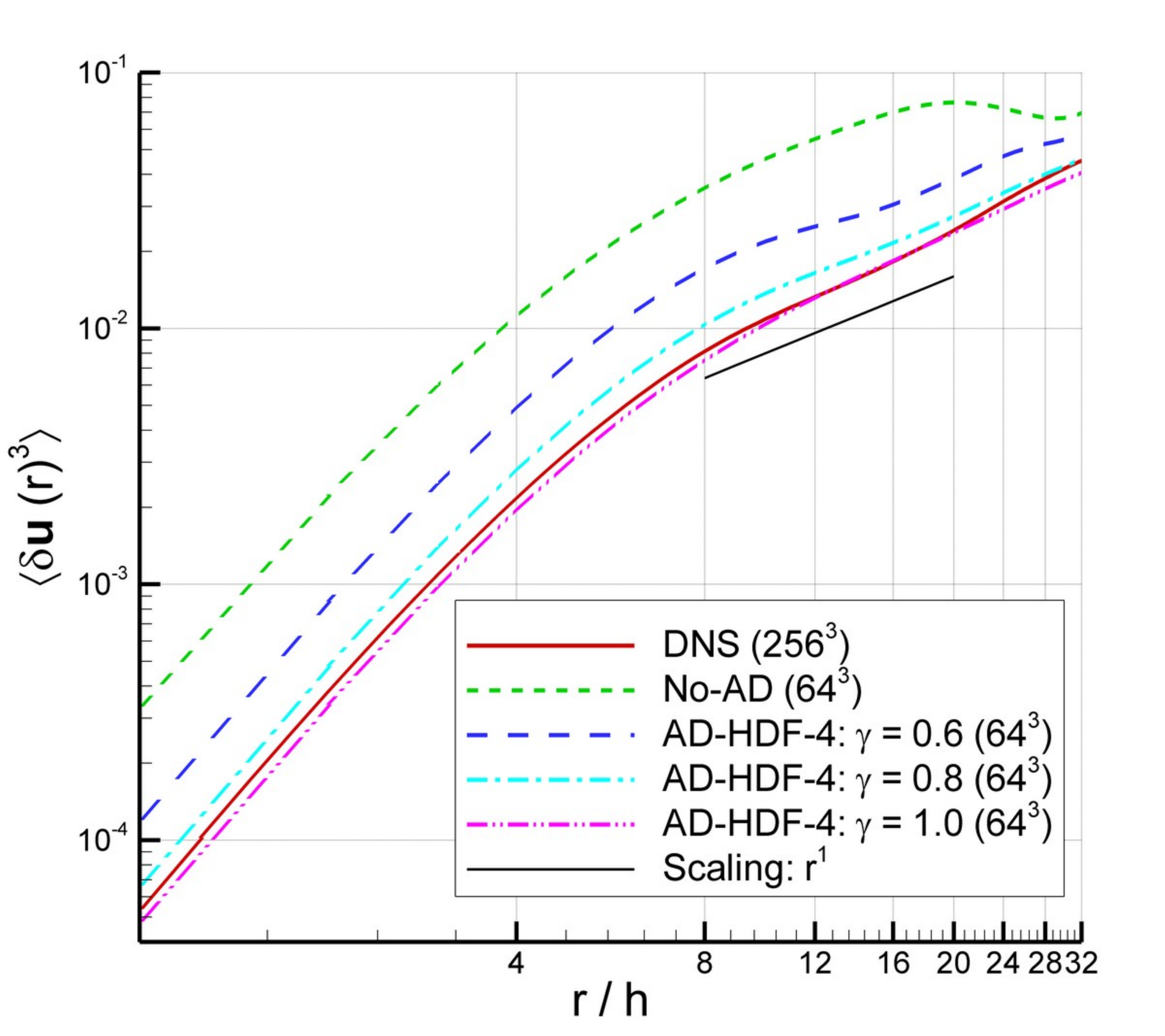}}
\subfigure[~$N=5$]{\includegraphics[width=0.5\textwidth]{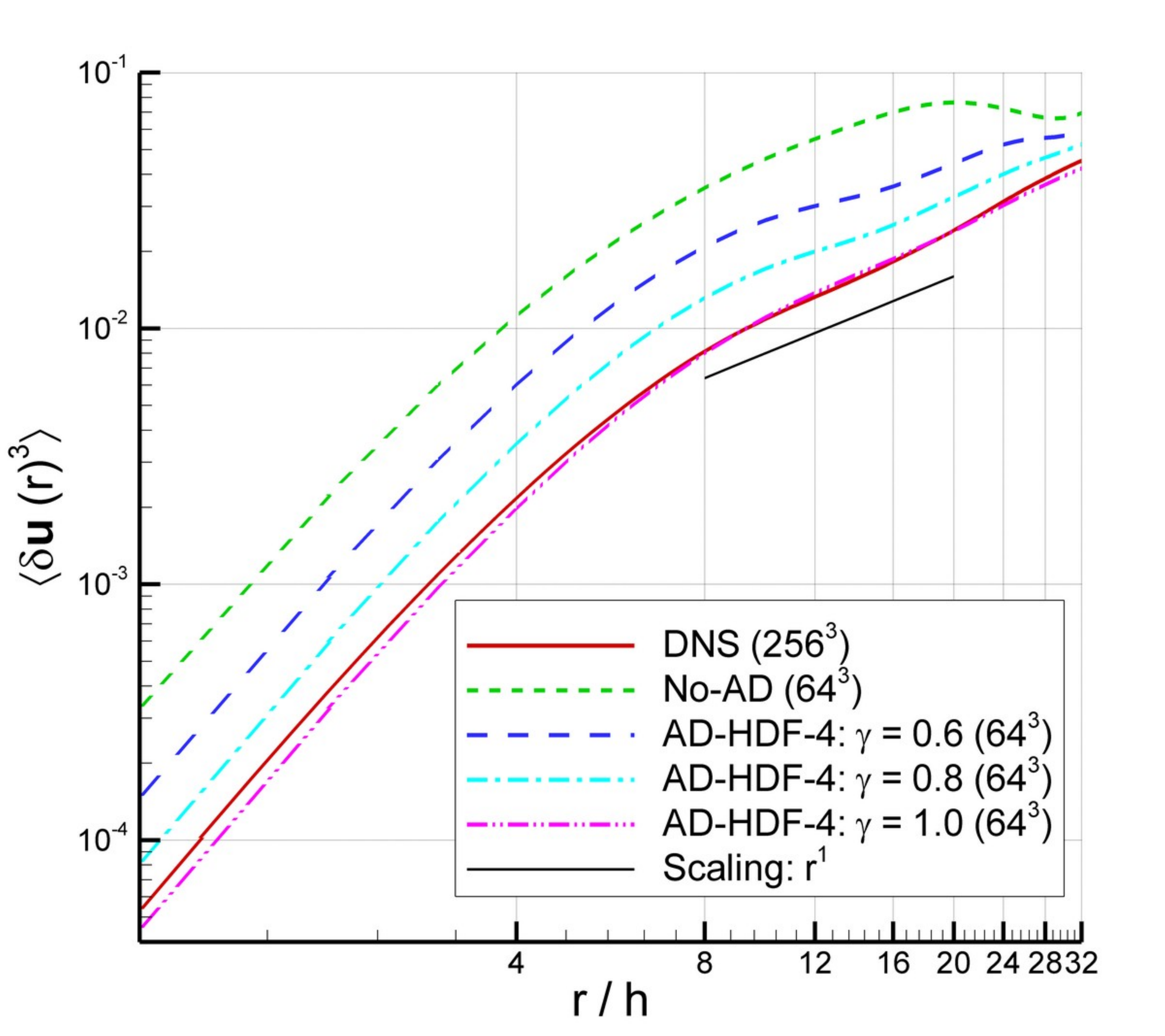}}
}
\caption{The third-order structure functions at $t=10$ for the hyper-differential filters ($m=4$).}
\label{fig:s-hyper4}
\end{figure}


\begin{table}
\centering
\caption{Discrete $L^2$-norms using the hyper-differential filters for $m=4$ (with resolutions of $64^3$) for ensemble averaging the data on a time interval between $t=8$ and $t=12$. The reference solution for computing the $L^2$-norm is the DNS data obtained with a resolution of $256^3$.}
\begin{tabular}{llllllll}
\hline
Field & No-AD & \multicolumn{2}{l}{\underline{HDF($m=4$, $\gamma=0.6$)}} & \multicolumn{2}{l}{\underline{HDF($m=4$, $\gamma=0.8$)}} & \multicolumn{2}{l}{\underline{HDF($m=4$, $\gamma=1.0$)}} \\
 &  & $N=2$ & $N=5$ & $N=2$ & $N=5$ & $N=2$ & $N=5$ \\
\hline
$u$         & 0.151 & 0.092 & 0.107 & 0.062 & 0.073 & 0.055 & 0.059 \\
$v$         & 0.169 & 0.093 & 0.110 & 0.066 & 0.075 & 0.061 & 0.063 \\
$w$         & 0.166 & 0.103 & 0.121 & 0.074 & 0.084 & 0.075 & 0.070 \\
$\omega_x$  & 1.233 & 0.659 & 0.738 & 0.496 & 0.538 & 0.481 & 0.463 \\
$\omega_y$  & 1.001 & 0.636 & 0.697 & 0.509 & 0.516 & 0.506 & 0.468 \\
$\omega_z$  & 1.307 & 0.553 & 0.637 & 0.401 & 0.414 & 0.373 & 0.385 \\
\hline
\end{tabular}
\label{tab:L2-hyper4-a}
\end{table}


\begin{table}
\centering
\caption{Correlation coefficient between the DNS data (with a resolution of $256^3$) and AD results with the hyper-differential filters (with resolutions of $64^3$) for the power $m=4$ for ensemble averaging on a time interval between $t=8$ and $t=12$. Correlation coefficients between the DNS and No-AD model (with a resolution of $64^3$) are also listed for comparison purposes.}
\begin{tabular}{llllllll}
\hline
Field & $C(\mbox{DNS, No-AD})$ & \multicolumn{2}{l}{\underline{$C(\mbox{DNS}, \gamma=0.6)$}} & \multicolumn{2}{l}{$\underline{C(\mbox{DNS}, \gamma=0.8)}$} & \multicolumn{2}{l}{$\underline{C(\mbox{DNS}, \gamma=1.0)}$} \\
 &  & $N=2$ & $N=5$ & $N=2$ & $N=5$ & $N=2$ & $N=5$ \\
\hline
$u$         & 0.692 & 0.877 & 0.831 & 0.937 & 0.923 & 0.950 & 0.940 \\
$v$         & 0.528 & 0.806 & 0.739 & 0.909 & 0.877 & 0.932 & 0.921 \\
$w$         & 0.552 & 0.776 & 0.731 & 0.836 & 0.822 & 0.805 & 0.827 \\
$\omega_x$  & 0.342 & 0.671 & 0.609 & 0.775 & 0.747 & 0.771 & 0.778 \\
$\omega_y$  & 0.410 & 0.639 & 0.582 & 0.732 & 0.729 & 0.727 & 0.750 \\
$\omega_z$  & 0.277 & 0.630 & 0.561 & 0.744 & 0.726 & 0.767 & 0.749 \\
\hline
\end{tabular}
\label{tab:corr-hyper4-a}
\end{table}

\subsection{The AD-LES model with hyper-differential filters ($m=4$)}
\label{sec:numerics_hyper_differential_4}

In this section, we numerically investigate the AD-LES model in conjunction with the hyper-differential filter given in \eqref{eq:10b} and discussed in Section \ref{sec:diff}, with $m=4$.
The resulting LES model is denoted as AD-HDF-4.
The following values for the parameter $\alpha$ are considered: $\gamma = 0.6$, $\gamma = 0.8$, and $\gamma = 1.0$.

Fig.~\ref{fig:q-hyper4} presents the time series of the integrated enstrophy $Q(t)$ defined in \eqref{eq:41} for the AD-HDF-4 with $N=2$ and $N=5$.
Results for the DNS and No-AD are also included for comparison purposes.
For $N=2$, $\gamma = 0.8$ yields the best results.
For $N=5$, $\gamma = 0.8$ yields again the best results.
Comparing the $N=2$ plot with the $N=5$ plot, the combination $\gamma = 0.8$ and $N=2$ yields the best results.
As expected, the No-AD performs the worst for both $N=2$ and $N=5$.

Fig.~\ref{fig:s-hyper4} presents the third-order structure function defined in \eqref{eq:str} for the AD-HDF-4 with $N=2$ and $N=5$ at $t=10$.
Results for the DNS and No-AD are also included for comparison purposes.
For $N=2$, $\gamma = 1.0$ yields the best results.
For $N=5$, $\gamma = 1.0$ yields again the best results..
Comparing the $N=2$ plot with the $N=5$ plot, $\gamma = 1.0$ together with $N=2$ or $N=5$ yields the best results.
As expected, the No-AD performs the worst for both $N=2$ and $N=5$.

Table \ref{tab:L2-hyper4-a} presents the $L^2$-norm of the error of the AD-HDF-4 for $N=2$ and $N=5$.
Results for the No-AD are also included for comparison purposes.
The errors are averaged over the time interval $8 \leq t \leq 12$.
For $N=2$, $\gamma = 1.0$ yields the best results.
For $N=5$, $\gamma = 1.0$ yields again the best results.
Comparing the $N=2$ results with the $N=5$ results, $\gamma = 1.0$ together with $N=2$ or $N=5$ yields the best results.
As expected, the No-AD performs the worst.
Increasing $N$ results in a consistent significant increase of the error for $\gamma = 0.6$ and $\gamma = 0.8$, but no clear trend is observed for $\gamma = 1.0$.
In general, the errors of the velocity components are lower than those of the vorticity components.

Table \ref{tab:corr-hyper4-a} presents the correlation coefficients for the AD-HDF-4 for $N=2$ and $N=5$.
Results for the No-AD are also included for comparison purposes.
The correlation coefficients are averaged over the time interval $8 \leq t \leq 12$.
For $N=2$, $\gamma = 0.8$ and $\gamma = 1.0$ yield the best results.
For $N=5$, $\gamma = 1.0$ yields the best results.
Comparing the $N=2$ results with the $N=5$ results, the combination $\gamma = 1.0$ and $N=5$ yields the best results.
As expected, the No-AD performs the worst for both $N=2$ and $N=5$.
Increasing $N$ results in a consistent significant decrease of the correlation coefficients for $\gamma = 0.6$ and $\gamma = 0.8$, but no clear trend is observed for $\gamma = 1.0$.
In general, the correlation coefficients of the velocity components are higher than those of the vorticity components.

Based on the results in Figs.~\ref{fig:q-hyper4} and \ref{fig:s-hyper4} and Tables~\ref{tab:L2-hyper4-a} and \ref{tab:corr-hyper4-a}, we conclude that the AD-HDF-4 with $\gamma = 1.0$ and $N=2$ or $N=5$ yields the best overall results.

\begin{figure}
\centering
\mbox{
\subfigure[~$N=2$]{\includegraphics[width=0.5\textwidth]{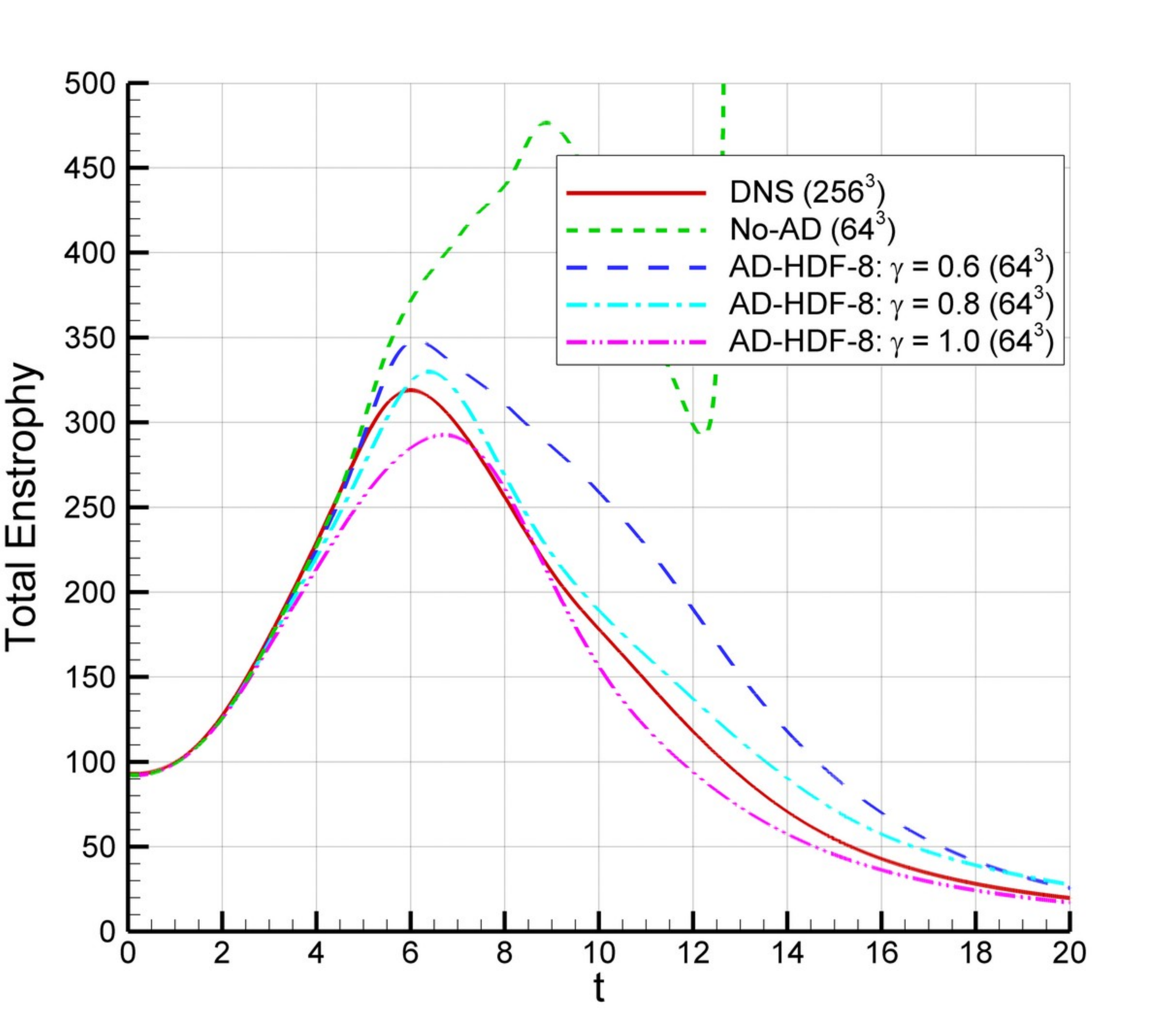}}
\subfigure[~$N=5$]{\includegraphics[width=0.5\textwidth]{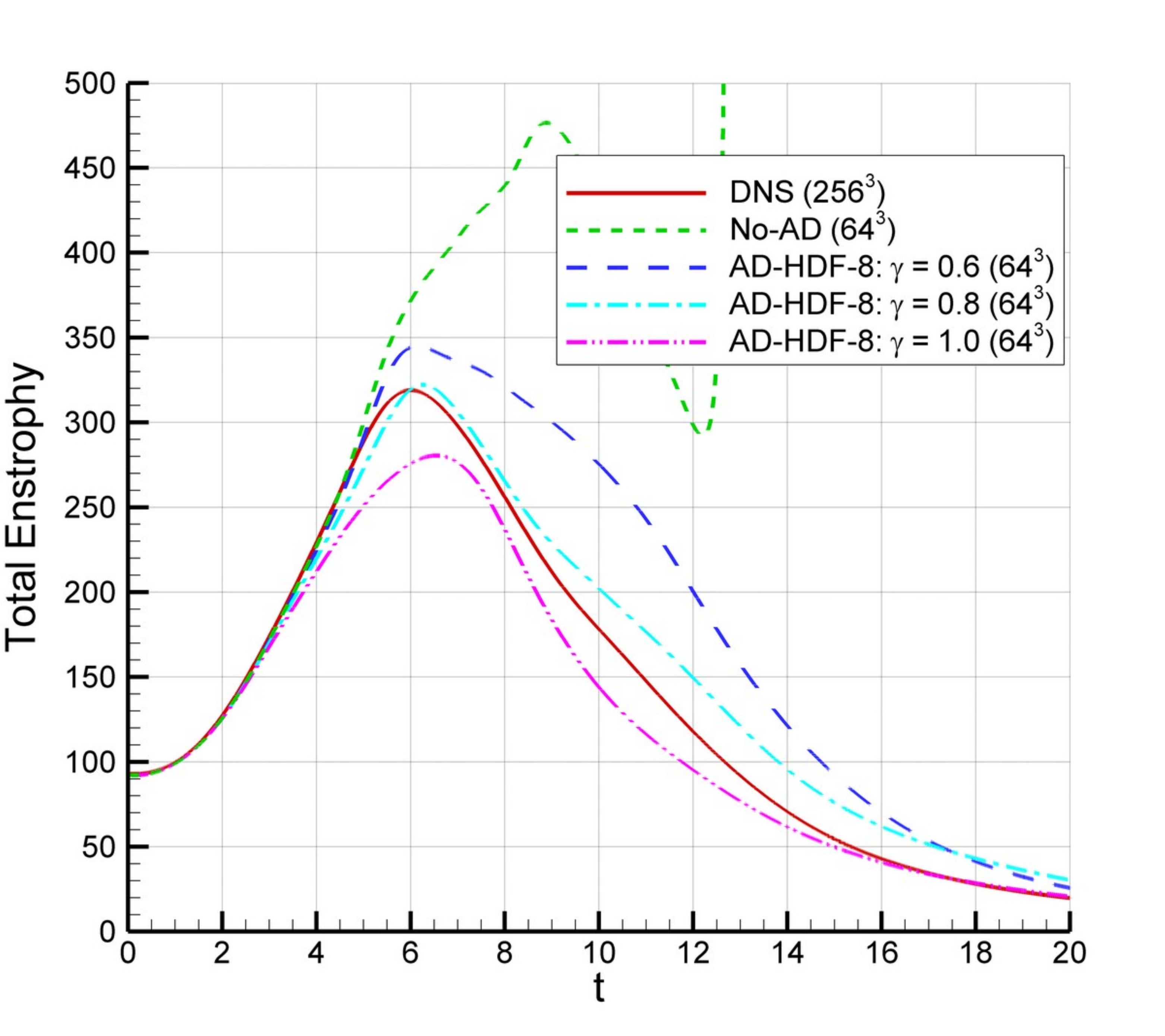}}
}
\caption{Time series of total enstrophy for the hyper-differential filters ($m=8$).}
\label{fig:q-hyper8}
\end{figure}

\begin{figure}
\centering
\mbox{
\subfigure[~$N=2$]{\includegraphics[width=0.5\textwidth]{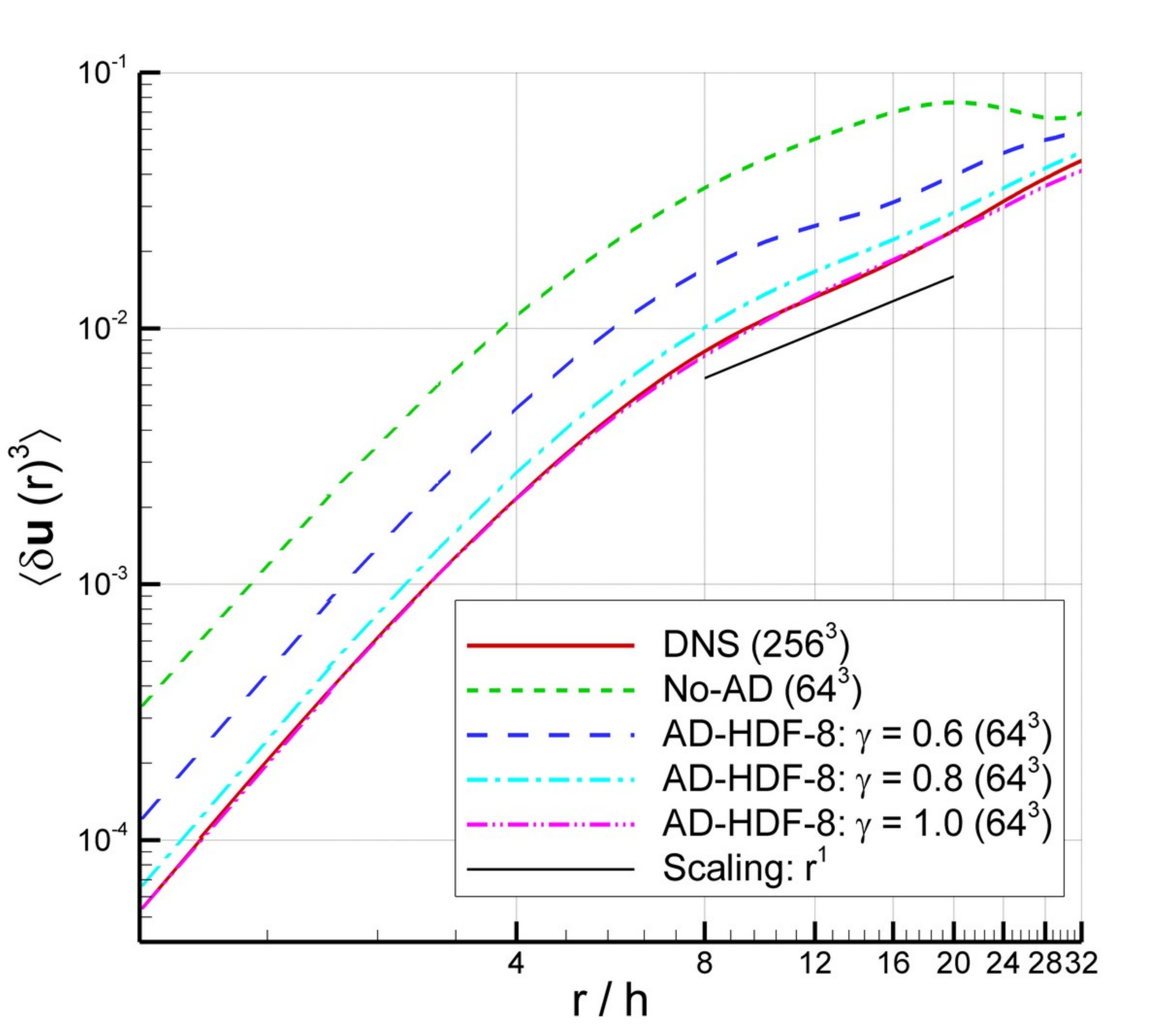}}
\subfigure[~$N=5$]{\includegraphics[width=0.5\textwidth]{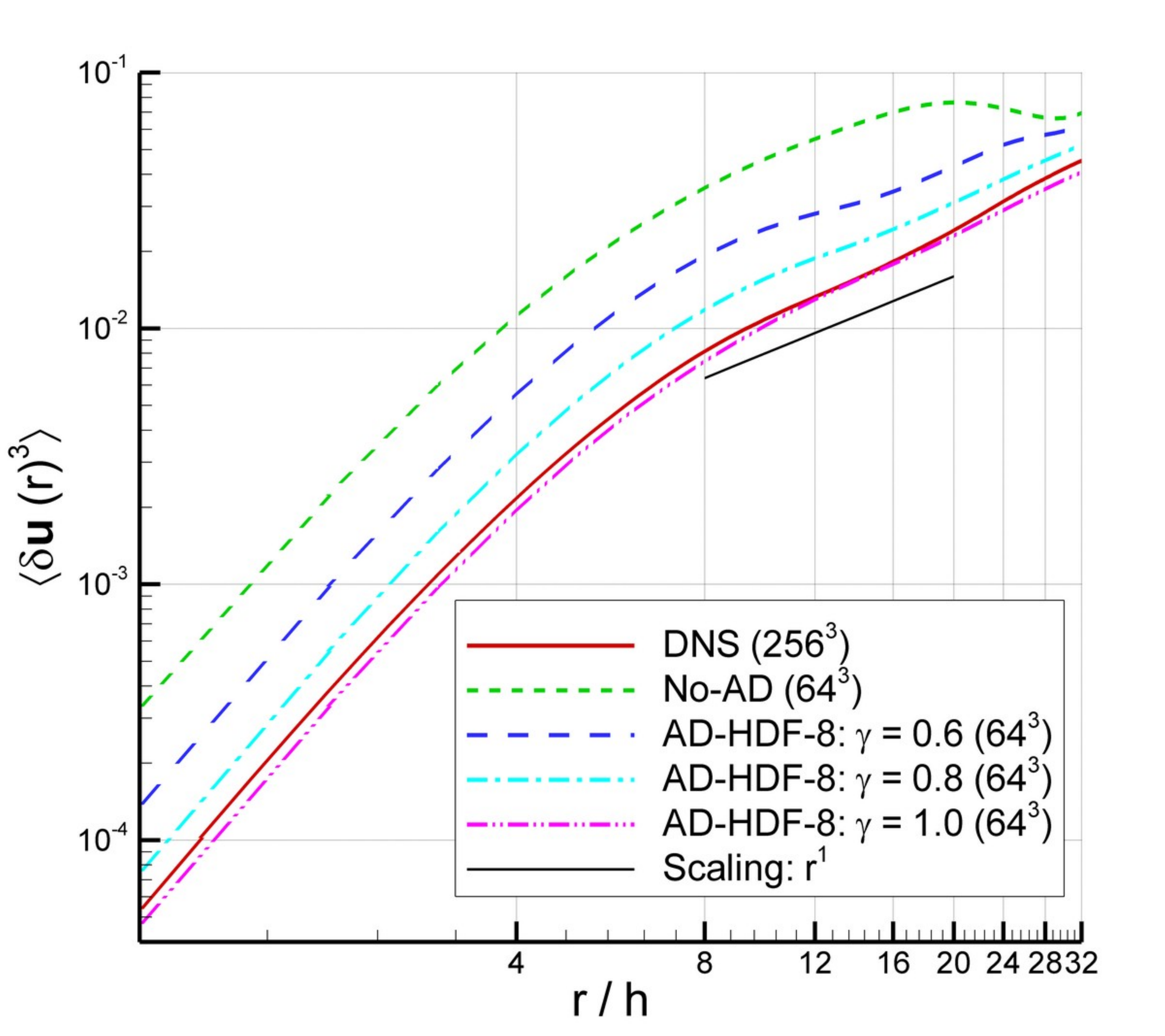}}
}
\caption{The third-order structure functions at $t=10$ for the hyper-differential filters ($m=8$).}
\label{fig:s-hyper8}
\end{figure}


\begin{table}
\centering
\caption{Discrete $L^2$-norms using the hyper-differential filters for $m=8$ (with resolutions of $64^3$) for ensemble averaging the data on a time interval between $t=8$ and $t=12$. The reference solution for computing the $L^2$-norm is the DNS data obtained with a resolution of $256^3$.}
\begin{tabular}{llllllll}
\hline
Field & No-AD & \multicolumn{2}{l}{\underline{HDF($m=8$, $\gamma=0.6$)}} & \multicolumn{2}{l}{\underline{HDF($m=8$, $\gamma=0.8$)}} & \multicolumn{2}{l}{\underline{HDF($m=8$, $\gamma=1.0$)}} \\
 &  & $N=2$ & $N=5$ & $N=2$ & $N=5$ & $N=2$ & $N=5$ \\
\hline
$u$         & 0.151 & 0.091 & 0.100 & 0.063 & 0.065 & 0.062 & 0.058 \\
$v$         & 0.169 & 0.091 & 0.101 & 0.064 & 0.067 & 0.065 & 0.061 \\
$w$         & 0.166 & 0.101 & 0.110 & 0.074 & 0.078 & 0.085 & 0.072 \\
$\omega_x$  & 1.233 & 0.658 & 0.699 & 0.518 & 0.532 & 0.535 & 0.490 \\
$\omega_y$  & 1.001 & 0.640 & 0.668 & 0.540 & 0.521 & 0.586 & 0.509 \\
$\omega_z$  & 1.307 & 0.557 & 0.600 & 0.417 & 0.405 & 0.412 & 0.397 \\
\hline
\end{tabular}
\label{tab:L2-hyper8-a}
\end{table}


\begin{table}
\centering
\caption{Correlation coefficient between the DNS data (with a resolution of $256^3$) and AD results with the hyper-differential filters (with resolutions of $64^3$) for the power $m=8$ for ensemble averaging on a time interval between $t=8$ and $t=12$. Correlation coefficients between the DNS and No-AD model (with a resolution of $64^3$) are also listed for comparison purposes.}
\begin{tabular}{llllllll}
\hline
Field & $C(\mbox{DNS, No-AD})$ & \multicolumn{2}{l}{\underline{$C(\mbox{DNS}, \gamma=0.6)$}} & \multicolumn{2}{l}{$\underline{C(\mbox{DNS}, \gamma=0.8)}$} & \multicolumn{2}{l}{$\underline{C(\mbox{DNS}, \gamma=1.0)}$} \\
 &  & $N=2$ & $N=5$ & $N=2$ & $N=5$ & $N=2$ & $N=5$ \\
\hline
$u$         & 0.692 & 0.883 & 0.858 & 0.938 & 0.939 & 0.939 & 0.943 \\
$v$         & 0.528 & 0.815 & 0.774 & 0.918 & 0.906 & 0.925 & 0.933 \\
$w$         & 0.552 & 0.784 & 0.761 & 0.828 & 0.831 & 0.755 & 0.822 \\
$\omega_x$  & 0.342 & 0.680 & 0.646 & 0.765 & 0.760 & 0.727 & 0.761 \\
$\omega_y$  & 0.410 & 0.644 & 0.620 & 0.714 & 0.736 & 0.663 & 0.732 \\
$\omega_z$  & 0.277 & 0.634 & 0.603 & 0.735 & 0.743 & 0.726 & 0.745 \\
\hline
\end{tabular}
\label{tab:corr-hyper8-a}
\end{table}

\subsection{The AD-LES model with hyper-differential filters ($m=8$)}
\label{sec:numerics_hyper_differential_8}

In this section, we numerically investigate the AD-LES model in conjunction with the hyper-differential filter given in \eqref{eq:10b} and discussed in Section \ref{sec:diff}, with $m=8$.
The resulting LES model is denoted as AD-HDF-8.
The following values for the parameter $\alpha$ are considered: $\gamma = 0.6$, $\gamma = 0.8$, and $\gamma = 1.0$.

Fig.~\ref{fig:q-hyper8} presents the time series of the integrated enstrophy $Q(t)$ defined in \eqref{eq:41} for the AD-HDF-8 with $N=2$ and $N=5$.
Results for the DNS and No-AD are also included for comparison purposes.
For $N=2$, $\gamma = 0.8$ yields the best results.
For $N=5$, $\gamma = 0.8$ yields again the best results.
Comparing the $N=2$ plot with the $N=5$ plot, the combination $\gamma = 0.8$ and $N=2$ yields the best results.
As expected, the No-AD performs the worst for both $N=2$ and $N=5$.

Fig.~\ref{fig:s-hyper8} presents the third-order structure function defined in \eqref{eq:str} for the AD-HDF-8 with $N=2$ and $N=5$ at $t=10$.
Results for the DNS and No-AD are also included for comparison purposes.
For $N=2$, $\gamma = 1.0$ yields the best results.
For $N=5$, $\gamma = 1.0$ yields again the best results..
Comparing the $N=2$ plot with the $N=5$ plot, $\gamma = 1.0$ together with $N=2$ yields the best results.
As expected, the No-AD performs the worst for both $N=2$ and $N=5$.

Table \ref{tab:L2-hyper8-a} presents the $L^2$-norm of the error of the AD-HDF-8 for $N=2$ and $N=5$.
Results for the No-AD are also included for comparison purposes.
The errors are averaged over the time interval $8 \leq t \leq 12$.
For $N=2$, $\gamma = 0.8$ yields the best results.
For $N=5$, $\gamma = 1.0$ yields the best results.
Comparing the $N=2$ results with the $N=5$ results, $\gamma = 1.0$ together with $N=5$ yields the best results.
As expected, the No-AD performs the worst.
Increasing $N$ results in a consistent significant increase of the error for $\gamma = 0.6$, no trend for $\gamma = 0.8$, and a consistent decrease for $\gamma = 1.0$.
In general, the errors of the velocity components are lower than those of the vorticity components.

Table \ref{tab:corr-hyper8-a} presents the correlation coefficients for the AD-HDF-8 for $N=2$ and $N=5$.
Results for the No-AD are also included for comparison purposes.
The correlation coefficients are averaged over the time interval $8 \leq t \leq 12$.
For $N=2$, $\gamma = 0.8$ yields the best results.
For $N=5$, $\gamma = 1.0$ and $\gamma = 0.8$ yield the best results.
Comparing the $N=2$ results with the $N=5$ results, the combination $\gamma = 1.0$ and $N=5$ yields the best results.
As expected, the No-AD performs the worst for both $N=2$ and $N=5$.
Increasing $N$ results in a consistent significant decrease of the correlation coefficients for $\gamma = 0.6$, no trend for $\gamma = 0.8$, and a consistent significant increase for $\gamma = 1.0$.
In general, the correlation coefficients of the velocity components are higher than those of the vorticity components.

Based on the results in Figs.~\ref{fig:q-hyper8} and \ref{fig:s-hyper8} and Tables~\ref{tab:L2-hyper8-a} and \ref{tab:corr-hyper8-a}, we conclude that the AD-HDF-8 with $\gamma = 1.0$ and $N=2$ or $N=5$ yields the best overall results.

\begin{figure}
\centering
\mbox{
\subfigure[~$N=2$]{\includegraphics[width=0.5\textwidth]{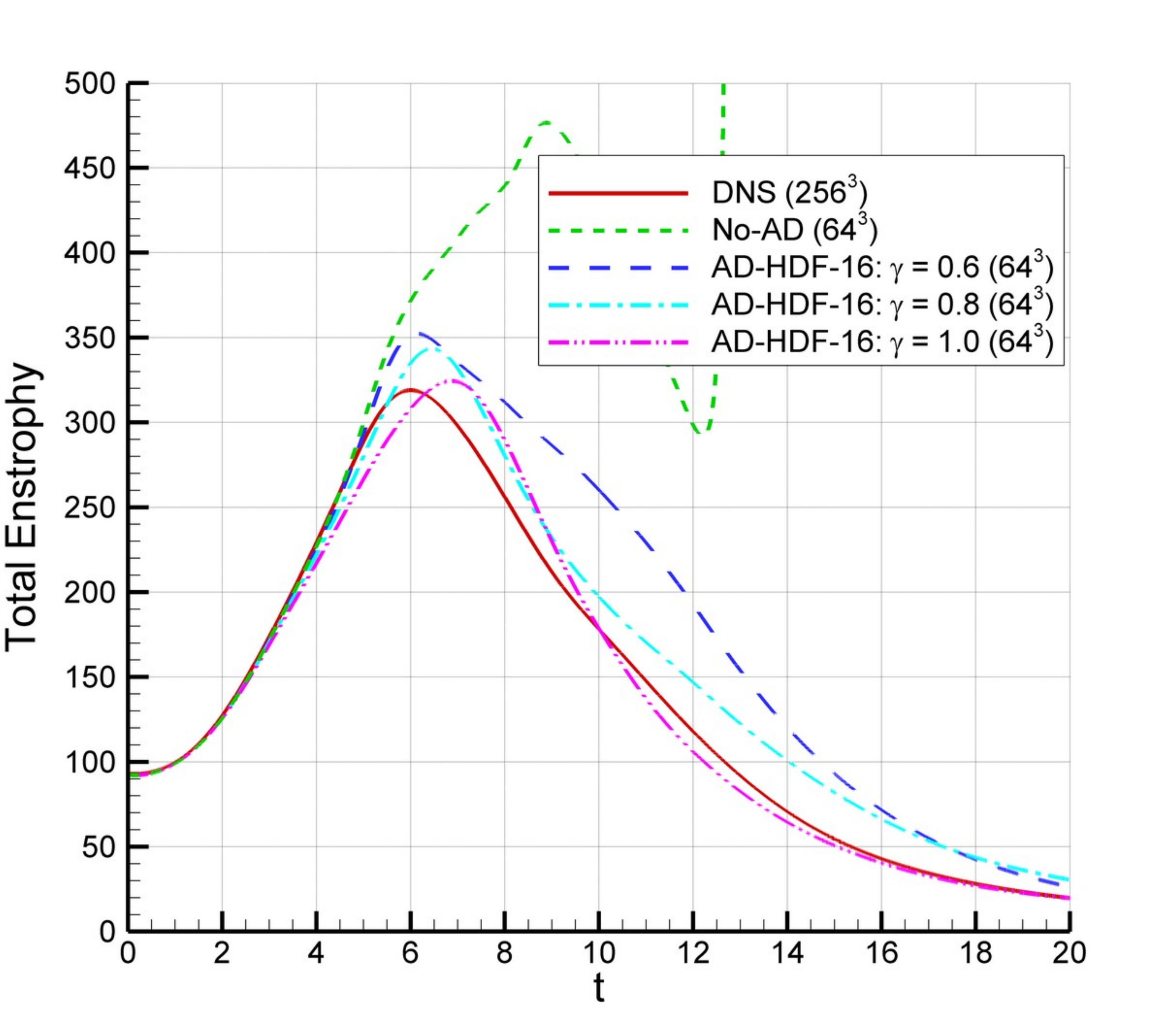}}
\subfigure[~$N=5$]{\includegraphics[width=0.5\textwidth]{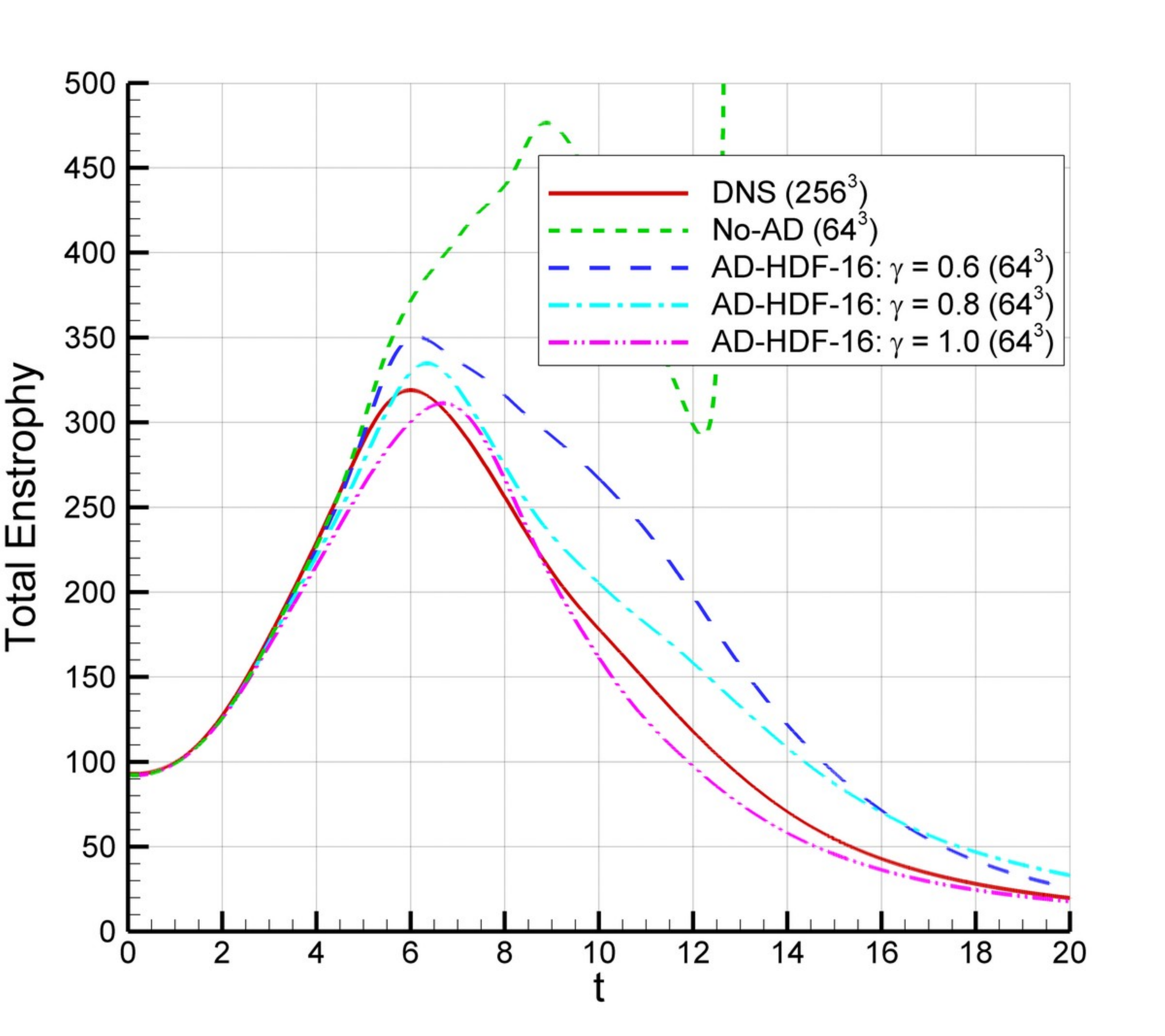}}
}
\caption{Time series of total enstrophy for the hyper-differential filters ($m=16$).}
\label{fig:q-hyper16}
\end{figure}

\begin{figure}
\centering
\mbox{
\subfigure[~$N=2$]{\includegraphics[width=0.5\textwidth]{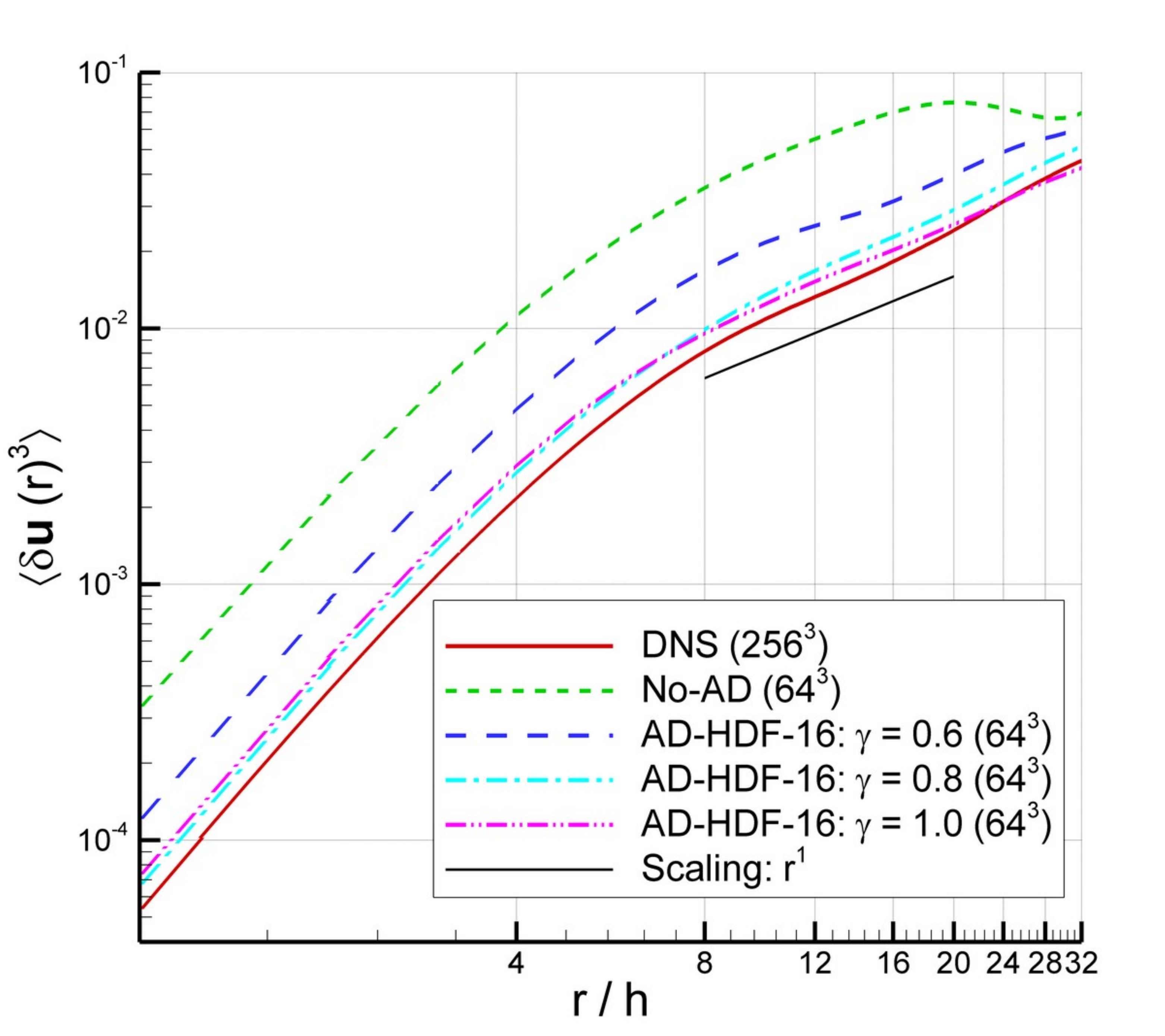}}
\subfigure[~$N=5$]{\includegraphics[width=0.5\textwidth]{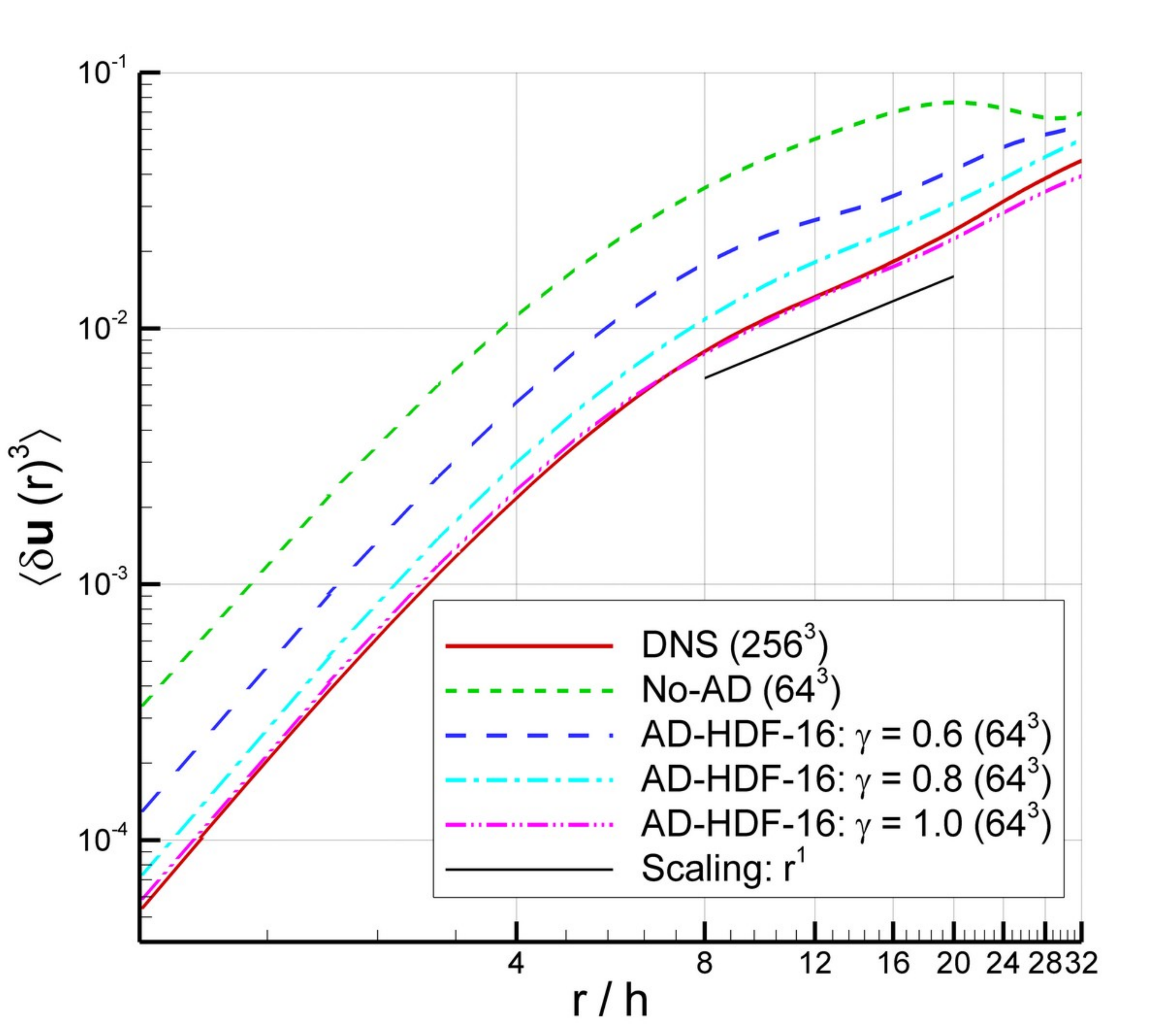}}
}
\caption{The third-order structure functions at $t=10$ for the hyper-differential filters ($m=16$).}
\label{fig:s-hyper16}
\end{figure}


\begin{table}
\centering
\caption{Discrete $L^2$-norms using the hyper-differential filters for $m=16$ (with resolutions of $64^3$) for ensemble averaging the data on a time interval between $t=8$ and $t=12$. The reference solution for computing the $L^2$-norm is the DNS data obtained with a resolution of $256^3$.}
\begin{tabular}{llllllll}
\hline
Field & No-AD & \multicolumn{2}{l}{\underline{HDF($m=16$, $\gamma=0.6$)}} & \multicolumn{2}{l}{\underline{HDF($m=16$, $\gamma=0.8$)}} & \multicolumn{2}{l}{\underline{HDF($m=16$, $\gamma=1.0$)}} \\
 &  & $N=2$ & $N=5$ & $N=2$ & $N=5$ & $N=2$ & $N=5$ \\
\hline
$u$         & 0.151 & 0.092 & 0.096 & 0.065 & 0.066 & 0.073 & 0.072 \\
$v$         & 0.169 & 0.090 & 0.094 & 0.063 & 0.065 & 0.078 & 0.073 \\
$w$         & 0.166 & 0.100 & 0.103 & 0.077 & 0.079 & 0.093 & 0.092 \\
$\omega_x$  & 1.233 & 0.656 & 0.669 & 0.549 & 0.560 & 0.620 & 0.597 \\
$\omega_y$  & 1.001 & 0.652 & 0.656 & 0.594 & 0.582 & 0.639 & 0.602 \\
$\omega_z$  & 1.307 & 0.569 & 0.585 & 0.437 & 0.434 & 0.465 & 0.459 \\
\hline
\end{tabular}
\label{tab:L2-hyper16-a}
\end{table}


\begin{table}
\centering
\caption{Correlation coefficient between the DNS data (with a resolution of $256^3$) and AD results with the hyper-differential filters (with resolutions of $64^3$) for the power $m=16$ for ensemble averaging on a time interval between $t=8$ and $t=12$. Correlation coefficients between the DNS and No-AD model (with a resolution of $64^3$) are also listed for comparison purposes.}
\begin{tabular}{llllllll}
\hline
Field & $C(\mbox{DNS, No-AD})$ & \multicolumn{2}{l}{\underline{$C(\mbox{DNS}, \gamma=0.6)$}} & \multicolumn{2}{l}{$\underline{C(\mbox{DNS}, \gamma=0.8)}$} & \multicolumn{2}{l}{$\underline{C(\mbox{DNS}, \gamma=1.0)}$} \\
 &  & $N=2$ & $N=5$ & $N=2$ & $N=5$ & $N=2$ & $N=5$ \\
\hline
$u$         & 0.692 & 0.885 & 0.874 & 0.938 & 0.939 & 0.907 & 0.909 \\
$v$         & 0.528 & 0.823 & 0.805 & 0.927 & 0.921 & 0.892 & 0.908 \\
$w$         & 0.552 & 0.786 & 0.780 & 0.819 & 0.809 & 0.733 & 0.729 \\
$\omega_x$  & 0.342 & 0.685 & 0.674 & 0.744 & 0.732 & 0.657 & 0.668 \\
$\omega_y$  & 0.410 & 0.639 & 0.636 & 0.685 & 0.696 & 0.624 & 0.650 \\
$\omega_z$  & 0.277 & 0.627 & 0.617 & 0.729 & 0.729 & 0.656 & 0.663 \\
\hline
\end{tabular}
\label{tab:corr-hyper16-a}
\end{table}

\subsection{The AD-LES model with hyper-differential filters ($m=16$)}
\label{sec:numerics_hyper_differential_16}

In this section, we numerically investigate the AD-LES model in conjunction with the hyper-differential filter given in \eqref{eq:10b} and discussed in Section \ref{sec:diff}, with $m=16$.
The resulting LES model is denoted as AD-HDF-16.
The following values for the parameter $\alpha$ are considered: $\gamma = 0.6$, $\gamma = 0.8$, and $\gamma = 1.0$.

Fig.~\ref{fig:q-hyper16} presents the time series of the integrated enstrophy $Q(t)$ defined in \eqref{eq:41} for the AD-HDF-16 with $N=2$ and $N=5$.
Results for the DNS and No-AD are also included for comparison purposes.
For $N=2$, $\gamma = 1.0$ yields the best results.
For $N=5$, $\gamma = 1.0$ yields again the best results.
Comparing the $N=2$ plot with the $N=5$ plot, the combination $\gamma = 1.0$ and $N=2$ or $N=5$ yields the best results.
As expected, the No-AD performs the worst for both $N=2$ and $N=5$.

Fig.~\ref{fig:s-hyper16} presents the third-order structure function defined in \eqref{eq:str} for the AD-HDF-16 with $N=2$ and $N=5$ at $t=10$.
Results for the DNS and No-AD are also included for comparison purposes.
For $N=2$, $\gamma = 1.0$ yields the best results.
For $N=5$, $\gamma = 1.0$ yields again the best results.
Comparing the $N=2$ plot with the $N=5$ plot, $\gamma = 1.0$ together with $N=5$ yields the best results.
As expected, the No-AD performs the worst for both $N=2$ and $N=5$.

Table \ref{tab:L2-hyper16-a} presents the $L^2$-norm of the error of the AD-HDF-16 for $N=2$ and $N=5$.
Results for the No-AD are also included for comparison purposes.
The errors are averaged over the time interval $8 \leq t \leq 12$.
For $N=2$, $\gamma = 0.8$ yields the best results.
For $N=5$, $\gamma = 0.8$ yields again the best results.
Comparing the $N=2$ results with the $N=5$ results, $\gamma = 0.8$ together with $N=2$ or $N=5$ yields the best results.
As expected, the No-AD performs the worst.
Increasing $N$ results in a consistent increase of the error for $\gamma = 0.6$, no trend for $\gamma = 0.8$, and a consistent decrease for $\gamma = 1.0$.
In general, the errors of the velocity components are lower than those of the vorticity components.

Table \ref{tab:corr-hyper16-a} presents the correlation coefficients for the AD-HDF-16 for $N=2$ and $N=5$.
Results for the No-AD are also included for comparison purposes.
The correlation coefficients are averaged over the time interval $8 \leq t \leq 12$.
For $N=2$, $\gamma = 0.8$ yields the best results.
For $N=5$, $\gamma = 0.8$ yields again the best results.
Comparing the $N=2$ results with the $N=5$ results, the combination $\gamma = 1.0$ and $N=2$ or $N=5$ yields the best results.
As expected, the No-AD performs the worst for both $N=2$ and $N=5$.
Increasing $N$ results in a consistent decrease of the correlation coefficients for $\gamma = 0.6$, no trend for $\gamma = 0.8$, and a consistent increase for $\gamma = 1.0$.
In general, the correlation coefficients of the velocity components are higher than those of the vorticity components.

Based on the results in Figs.~\ref{fig:q-hyper16} and \ref{fig:s-hyper16} and Tables~\ref{tab:L2-hyper16-a} and \ref{tab:corr-hyper16-a}, we conclude that the AD-HDF-16 with $\gamma = 0.8$ and $N=2$ or $N=5$ yields the best overall results.

\subsection{Computational efficiency}
\label{sec:computational_efficiency}

This section presents the computational efficiencies of the various methods for the AD-LES model.
The CPU times for the AD-LES model with the box filters (TF, SF, 7PF), Pad\'{e}-type filter (PF), differential filter (DF), and hyper-differential filter (HDF) are listed in Table~\ref{tab:eff}. The CPU times for the DNS and No-AD models are also included. The following conclusions can be drawn. The CPU times of all the AD-LES runs are significantly lower than that of the DNS, and higher than the CPU time of the No-AD. We note here that the DNS data are obtained using a computation with a resolution of $256^3$, which requires a CPU time of $171$ hours, while all other computations are performed using a resolution of $64^3$, which require much lower CPU times, on the order of $3$-$10$ hours. We also note that increasing the AD order $N$ from 2 to 5 results in an increase by a factor of $2$-$3$ of the CPU time for the PF, DF, and HDF. The increase for the box filters is lower (about 40\%). Among all the filters, the box filters are the most efficient, followed by the PF, and then the DF and HDF. The HDF is as efficient as the DF due to the FFT-based inversion method for solving the elliptic system.


\begin{table}
\centering
\caption{Computational efficiencies of the DNS, No-AD, and AD-LES models.}
\begin{tabular}{p{0.42\textwidth}p{0.42\textwidth}}
\hline
Method & CPU time \\
\hline
DNS ($256^3$)   & 171.5 hrs \\
No-AD ($64^3$)  & 1.8 hrs \\
AD-TF ($64^3$)  & 2.0 hrs ($N=2$); 2.8 hrs ($N=5$) \\
AD-SF ($64^3$)  & 2.0 hrs ($N=2$); 2.8 hrs ($N=5$) \\
AD-7PF ($64^3$) & 2.1 hrs ($N=2$); 2.9 hrs ($N=5$) \\
AD-PF ($64^3$) ($\alpha=0.25$)          & 3.1 hrs ($N=2$); 9.1 hrs ($N=5$) \\
AD-DF ($64^3$) ($\gamma=0.8$)           & 4.0 hrs ($N=2$); 9.5 hrs ($N=5$) \\
AD-HDF ($64^3$) ($\gamma=0.8$, $m=8$)   & 4.0 hrs ($N=2$); 9.5 hrs ($N=5$) \\
\hline
\end{tabular}
\label{tab:eff}
\end{table}

\begin{figure}
\centering
\mbox{
\subfigure[~DNS]{\includegraphics[width=0.27\textwidth]{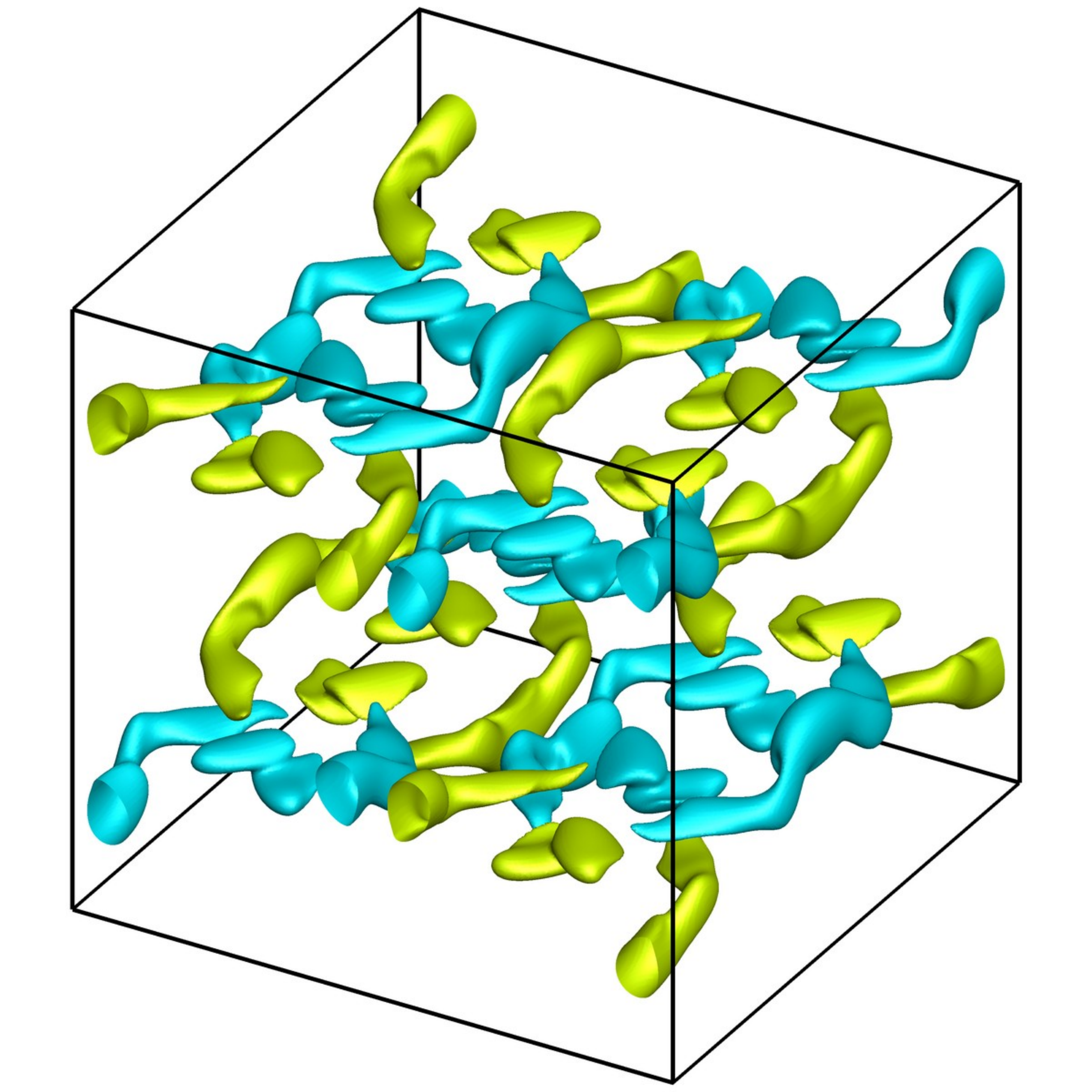}}
\subfigure[~No-AD]{\includegraphics[width=0.27\textwidth]{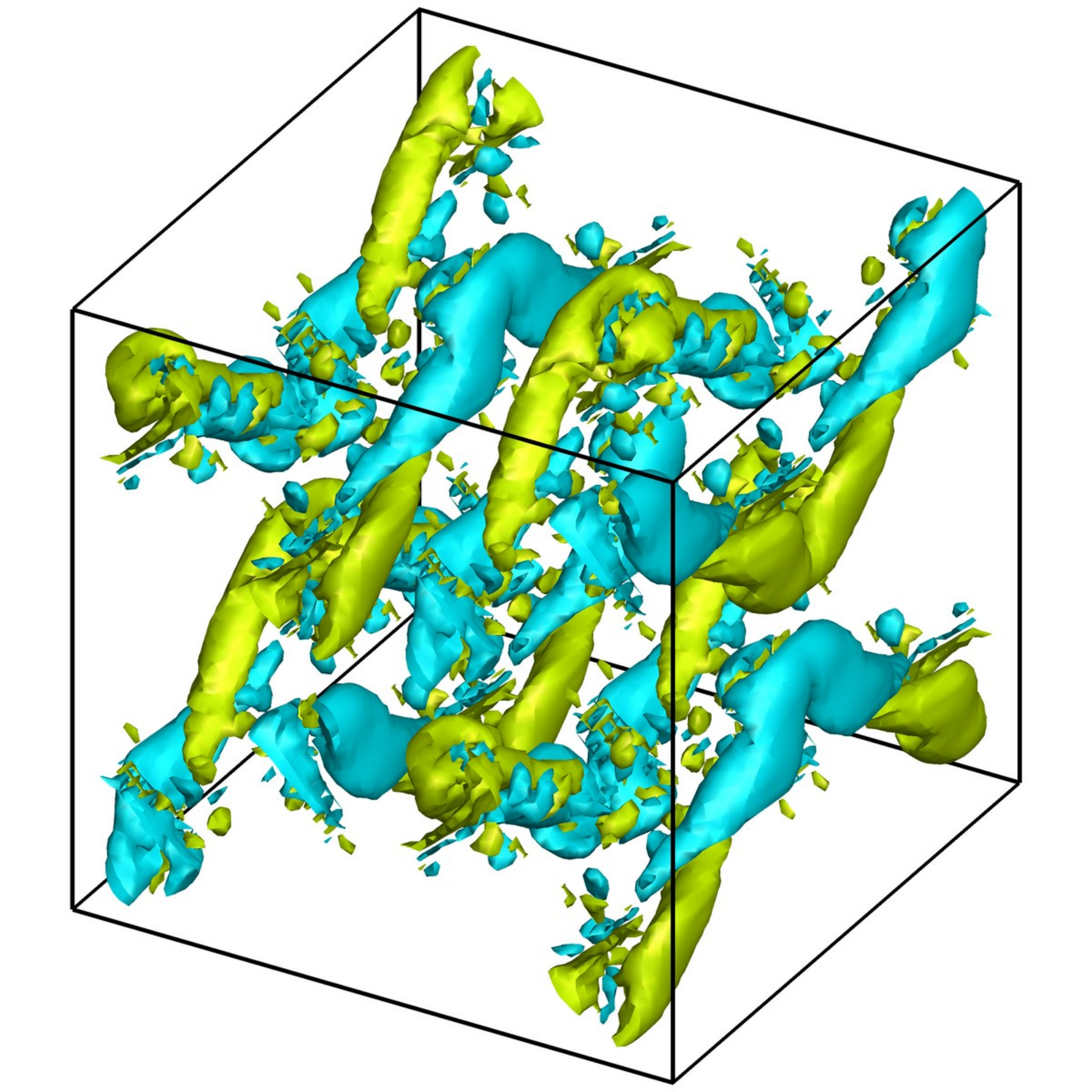}}
\subfigure[~AD-HDF-8 ($\gamma=0.8$)]{\includegraphics[width=0.27\textwidth]{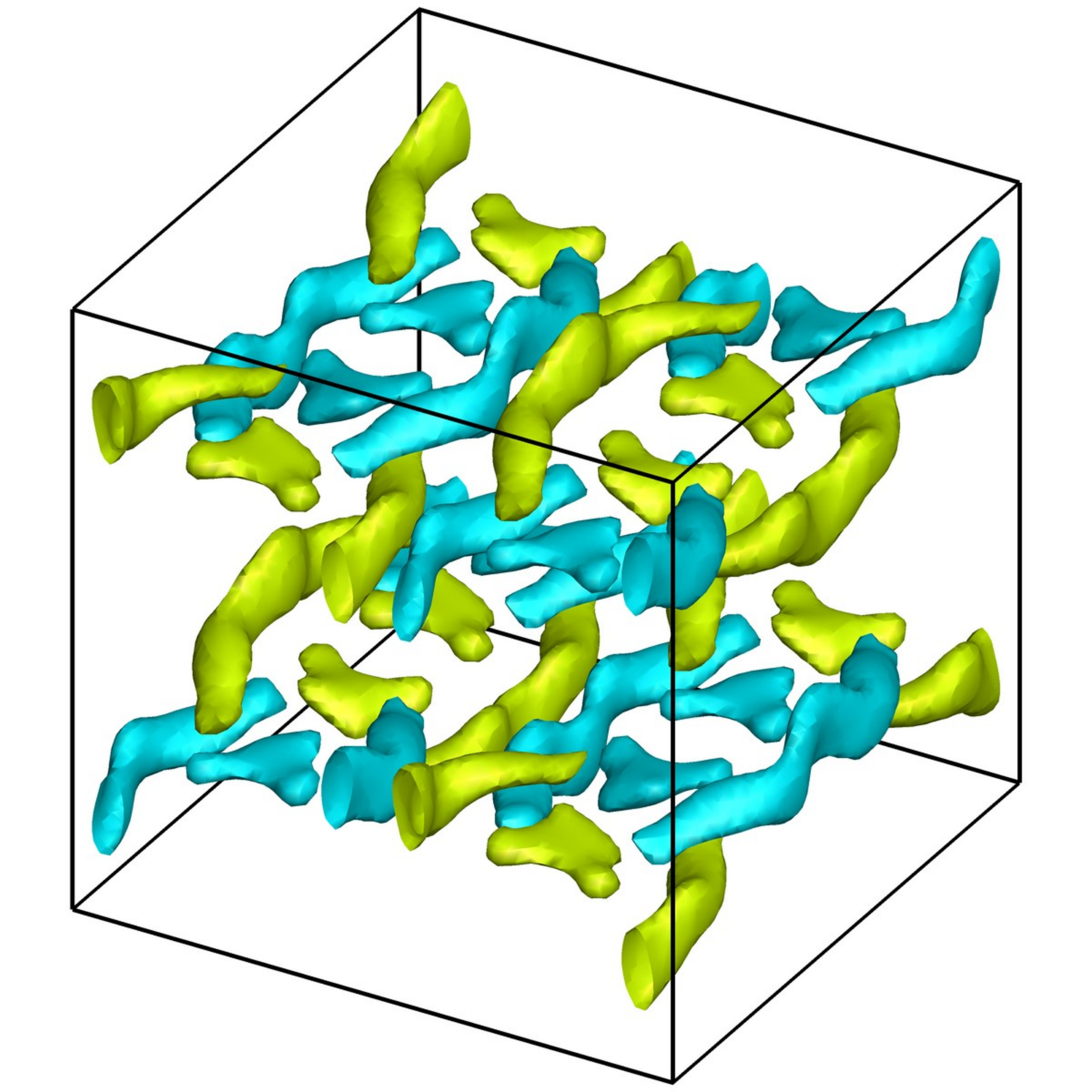}}
}
\mbox{
\subfigure[~AD-TF]{\includegraphics[width=0.27\textwidth]{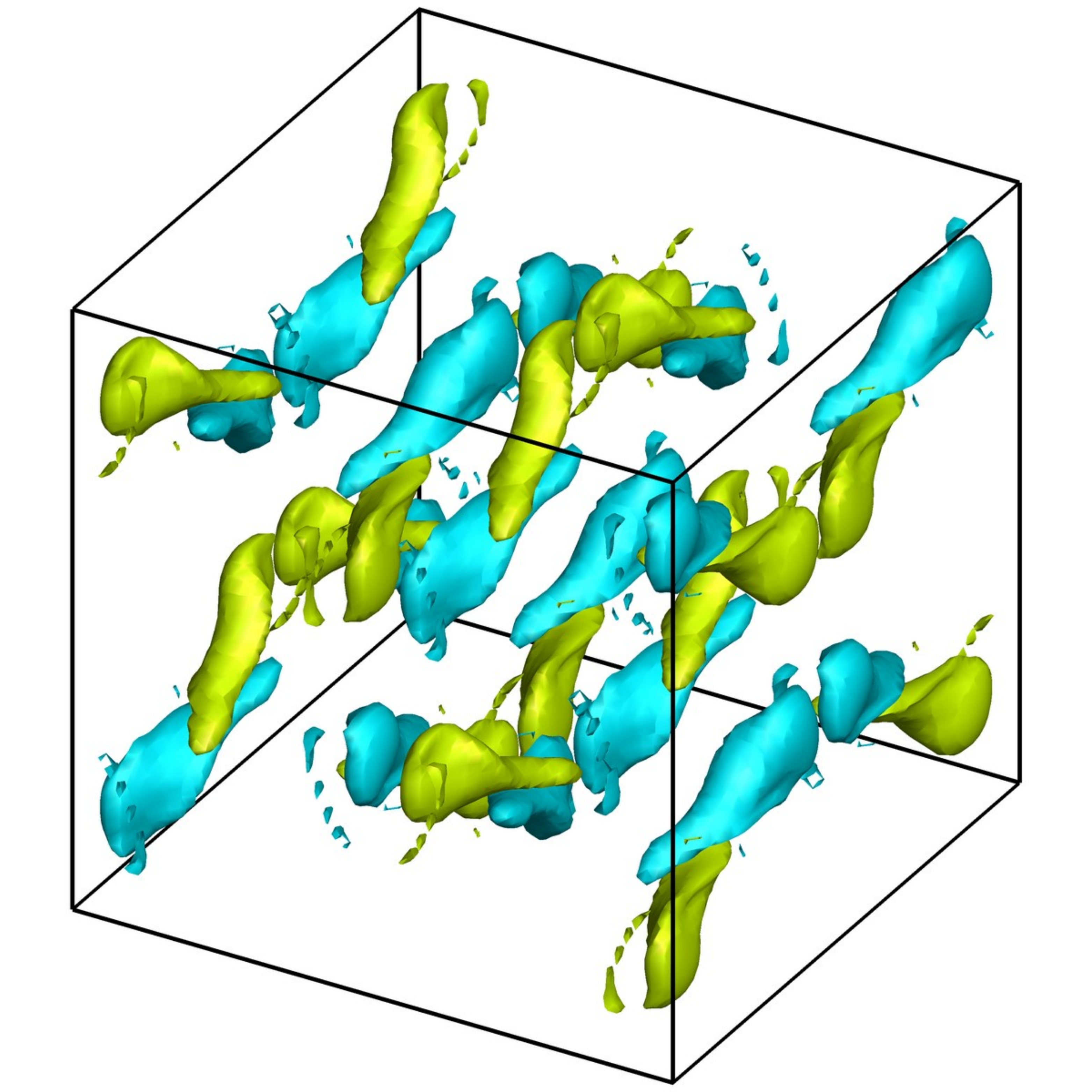}}
\subfigure[~AD-SF]{\includegraphics[width=0.27\textwidth]{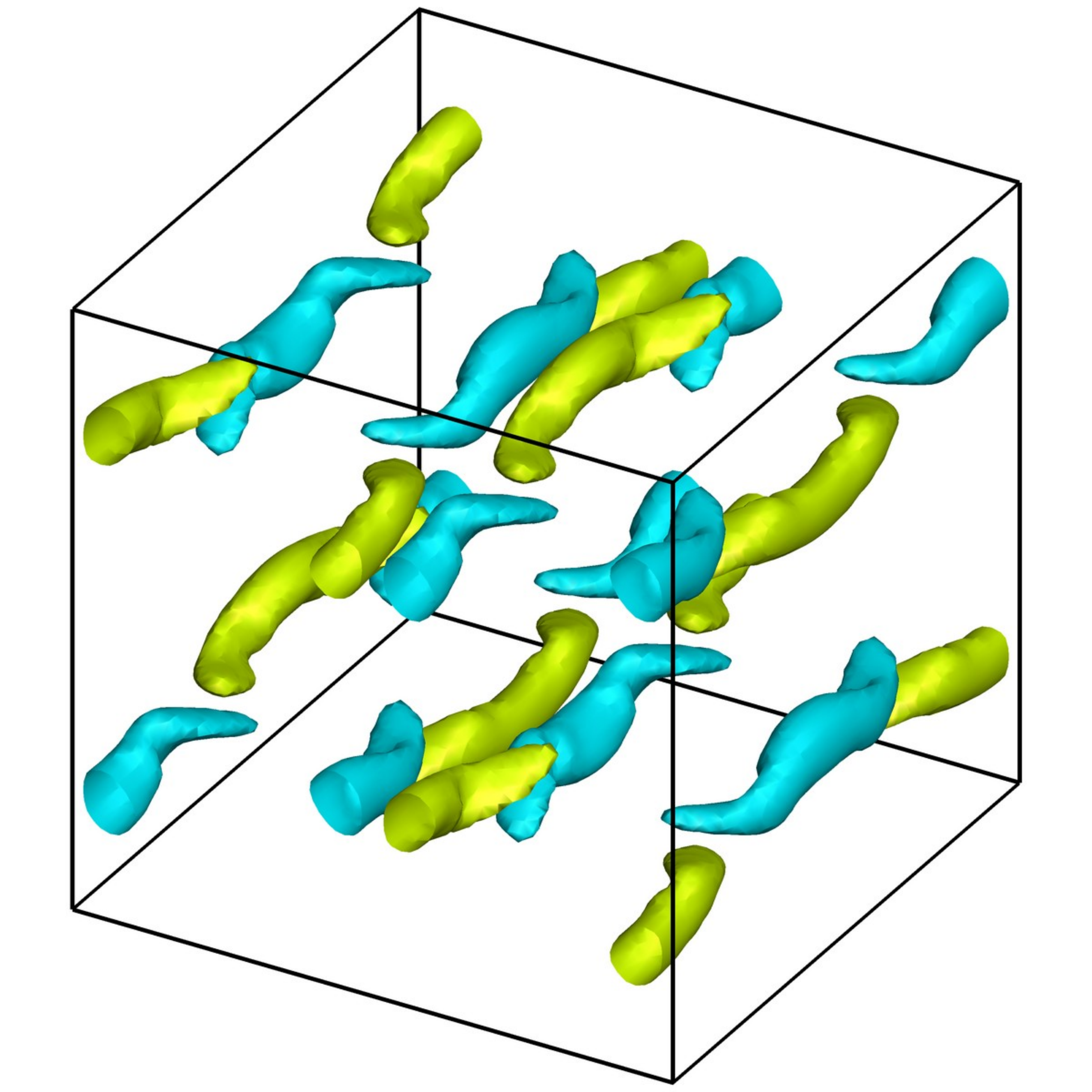}}
\subfigure[~AD-7PF]{\includegraphics[width=0.27\textwidth]{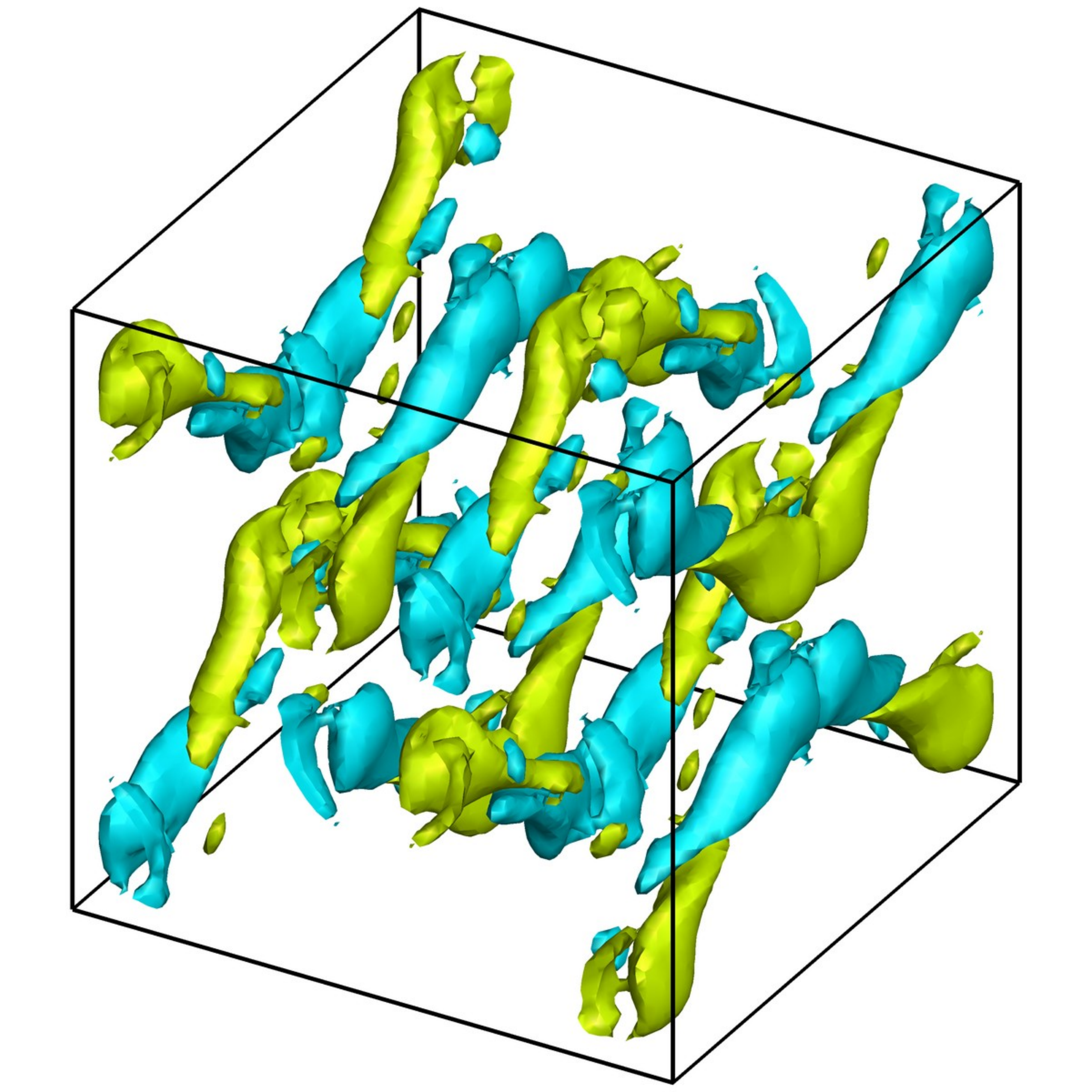}}
}
\mbox{
\subfigure[~AD-PF ($\alpha=0.25$)]{\includegraphics[width=0.27\textwidth]{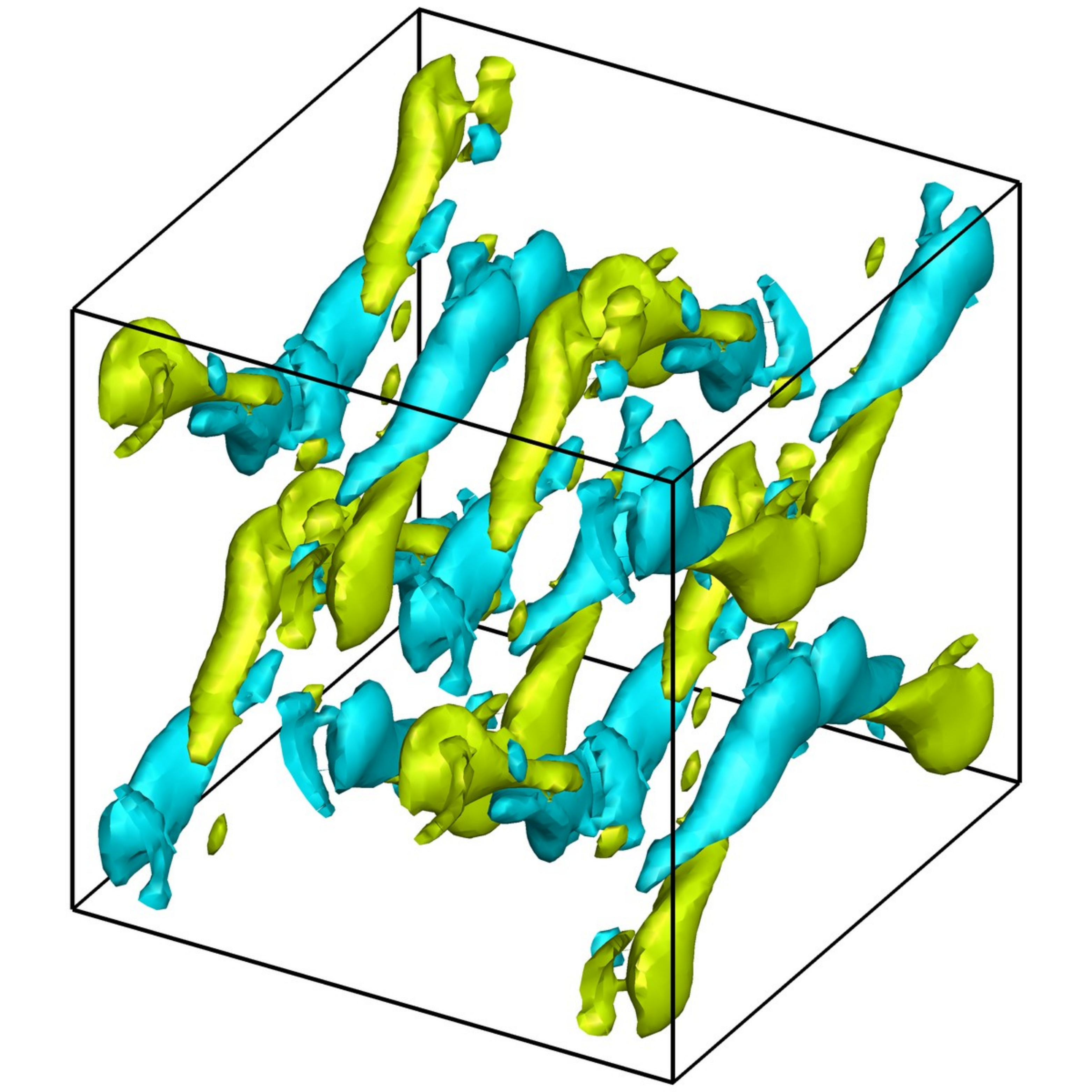}}
\subfigure[~AD-PF ($\alpha=0.15$)]{\includegraphics[width=0.27\textwidth]{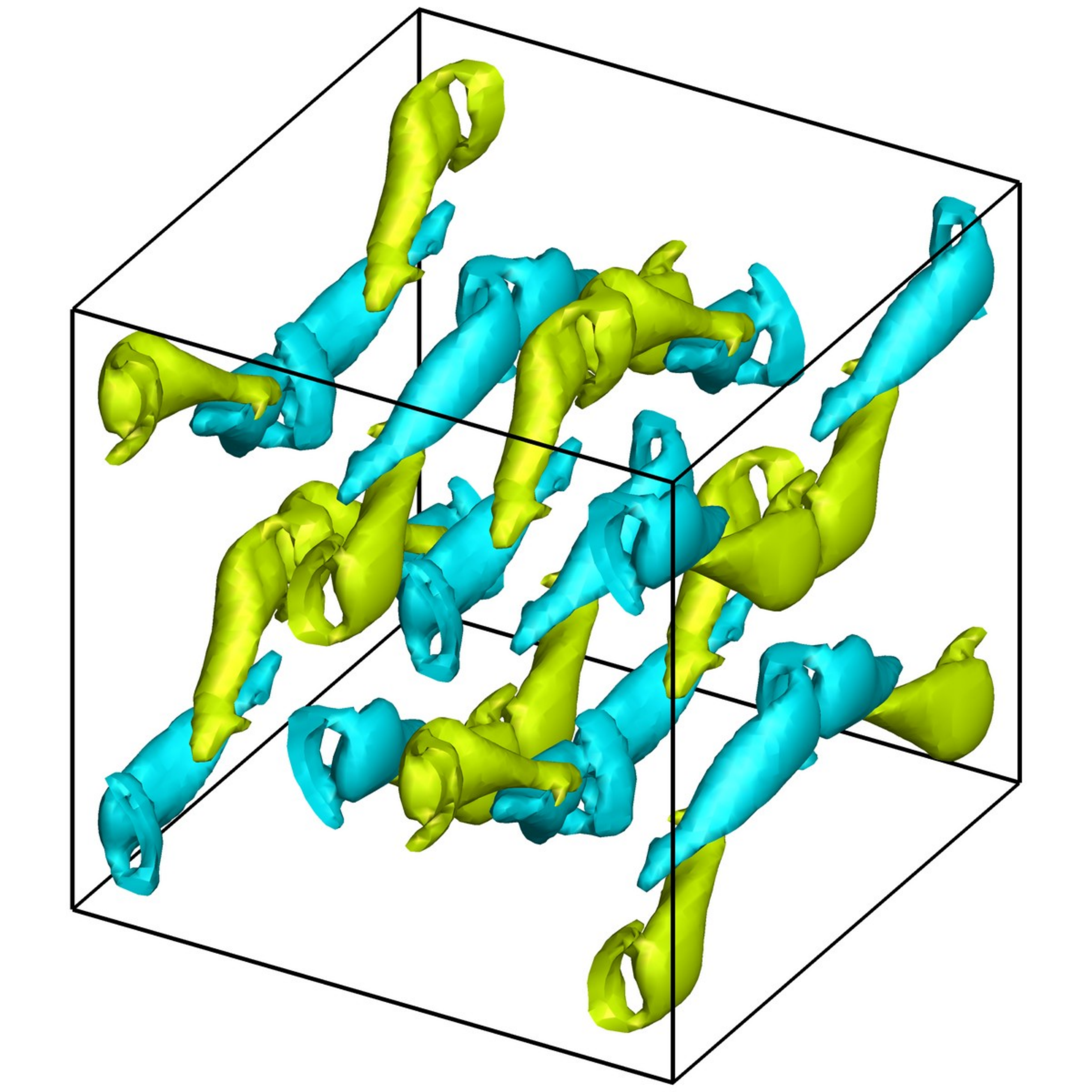}}
\subfigure[~AD-PF ($\alpha=-0.15$)]{\includegraphics[width=0.27\textwidth]{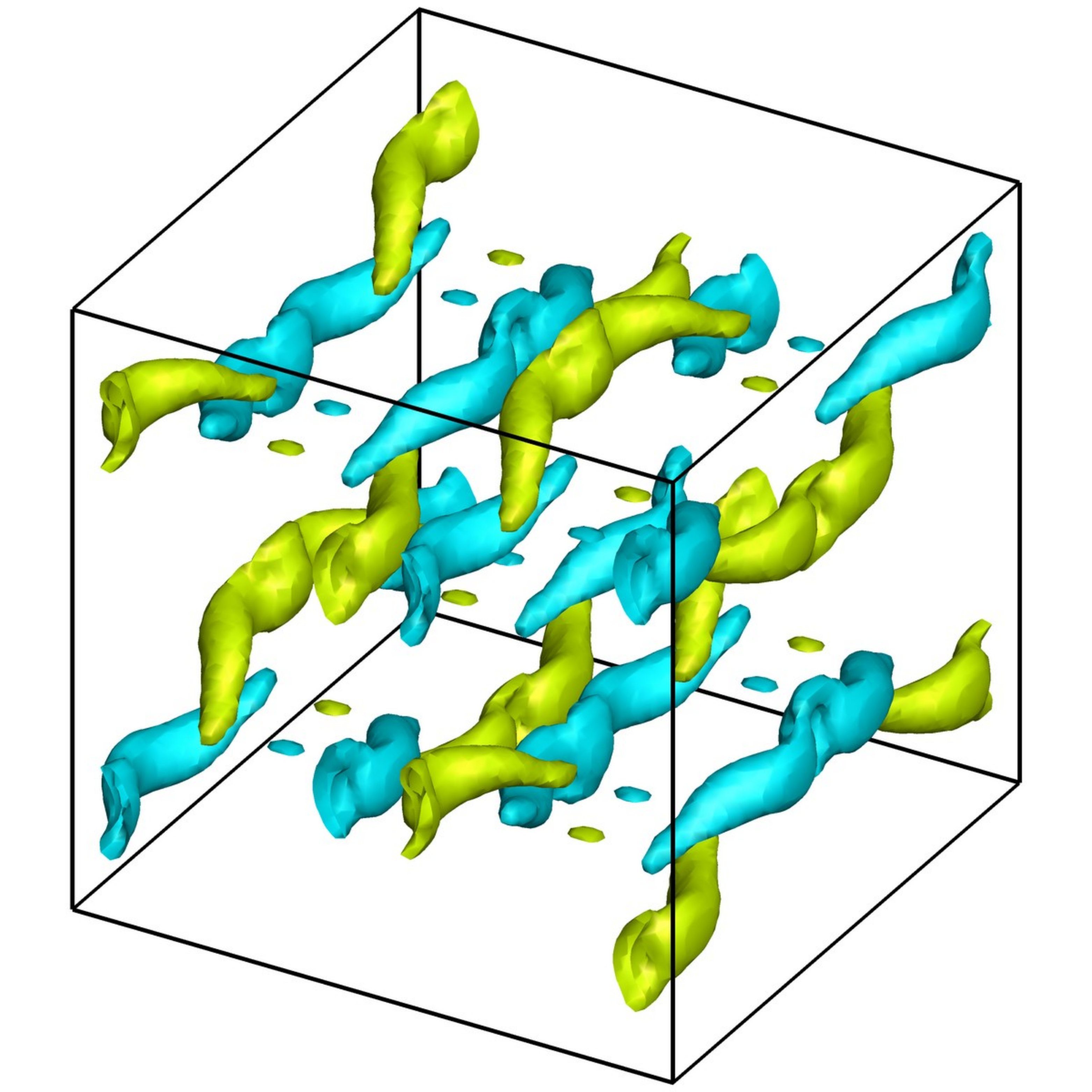}}
}
\mbox{
\subfigure[~AD-DF ($\gamma=0.5$)]{\includegraphics[width=0.27\textwidth]{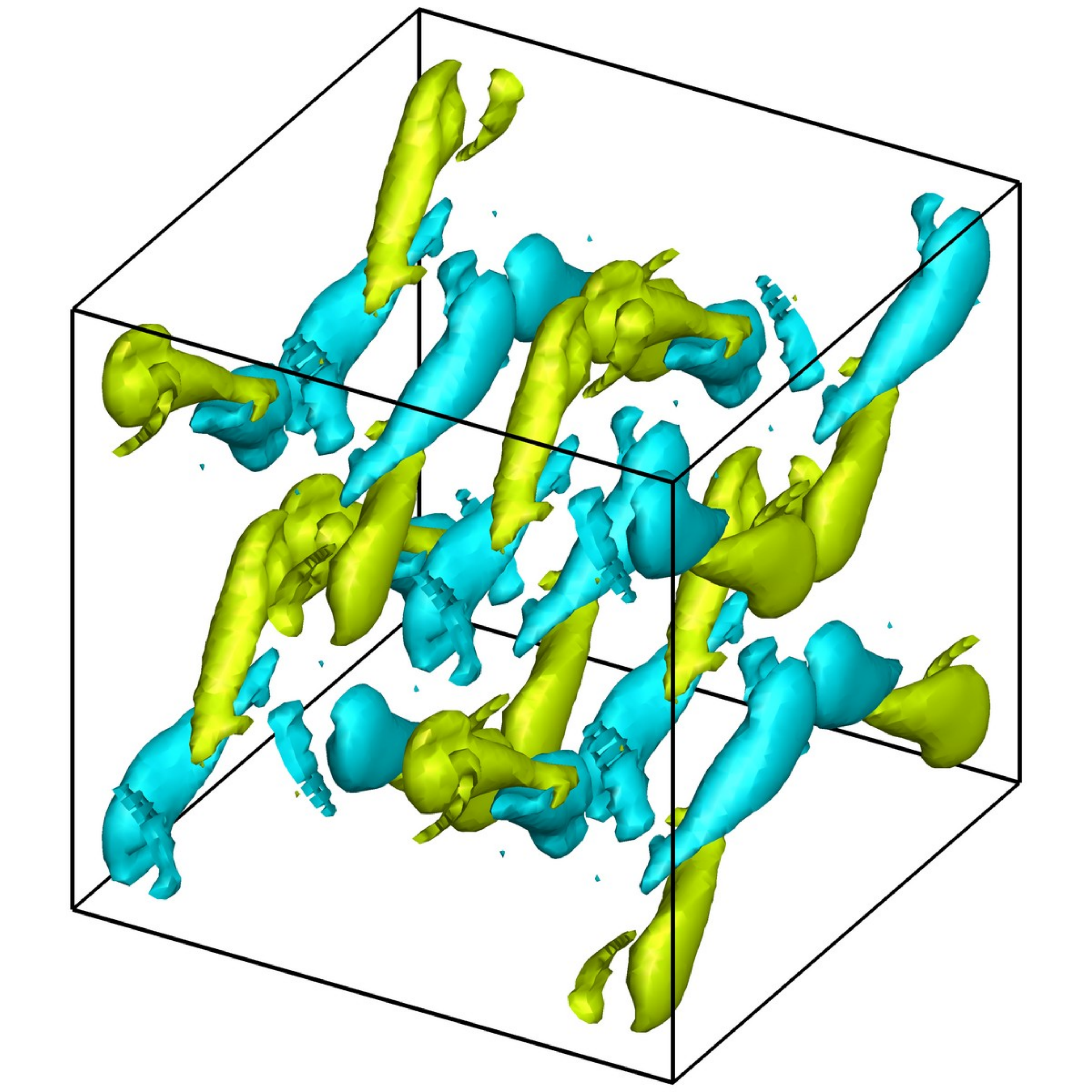}}
\subfigure[~AD-DF ($\gamma=0.8$)]{\includegraphics[width=0.27\textwidth]{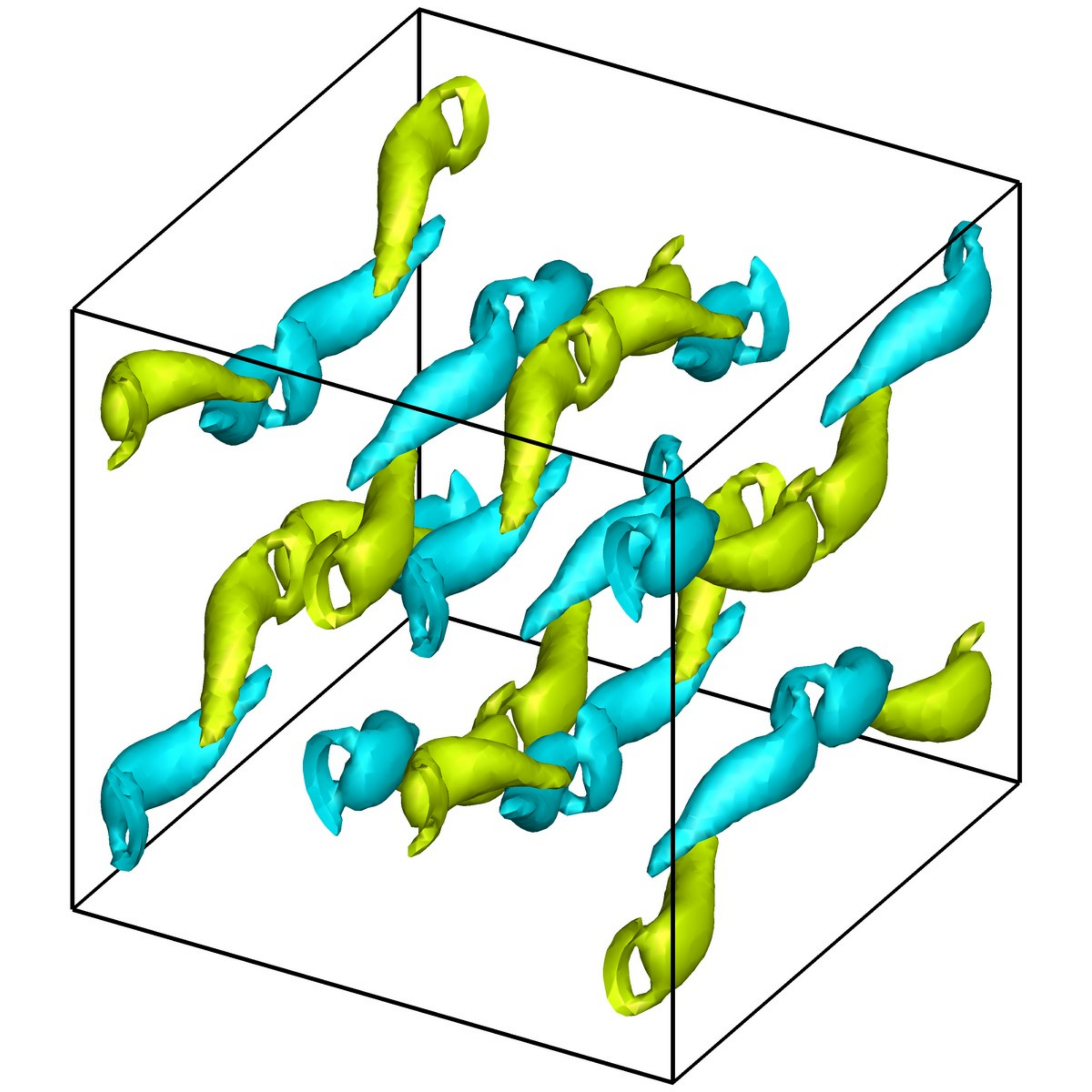}}
\subfigure[~AD-DF ($\gamma=1.0$)]{\includegraphics[width=0.27\textwidth]{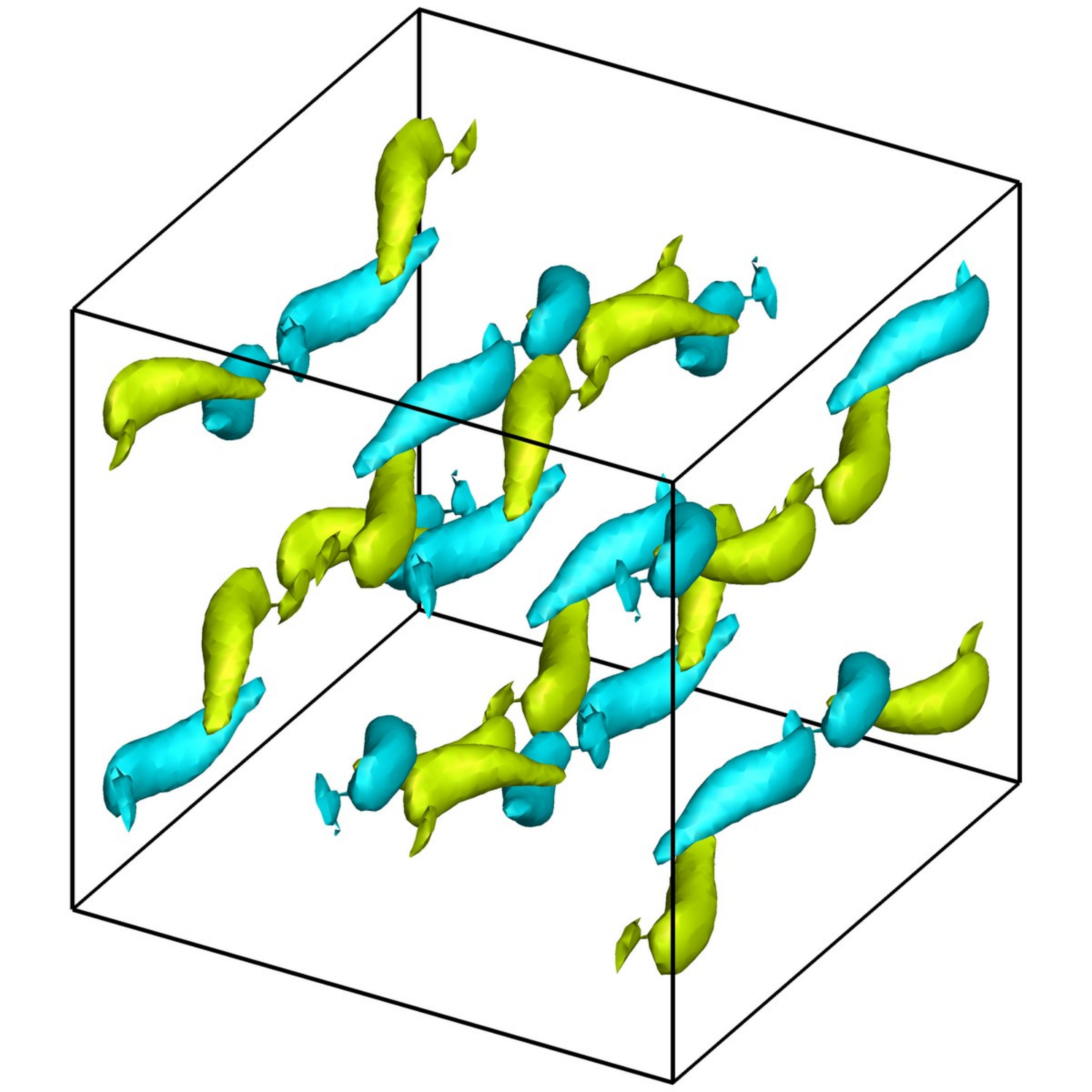}}
}
\caption{Iso-surfaces of $\omega_x= \pm 2.0$ at time $t=10$ using $N=5$ for the AD-LES models.}
\label{fig:time-tgv38n}
\end{figure}

\section{Summary and Conclusions}
\label{sec:sum}


A selection of discrete low pass spatial filters has been evaluated for the approximate deconvolution large eddy simulation (AD-LES) of homogeneous incompressible three-dimensional flows.
Four families of filters were considered: (i) box filters, (ii) Pad\'{e}-type filters, (iii) differential filters, and (iv) hyper-differential filters. Fourier analyses have been performed to compute the filter transfer functions, which relate the resolved quantities to unresolved quantities. The AD-LES model equipped with these four classes of spatial filters was tested on the three-dimensional Taylor-Green vortex problem, and the results were compared with direct numerical simulation (DNS) data for the same problem. An under-resolved numerical simulation (No-AD) was also used for comparison purposes. Detailed sensitivity analyses of the filter parameters have been presented. Four criteria were used to assess the numerical results:
(i) the time series of the volume-averaged enstrophy;
(ii) the volume-averaged third-order structure function;
(iii) the $L^2$-norm of the velocity and vorticity errors; and
(iv) the volume-averaged velocity and vorticity correlation coefficients.


The numerical results yielded the following conclusions.
For all cases, the AD-LES model produced more accurate results than No-AD and had a computational cost that is significantly lower than the DNS cost.
The numerical results with the AD-LES model displayed a significant sensitivity with respect to the spatial filter employed, the filter parameters, and the order of the AD procedure, $N$.
Besides these conclusions, it is hard to draw any other general conclusions that cover all cases.
Thus, we first discuss each type of filter separately.
Among the box filters, the SF with a high AD order ($N=5$) consistently performs the best.
For the Pad\'{e} filters, a negative value of the parameter ($\alpha = -0.15$) and a low AD order ($N=2$) consistently yield the best results.
For the differential filters, a high value of the parameter ($\gamma = 1.0$) and a low AD order ($N=2$) consistently yield the best results.
For the HDF, a low HDF order ($m=4$), a high value of the parameter ($\gamma = 1.0$), and a low AD order ($N=2$) yield the best results.

Among the filters considered, the HDF yields the most accurate results, although the DF also produces accurate results.
This can been seen in Fig.~\ref{fig:time-tgv38n}, which presents results for the instantaneous iso-surfaces of the $x$-component of the vorticity, $\omega_x$.
All the AD-LES models are more accurate than the No-AD run, yielding results that are relatively close to the DNS data.
Fig.~\ref{fig:time-tgv38n} clearly shows that the most accurate results are obtained with the AD-HDF model.
All the other models are either inaccurate (e.g., AD-7PF and AD-PF), or overly-dissipative (e.g., AD-SF and AD-DF).
Given that both the the AD-HDF and AD-DF have a computational cost that is much lower than that of a DNS, the HDF and the DF appear as appropriate choices in the AD-LES framework.
We note that there is no surprise that the DF's performance is close to that of the HDF, since the latter in fact reduces to the former when $m=1$.
The numerical results also yield the following general conclusion:
Although a careful parameter choice makes each class of filters competitive, it seems that filters whose transfer function resembles that of the Fourier cut-off filter (such as the HDF) tend to perform best.

Finally, we emphasize that the entire numerical study was centered around a second-order finite difference discretization used in both the DNS and the AD-LES model.
Although this type of discretizations can be encountered in numerous practical applications, we emphasize that higher-order discretizations could (and probably should) be used in the AD-LES framework (see, e.g., Drikakis et al. \cite{drikakis2009large} and Habisreutinger et al. \cite{habisreutinger2010grid} for first steps in this direction).
Thus, a natural question is whether the qualitative conclusions drawn from this numerical study extend to the higher-order numerical discretization case, and if they do, to what extent these conclusions carry over.
We plan to investigate these issues in a follow-up study.



\bibliographystyle{abbrv}
\bibliography{references}







\end{document}